\documentclass[preprint]{aastex}

\usepackage{amssymb,amsmath}
\usepackage{morefloats}
\usepackage{graphicx}
\usepackage{natbib}

\newcommand{\valf}{v_\text{A}}

\newcommand{\be}{\begin{equation}}
\newcommand{\ee}{\end{equation}}

\newcommand{\gray}{$\gamma$-ray}

\newcommand{\hi}{H~{\sc i}}
\newcommand{\hii}{H~{\sc ii}}

\newcommand{\Xco}{$X_{\rm CO}$}

\newcommand{\adv}{Adv.\ Space Res.}

\newcommand{\fermilat}{{\it Fermi}--LAT}
\newcommand{\fermi}{{\it Fermi}}
\newcommand{\hess}{{\it H.E.S.S.}}
\newcommand{\gardian}{{\it GaRDiAn}}
\newcommand{\GP}{\mbox{\it GALPROP}}

\newcommand{\ebv}{{\it E(B$-$V)}}
\newcommand{\model}[5]{$^{\rm S}$#1$^{\rm Z}#2^{\rm R}#3^{\rm T}#4^{\rm C}#5$}

\shorttitle{\fermilat{} Observations of the Diffuse \gray{} Emission}
\shortauthors{Ackermann~et~al.}

\begin{document}



\title{\fermilat{} Observations of the Diffuse \gray{} Emission: Implications for Cosmic Rays and the Interstellar Medium}

\author{
M.~Ackermann\altaffilmark{1}, 
M.~Ajello\altaffilmark{2}, 
W.~B.~Atwood\altaffilmark{52},
L.~Baldini\altaffilmark{3}, 
J.~Ballet\altaffilmark{4}, 
G.~Barbiellini\altaffilmark{5,6}, 
D.~Bastieri\altaffilmark{7,8}, 
K.~Bechtol\altaffilmark{2}, 
R.~Bellazzini\altaffilmark{3}, 
B.~Berenji\altaffilmark{2}
% ...
}
\altaffiltext{1}{Deutsches Elektronen Synchrotron DESY, D-15738 Zeuthen, Germany}
\altaffiltext{2}{W. W. Hansen Experimental Physics Laboratory, Kavli Institute for Particle Astrophysics and Cosmology, Department of Physics and SLAC National Accelerator Laboratory, Stanford University, Stanford, CA 94305, USA}
\altaffiltext{3}{Istituto Nazionale di Fisica Nucleare, Sezione di Pisa, I-56127 Pisa, Italy}
\altaffiltext{4}{Laboratoire AIM, CEA-IRFU/CNRS/Universit\'e Paris Diderot, Service d'Astrophysique, CEA Saclay, 91191 Gif sur Yvette, France}
\altaffiltext{5}{Istituto Nazionale di Fisica Nucleare, Sezione di Trieste, I-34127 Trieste, Italy}
\altaffiltext{6}{Dipartimento di Fisica, Universit\`a di Trieste, I-34127 Trieste, Italy}
\altaffiltext{7}{Istituto Nazionale di Fisica Nucleare, Sezione di Padova, I-35131 Padova, Italy}
\altaffiltext{8}{Dipartimento di Fisica ``G. Galilei", Universit\`a di Padova, I-35131 Padova, Italy}
% ...
\altaffiltext{52}{Santa Cruz Institute for Particle Physics, Department of Physics and Department of Astronomy and Astrophysics, University of California at Santa Cruz, Santa Cruz, CA 95064, USA}
% ...

\begin{abstract}

The \gray{} sky $>100$ MeV is dominated by the diffuse emissions 
from interactions of cosmic rays with the interstellar gas and 
radiation fields of the Milky Way.
Observations of these diffuse emissions provide a tool to study cosmic-ray 
origin and propagation, and the interstellar medium.
We present measurements from the first 21 months of the \fermilat{} mission and 
compare with models of the diffuse \gray{} emission generated using 
the \GP{} code.
The models are fitted to cosmic-ray data and incorporate astrophysical
input for the distribution of cosmic-ray sources, interstellar 
gas and radiation fields.
To assess uncertainties associated with the astrophysical input, a grid of 
models is created by varying within observational limits the distribution 
of cosmic-ray sources, the size of the cosmic-ray 
confinement volume (halo), and the distribution of interstellar gas.
An all-sky maximum-likelihood fit is used to determine the \Xco-factor, 
the ratio between integrated CO-line intensity and H$_2$ column density, 
the fluxes and spectra of the \gray{} point 
sources from the first \fermilat{} catalogue, and 
the intensity and spectrum of the isotropic background including residual 
cosmic rays that were misclassified as \gray{s}, all of which 
have some dependency on the assumed diffuse emission model.
The models are compared on the basis of their maximum likelihood ratios
as well as spectra, longitude, and latitude profiles. 
We also provide residual maps for the data following subtraction of the 
diffuse emission models.
The models are consistent with the data at high and intermediate
latitudes but
under-predict the data in the inner Galaxy for energies above a few GeV.
Possible explanations for this discrepancy are discussed, 
including the contribution by undetected point source populations and spectral
variations of cosmic rays throughout the Galaxy.
In the outer Galaxy, we find that the data prefer models with
a flatter distribution of cosmic-ray sources, a 
larger cosmic-ray halo, or greater gas density than is usually assumed.
Our results in the outer Galaxy are consistent with other 
\fermilat{} studies of this region that used different analysis methods 
than employed in this paper.

\end{abstract}


\section{Introduction}

The diffuse Galactic \gray{} emission (DGE) is produced by 
cosmic-ray (CR) particles interacting with the gas and radiation fields
in the interstellar medium (ISM). 
As has been recognised since the late 1950s \citep{Morrison1957},
measurements of the DGE can be used
to study CR origin and propagation in the Galaxy, and also to probe the 
content of the ISM itself, independent of other methods. 
They are complementary to direct measurements of CRs by balloons and 
satellites, and to radio astronomical surveys of synchrotron radiation that
is produced by CR electrons and positrons losing energy in the Galactic 
magnetic field.

The first \gray{} observations made by the OSO-3 
satellite \citep{1972ApJ...177..341K} showed emission in the inner Galaxy.
The breakthrough came with the SAS-2 \citep{Fichtel:1975} 
and COS-B \citep{Bignami1975a} instruments,  
whose Galactic plane surveys above 100~MeV allowed testing of DGE 
models based on CRs and their interactions in the ISM \citep[e.g.,][]{PugetStecker1974,Bignami1975b,Stecker1975,Bloemen1986,Strong1988}.
The COMPTEL and EGRET instruments on the {\it Compton Gamma-Ray Observatory} 
provided higher-quality data covering the entire sky in the energy 
range 1~MeV to 10~GeV, which stimulated more detailed modelling \citep{1997ApJ...481..205H,SMR00,2004ApJ...613..962S,2004A&A...422L..47S}.
Recently, the SPI instrument on the {\it INTErnational Gamma-Ray Astrophysics
Laboratory (INTEGRAL)} observatory has extended the 
observations of CR-induced diffuse emissions into the 
hard X-ray range \citep{2008ApJ...679.1315B,Bouchet2011}, 
while ground-based instruments \citep{HessGC, MilagroPlain} 
have detected emission at TeV energies from the Galactic plane and 
Galactic centre regions that are likely to have a CR-induced origin.
For a recent pre--\fermi{} review of the subject 
see \citet{2007ARNPS..57..285S}.

The \fermi{} Large Area Telescope (\fermilat) 
provides a view of the entire \gray{} sky from 30 MeV to beyond 
several hundred GeV with a sensitivity surpassing its predecessor instrument, 
EGRET, by more than an order of magnitude.
Studies of the DGE by the \fermilat\
collaboration have so far concentrated on specific regions of the sky.
At intermediate Galactic latitudes, 
\fermilat{} observations did not confirm the EGRET ``GeV excess'' and 
showed that the spectrum was consistent with a DGE model based on 
measured CR spectra \citep{LAT:GeVExcess}. 
The emissivity of nearby \hi{} gas was derived from analysis of a selected
high-latitude region \citep{LAT:HIEmissivity} and found to agree with the
emissivity calculated assuming measured CR spectra.
Analysis of data for the 2$^{\rm nd}$ and 3$^{\rm rd}$ Galactic 
quadrants \citep{2010ApJ...710..133A,2011ApJ...726...81A} showed a 
higher-than-expected \hi{} emissivity in the outer Galaxy with respect to 
the DGE 
model used by \citet{LAT:HIEmissivity}.
These studies of the outer Galaxy also indicated a lower \Xco-factor 
(the ratio between integrated CO-line intensity and H$_2$ column density) 
compared
to that used by \citet{Strong:2004}.
The early studies with the \fermilat{} data 
have systematically moved from understanding the DGE 
produced by CRs interacting with the relatively nearby ISM to
progressively larger regions of the Galaxy.

Modelling the DGE requires knowledge of 
CR intensities and spectra, along with the distributions of interstellar gas
and radiation fields, throughout the Galaxy.
Starting from the distribution of CR sources and the injection 
spectra of CRs, the CR intensities and spectra can be estimated by 
modelling their propagation in the Galaxy, taking into account relevant 
energy losses and gains.
The resulting CR distributions are then folded with the target 
distributions of the interstellar gas and radiation fields 
to calculate the DGE 
\citep[e.g.,][]{2004A&A...422L..47S}.
Defining the input distributions and calculating the models is not a 
trivial task and involves analysis of data from a broad range of 
astronomical and astroparticle observatories.

Cosmic-ray propagation models can be constrained to a certain extent 
using measurements of 
CRs in the solar system 
\citep[see e.g.,][for a recent review]{2007ARNPS..57..285S}.
For an assumed propagation model, e.g., plain diffusion, 
diffusive-reacceleration, diffusion-convection, etc., the propagation 
parameters of the model and size of the CR confinement region can be derived 
by comparing the modelled secondary-to-primary CR nuclei ratios with data.
However, some key components are difficult to constrain with this method.
An example is the CR source distribution because the CR data essentially probe
only relatively nearby sources, and not the Galaxy-wide distribution.
Because \gray{s} are undeflected by magnetic fields and absorption in
the ISM is negligible below $\sim 10$ TeV \citep{2006ApJ...640L.155M}, 
the DGE is a direct probe of the CR intensities 
and spectra in distant locations, allowing the study of the Galactic
distribution of CR sources.

In this paper, we analyse the DGE from the full sky observed with the \fermilat.
We use data from the first 21 months of observations with \fermilat{} 
for energies 200 MeV to 100 GeV
that contain the lowest fraction of background events 
in the \fermilat{} data. 
The DGE is modelled using the \GP{} code \citep[see e.g.,][and references therein]{1998ApJ...493..694M, SMR00,2004ApJ...613..962S, 2006ApJ...642..902P, Porter:2008, Vladimirov:2011}.
We create a grid of DGE models by varying the CR source distribution, the 
CR halo size, and the distribution of interstellar gas.
The models are constrained to reproduce directly measured CR data, and 
then compared to the \gray{} data using an iterative maximum-likelihood 
fitting procedure. 
Our fits allow for variations in the \Xco--factor, the fluxes and spectra 
of the point sources in the first \fermilat\ 
catalogue \citep[1FGL;][]{FermiCatalog}, the intensity and 
spectrum of an isotropic \gray{} 
background component, and scaling of the optical and infrared component of the 
interstellar radiation field (ISRF).
We compare the likelihood of the models, and match observed and predicted 
intensities and spectra for various regions of the sky.
Maps of the residual \gray{} emissions after subtraction of the DGE 
models are presented with the discussion of 
possible interpretations.
 
Our study is necessarily limited due to the (potentially) large number of 
DGE model parameters.
Only diffusive-reacceleration models are considered for the CR 
propagation, even though 
models with convection and even plain diffusion models can in some cases 
provide an equivalent fit to the CR 
data~\citep[][]{2007ARNPS..57..285S}.
We assume a smooth distribution of CR sources with homogeneous 
injection spectra, although we expect CRs to originate in discrete 
sources and show 
variability in their emission spectra \citep{2010ApJ...722.1303A, 2010ApJ...718..348A, 2010ApJ...712..459A, 2010Sci...327.1103A}.
Because the homogeneity assumption tends to mainly affect the CR 
electrons, this is 
a very good approximation for the source distributions of the CR nuclei 
producing the bulk of the DGE in the energy range
considered in this paper. 
Our propagation and emissivity calculations are limited to two dimensions (2D).
Depending on the correlation of CR sources and gas densities, 
three-dimensional (3D) calculations 
taking the spiral arm structure of the Galaxy into account might 
quantitatively change the results of the paper
even though the azimuthal averaged distributions are not changed.  
Our
qualitative conclusions are, however, independent of these assumptions.

The intent of this paper is not to find the perfect DGE model, 
but rather to test a selection of astrophysically motivated 
models and their compatibility with \gray{} observations.
An essential aspect of this study is also assessing the impact of 
uncertainties involved in the many astrophysical inputs needed for a 
proper calculation of the DGE.
Finally, we compare our results with the earlier mentioned analyses by
the \fermilat{} collaboration that have used different methods to this 
paper.

\section{Data Preparation}

The \fermilat{} instrument, event reconstruction, and response are 
described by \citet{Atwood:2009}.
In this paper, 
we use the Pass 6 DataClean event selection and instrument response functions (IRFs) employed in 
\citet{EGBpaper} to derive the isotropic \gray{} intensity and spectrum.
The DataClean event selection is used for our analysis because
of the greatly reduced CR background, particularly in the 10--100 GeV energy
range, compared to the standard low-background Diffuse class 
event selection \citep{Atwood:2009}.
The slight reduction in effective area compared to the Diffuse class IRFs 
is not a limitation for this analysis because the data are not limited by 
counting statistics for energies $<100$ GeV.
The reduced CR background is especially important at high latitudes 
where the \gray{} signal is weakest.
For the description of the procedure to select this data and generate the 
IRFs, we refer to \citet{EGBpaper}.
Note, that we restrict the upper energy range of our analysis to 
100~GeV because the CR background for the DataClean event 
selection is determined only up to this energy.

We use 21 months of data, starting from 2008 August 5 to 
2010 May 4.
To minimise the contribution from the very bright Earth limb, we apply a 
maximum zenith angle cut of $100^\circ$.  
In addition, we also 
limit our data set to include only photons with an incidence angle 
from the instrument $z$-axis of $<65^\circ$.  
The rejection power for
CRs is reduced at large incident angles and the fraction of events
converting in the thick tracker layers increases causing the effective
PSF to worsen significantly \citep{Atwood:2009}.  
The signal loss is minimal
because the effective area for \gray{s} is reduced significantly 
at high incidence 
angles.
Exposure maps and the point spread function (PSF) for the pointing 
history of the observations were generated using the standard 
\fermilat{} ScienceTools package available from the \fermi{} Science 
Support Center\footnote[1]{http://fermi.gsfc.nasa.gov/ssc/data/analysis/}.
We use the \gardian{} package (see Appendix~\ref{appendix:gardian} 
for a description)
to process the data and exposure maps to produce all-sky intensity maps, and 
the same package for the maximum-likelihood fitting procedure.
The counts are spatially binned with a 
HEALPix\footnote[2]{http://healpix.jpl.nasa.gov} 
order-7 isopixelisation scheme giving an angular 
resolution $\approx 0.5^\circ$ \citep{Gorski:2005}.
For the fitting procedure, the data are binned with 9 equally-spaced 
logarithmic bins between 200 MeV and 100 GeV.
This relatively coarse energy binning is used to ensure stable fits for 
the point sources that have a free scaling parameter for each energy bin.
The intensity maps are created by dividing the counts map with the exposure
that is weighted using a DGE model.
The model weighting of the exposure is required because of the strong energy 
dependence of the exposure at energies below $\sim1$~GeV
\citep{EGBpaper}. 
However, this causes insignificant variations between the intensities for 
the different DGE models considered in this paper due to 
the similarity of their spectra.
Note, for our plots of the intensities for the DGE models and 
data, we use 15 equally-spaced logarithmic energy bins for the same energy 
range to better utilise the energy resolution of the \fermilat\
instrument.

When comparing the data and models we perform the maximum likelihood fit in 
photon space, forward folding the models to create the expected counts, 
using the exposure maps and PSF as described in 
Appendix~\ref{appendix:gardian}.
To ensure we properly take the PSF into account for the spectra and 
longitude/latitude profiles, 
we calculate the model intensity maps from their expected counts in the 
same way as the intensity maps of the observed counts.
This ensures that the comparison of intensity and photon counts is consistent. 
When selecting special regions of the sky, all pixels whose 
centres are within the region are used.
Because the pixel size is significantly smaller than these regions 
the edge effects are minimal.

The systematic error in the effective area of the \fermilat{} is estimated 
to be 10\%
below 100~MeV, 5\% at 562~MeV, and 20\% above 10~GeV with linear interpolation
in logarithm of energy between the values \citep{EGBpaper}.  
The systematic error is not taken
into account in the parameter estimates but is included in the figures
below which compare
the spectra and profiles of the models to the data.

The selection corresponding to the 
Pass 7\footnote[3]{http://fermi.gsfc.nasa.gov/ssc/} data used for the 
second \fermilat{} source catalogue (2FGL) \citep{FermiCatalogue2} was not 
available at the start of the analysis presented in this paper.  
The effect of the improved IRFs
in the several-GeV energy range was tested by repeating the last step of the
analysis for a single model using the Pass 7 clean photon class.  
The data was prepared in the same way as the Pass 6 DataClean photons
described above.  
The results of the test are described in Section~\ref{sec:spectra}.

\section{Grid Setup and Analysis Procedure}

We use a ``conventional'' CR propagation 
model paradigm where a set of 
CR propagation models is created that reproduce locally measured CR 
intensities and spectra.  
All of the models investigated in this paper are based on, and constrained 
by, a variety of non-\gray{} data: CRs measured near the Earth, 
the distribution of potential CR sources in the Galaxy derived from 
observations, and the 
distributions of interstellar gas and radiation fields from survey data and 
modelling.
We summarise these details below.

\subsection{Models of the Diffuse Galactic \gray{} Emission}

We use the recently released
version~54 (v54) of the \GP{} code \citep{Vladimirov:2011}
to create models of the DGE. 
We limit ourselves to models using diffusive reacceleration with no
convection for a Kolmogorov spectrum of interstellar turbulence as has
been successfully used to explain CR data and the EGRET \gray{}-sky
\citep{2004ApJ...613..962S} as well as
INTEGRAL data \citep{Porter:2008}.  
For a detailed description of the \GP{} code and the improvements in v54 with
respect to earlier versions we refer the reader
to the dedicated website\footnote[4]{http://galprop.stanford.edu}.

The parameter files of the \GP{} models used in this paper are available in
the Supplementary Material to this paper.  
These give a precise definition of
the models used which can be reproduced as required.  
Note that only one
scaling factor for the ISRF is given in each file which is the average of the
local and inner scaling factors found from the fit (see
Section~\ref{sec:ISRF} and Section~\ref{sec:Regions}).

The collection of models used in this analysis is created using the CR source 
distributions and propagation parameters as described in 
Section~\ref{sec:CRs} for different sizes
for the CR confinement region: for an assumed cylindrical geometry where 
the Sun is located in the Galactic plane 8.5~kpc from the Galactic centre, we 
use radial boundaries, $R_h$, of 20~kpc and 30~kpc, and vertical
boundaries (halo size), $z_h$, of $4, 6, 8,$ and 10~kpc, respectively.
In addition, we use two assumptions for the optical depth
correction of the \hi{} component (see Section~\ref{sec:hi}) and also two
values for the cut at
which dust emission is no longer used to correct the total column density (see
Section~\ref{sec:dust}).  
This results in a total of 128 models.

\subsection{Cosmic-Ray Injection and Propagation}
\label{sec:CRs}

\begin{figure}
\plotone{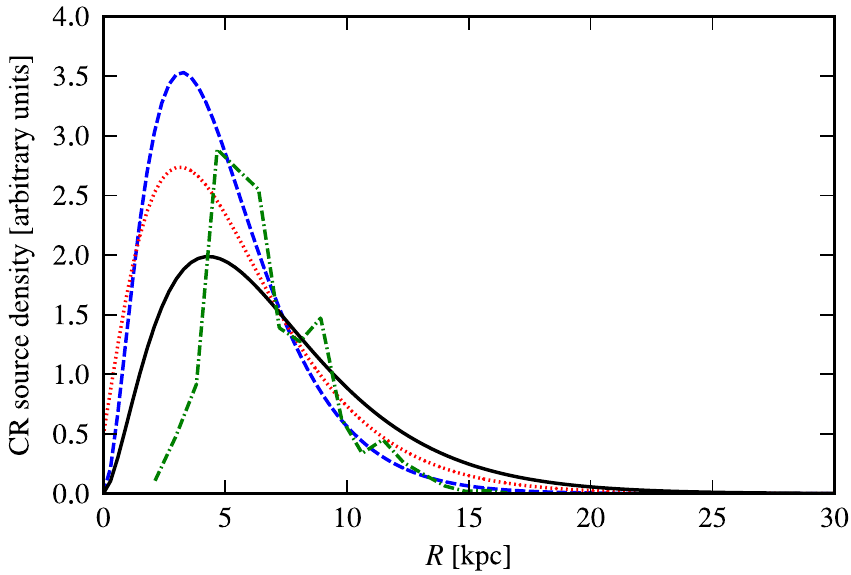}
\caption{CR source density at $z=0$ in arbitrary units as a
function of Galactocentric radius.
Solid black curve: SNRs \citep{1998ApJ...504..761C}. 
Dashed blue curve: Pulsars \citep{2006MNRAS.372..777L}. 
Dotted red curve: Pulsars \citep{2004A&A...422..545Y}. 
Dash-dotted green curve: OB stars \citep{2000A&A...358..521B}.
While the units are arbitrary, the relative normalisations of the curves 
in the figure match those found in the \GP{} models used in this analysis.  The
CR flux of the models is
normalised to the observed CR flux at the solar circle after propagation.  The
normalisation is done at 100 GeV and is therefore unaffected by modulation.
\label{fig:CRsources}}
\end{figure}

Supernova remnants (SNRs) are believed to be the principal
sources of CRs.
However, their Galactic distribution is not well 
determined \citep{1998ApJ...504..761C,Green2005}.
Therefore we consider, in addition to the measured SNR distribution 
from \citet{1998ApJ...504..761C}\footnote[5]{The validity of 
the \citet{1998ApJ...504..761C} $\sigma$-D relation
has been criticised \citep[e.g.][]{Green2005} but it is used in this paper as an 
alternative to probe the effect of changing the CR distribution.}
(hereafter SNR distribution), other 
tracers of supernovae explosions.
The pulsar distribution is a prime candidate as a proxy distribution, 
because pulsars are a SN explosion end state.
The distribution of pulsars is also better determined than SNRs.
Still, it suffers from observational biases and for that reason we test 
two different pulsar distributions, one from \citet{2004A&A...422..545Y}
(hereafter, Yusifov distribution) and 
another from \citet{2006MNRAS.372..777L} (hereafter, Lorimer distribution).
One of the main differences between the two distributions is the functional 
form used to fit the observational points.
Both have a maximum between $R=0$ and $R=R_\odot$ and fall off exponentially
in the outer Galaxy.  
However, \citet{2006MNRAS.372..777L} force the source spatial distribution 
to zero at $R=0$, whereas it is non-zero in \citet{2004A&A...422..545Y}.
As an additional proxy for the distribution of CR sources, we consider the 
distribution of OB stars from \citet{2000A&A...358..521B} (hereafter, OB stars
distribution). 
OB associations are putative CR acceleration regions and 
these stars are also the progenitors of core collapse supernovae that can 
leave compact objects, such as pulsars.
The four CR source distributions used in this paper 
are plotted in Figure~\ref{fig:CRsources}.

To determine the CR injection spectra and propagation parameters 
we perform a $\chi^2$ fit to CR nuclei, electron, and positron data, using 
the Minuit2\footnote[6]{http://seal.web.cern.ch/seal/snapshot/work-packages/mathlibs/minuit/} minimiser.
To reduce computation time, the fit is done in two parts.
The propagation parameters and CR nuclei injection spectrum is 
found from a fit to the CR nuclei data first.
The electron injection spectrum is then found by fitting to the total 
CR electron and positron spectrum, including the contribution by secondary 
electrons 
and positrons from CR protons and He interacting with the interstellar gas.
Fitting the propagation parameters in the first step 
decreases the computation time because 
nuclei up to $_{14}$Si must be included in the propagation calculation 
for an accurate B/C ratio determination, while for the 
secondary electrons and positrons only protons and He are important. 
Not calculating the secondary electrons and positrons in the fit to the 
propagation parameters 
also saves a considerable amount of time.
We use the CR database created by \citet{2009arXiv0907.0565S} and use 
the datasets and parameters as discussed below.
When comparing the models to the CR data we account for solar modulation 
using the force-field approximation \citep{1968ApJ...154.1011G}.
In addition to the propagation parameters, the modulation 
potential for each experiment
that has data below a few GeV (AMS, BESS, and ACE) is a free parameter during
the fit.  
For the \fermilat{} and HEAO--3 we fixed the modulation at 
300~MeV and 600~MeV, respectively, appropriate for the low solar 
activity during observations by each instrument as observed with the ACE
satellite \citep{CRISModulation}.
Solar modulation is unimportant for JACEE data.

\subsubsection{Protons and Heavier Nuclei}

We assume that the injection spectra for all CR nuclei species are 
described by the same rigidity-dependent function

\begin{equation}
n_p(\rho) \propto
\begin{cases}
\rho^{\gamma_{p,1}} & \rho < \rho_p, \\
\rho^{\gamma_{p,2}} & \rho_p \leq \rho,
\end{cases}
\end{equation}

\noindent
where the indices and break rigidities
are obtained by tuning the model to the observed spectrum of 
CR protons as well as the 
He, C, and O nuclei spectra.
The low-energy intensity and spectrum 
is affected by solar modulation so we use data taken at low periods 
in the solar activity cycle.
The inclusion of nuclei up to O is to ensure 
the major contributor species to the production of the secondaries B and Be 
are properly included.
For the proton and He spectra we use low-energy data 
from BESS \citep{2000ApJ...545.1135S} and high-energy data 
from JACEE \citep{1998ApJ...502..278A}.
For the C and O spectra we use low-energy data 
from ACE \citep{CRISwebsite} and high-energy data from 
HEAO--3 \citep{1990A&A...233...96E}.
To determine the diffusion coefficient, $D_0$, and Alfv\'{e}n 
speed, $\valf$, for an assumed halo size, 
we use the B/C ratio because it is the most accurately 
observed secondary-to-primary ratio.
Low-energy ACE \citep{2000AIPC..528..421D} and high-energy 
HEAO--3 \citep{1990A&A...233...96E} data are used for this ratio.

The break in the CR proton and He spectrum observed by ATIC-2, 
CREAM, and PAMELA
\citep{2008ICRC....2...31W,2010ApJ...715.1400A,2011ApJ...728..122Y,Pamela:protons} is not taken into account in this modelling and
neither are the different spectral indices for protons and He. 
\citet{Vladimirov:Pamela} explore different scenarios for the break and
different indices and find that the \gray{} intensities and 
spectra for their models are smaller than the
systematic uncertainty of the \fermilat{} effective area.

Because we derive constraints on the halo size from the \gray{} data, 
the radioactive secondary ratios are not directly used to fix the 
propagation conditions, as is usually the case in propagation model 
studies.
However, the models are compared to the $^{10}$Be/$^9$Be ratio to 
check the consistency of the constraints derived from the \gray{s}.
The $^{10}$Be/$^9$Be data uncertainties  
are large enough to allow a halo size range from 4 kpc 
to 10 kpc \citep{2007ARNPS..57..285S}, depending on the assumed 
propagation model.
We keep the source abundances of nuclei fixed to values determined 
for ACE data \citep{Moskalenko2008}.

\subsubsection{Electrons and Positrons}

We assume the injection spectrum of primary CR electrons 
is described by the rigidity-dependent function

\begin{equation}
n_e(\rho) \propto 
\begin{cases}
\rho^{\gamma_{e,1}} & \rho < \rho_{e,1}, \\
\rho^{\gamma_{e,2}} & \rho_{e,1} \leq \rho < \rho_{e,2}, \\
\rho^{\gamma_{e,3}} & \rho_{e,2} \leq \rho,
\end{cases}
\end{equation}

\noindent
and use data from AMS \citep{Aguilar2002}, 
\fermilat{} \citep{2009PhRvL.102r1101A,Ackermann:electrons} and 
\hess{} \citep{2008PhRvL.101z1104A, 2009A&A...508..561A} to 
determine the spectral indices and break rigidities\footnote[7]{The 
CR electron spectrum measured by PAMELA~\citep[][]{2011PhRvL.106t1101A} 
is consistent with the \fermilat{} data.}.
Unfortunately, the data are insufficient to constrain the entire parameter
set and unphysical values are obtained if all are freely fit.
We therefore fix the values of 
$\gamma_{e,1} = 1.6$ and $\gamma_{e,3} = 4$, allowing only for freedom in
$\rho_{e,1}$, $\rho_{e,2}$, and $\gamma_{e,2}$.
The $\gamma_{e,1}$ used in this paper is consistent with that employed by
\citet{StrongElena:2011} for modelling the Galactic synchrotron emission, 
while the exact value of the 
high-energy index does not significantly affect our results.
There is a strong correlation between $\gamma_{e,1}$ and the modulation
potential for the AMS data in the fits.  We therefore caution that the derived values for 
the modulation potential are biased, although they are in reasonable
agreement with the values derived by ACE for the same period
\citep{CRISModulation}.
The primary electron injection spectrum 
is dependent on that obtained from the CR nuclei fits
through the size of the confinement volume and corresponding 
propagation parameters, as well as the contribution of
secondary CR electrons and positrons produced by the 
CR nuclei interacting with the interstellar gas.

The overall fit is made to the data as described above. 
The normalisation is essentially determined from the \fermilat{} total 
electron and positron data.
No attempt is made to fit the increasing positron fraction reported by
\citet{PAMELApositrons}.  
These can be neglected because 
the contribution from the excess positrons at high energies to
the \gray{} emission is small.

\subsection{Interstellar Gas and its Tracers}

The DGE in the energy range
considered in this paper has a strong contribution from $\pi^0$-decay 
emission\footnote[8]{See Section~\ref{sec:visuals} 
for a comparison of the contribution of components.}.
The treatment in \GP{} is described in detail by \citet{1998ApJ...493..694M}.  
For proton-proton
interactions we use the formulation described in
\citet{Kamae:2006} for the calculation of the production cross sections.  
The production of pions from interactions of He nuclei with the 
interstellar hydrogen, as well as from collisions of CRs with 
interstellar He, are explicitly included, where we assume in this 
paper an interstellar He/H ratio of 0.11 by number \citep{1998ApJ...509..212S}.
This is slightly higher than the canonical value of 0.1 found by observations
of \hii{} regions \citep{Deharveng_He:2000}, but is within systematic
uncertainty of those observations.
We also ignore production of pions from CR and interstellar gas nuclei 
heavier than He while their contribution could be as high as $\sim$10\%
\citep{2009APh....31..341M}.
It is assumed that the distribution of
interstellar He follows that of interstellar hydrogen, detailed below.

\subsubsection{Atomic Hydrogen}
\label{sec:hi}

The atomic hydrogen (\hi) is the most massive component of the ISM 
and has a large filling factor, being observed in all directions.
A recent comprehensive review of the \hi{} content of the Galaxy can be 
found in \citet{2009ARA&A..47...27K}. 
For the CR propagation, 
the \GP{} code uses a 2D analytical gas model for the \hi{} 
distribution~\citep[][]{2002ApJ...565..280M}.
The radial distribution is taken from \citet{GordonBurton1976} while 
the vertical distribution is from \citet{DickeyLockman1990} 
for $0\leq R \leq 8\,{\rm kpc}$ and \citet{Cox1986} for $R \geq 10\,{\rm kpc}$ 
with linear interpolation between the two ranges.
For the evaluation of the diffuse \gray{} intensity, the spatial structure of 
the gas is essential and we renormalise
the column densities of the analytical gas model with those found from the
Leiden-Argentine-Bonn (LAB)
21-cm \hi{} line survey of \citet{2005A&A...440..775K}.
Using 
the distance information derived from the radial velocity of the gas and 
the Galactic rotation curve of \citet{1985ApJ...295..422C}, we assign 
the gas to Galactocentric annuli, generating column density maps for each 
annulus (see Appendix~\ref{appendix:gasRings} for a detailed description of the 
procedure and Table~\ref{tab:AnnuliBoundaries} for annuli boundaries
used in this analysis.).

The main uncertainty when deriving \hi{} column densities, N(\hi), 
from 21-cm \hi{} line surveys is the 
assumed spin temperature $T_S$ used to correct for the 
opacity of the 21-cm line (see Appendix~\ref{appendix:gasRings} for 
the definition of $T_S$).
The value $T_S$ = 125~K has been almost universally adopted in 
previous \gray{} studies but the quality of the \fermilat{} data require 
that this assumption be critically examined. 
\hi{} in the ISM exists in a mixture of phases, with $T_S$
ranging from 40~K to a few thousand~K.
A recent study using \hi{} absorption in the outer Galaxy 
\citep{2009ApJ...693.1250D}  
suggests that $\sim15-20$\% is in the cold (40--60~K) phase, 
while $\sim80-85$\% is in the warm phase, 
resulting in an average $T_S$ value in the range 250--400~K.
To limit the scope of the present paper, we 
gauge the uncertainty of the assumed $T_S$ value by using results 
for $T_S$ = 150~K and the optically thin assumption, which is 
suitable for a $T_S$ many times larger than the observed brightness 
temperature of the 21-cm spectral line.
These two values should encompass the real $T_S$ value for most of the sky.
Our choice of $T_S = 150$~K over 125~K is motivated by the fact that the 
maximum observed brightness temperature in the LAB survey is around 
150~K and $T_S$ must be greater than the observed brightness temperature.
Note that we are not trying to determine the value of $T_S$ from the \gray{}
data, only probing the uncertainty associated with using a single $T_S$ value
over the entire sky.
Due to the nonlinearity of the optical depth correction, no attempt is 
made to correct the analytical model of the \hi{} distribution used in 
the \GP\ code, which assumes $T_S = 125$~K.  
Because we renormalise the analytical gas model
when generating the \gray{} sky maps, the uncertainty associated with this 
inconsistency is minor and mostly affects the CR propagation parameters.

For a large region of the sky, N(\hi) is replaced by the dust-reddening corrected
column density.
(The region depends on the
actual magnitude cut used, see Section~\ref{sec:dust}).
Changing $T_S$ affects the inferred dust-to-gas ratio and hence the 
column density estimate from dust-reddening.
Because the latter replaces
that of CO and \hi{} combined (see Section~\ref{sec:dust}), the $T_S$ value 
has an effect only through
the gas-to-dust ratio in this region. 
In addition, there is a small secondary effect
caused by a slightly different distribution of N(\hi) that
is used to distribute the dust-reddening correction. 
For these reasons the assumed $T_S$ value should be interpreted with care.

\subsubsection{Molecular Hydrogen}

The molecular hydrogen (H$_2$) has less mass overall than the \hi{} but is 
concentrated in massive cloud complexes with large column densities.
For typical cold interstellar conditions it cannot be directly observed in
emission.
Instead, the 2.6~mm 
line of the $^{12}$CO molecular $J=1\rightarrow0$ transition is used as a 
tracer of H$_2$, assuming a proportionality between the integrated line
intensity of CO, W(CO), and the column density of H$_2$, N(H$_2$), given
by the factor \Xco.
For the CR propagation, 
the \GP{} code uses a 2D analytical gas model for the CO 
distribution.
The model described in~\citet{2002ApJ...565..280M}
uses the gas distribution from \citet{Bronfman1988} 
for $1.5\,{\rm kpc} < R < 10\,{\rm kpc}$, 
and that from \citet{Wouterloot1990} for $R \geq 10\,{\rm kpc}$, 
and is augmented with the \citet{2007A&A...467..611F} model 
for $R \leq 1.5\, {\rm kpc}$.
We use the 2.6~mm CO-line survey of \citet{2001ApJ...547..792D} for the 
spatial structure of the gas.  
To reduce noise the data has been filtered with the moment masking
technique \citep{2001ApJ...547..792D}.  
This technique uses a smooth
version of the map to create a mask selecting regions of the sky 
that have a large signal-to-noise ratio.  
This way the noise is reduced but the resolution of the original survey 
is preserved.
As with \hi, we
use the distance information from the line-of-sight (LOS) velocity 
together with a rotation curve to assign the gas to Galactocentric annuli.
These are used for the evaluation of the diffuse \gray{} intensity where the 
spatial structure of 
the gas is essential, and we renormalise
the column densities of the analytical gas model using survey data.

The \Xco{} factor may change with Galactocentric radius
\citep[e.g.,][]{2004A&A...422L..47S}.  
However, the spatial distribution is not well known and therefore we 
allow it to vary in this analysis.
This is done using the Galactocentric annuli output from \GP,
where 
each W(CO) annulus (see Table~\ref{tab:AnnuliBoundaries}) can be 
scaled freely in the fit.  
To decrease cross correlation in the derived \Xco{} values, we
reduce the number of scaled annuli in the fit to 7, putting annular 
boundaries at 0, 1.5
kpc, 3.5 kpc, 5.5 kpc, 8 kpc, 10 kpc, 16.5 kpc, and 50 kpc.

\subsubsection{Ionised Hydrogen}

Ionised hydrogen (\hii), although averaging only a few percent of the 
density of the neutral gas, contributes significantly to the \gray{} emission
at high latitudes because of its extended spatial distribution.
The extended warm ionised medium (WIM) is probed using pulsar 
dispersion measures. 
The most widely used model for the distribution of the WIM is 
NE2001 \citep{2002astro.ph..7156C,2003astro.ph..1598C,2004ASPC..317..211C}, 
but this model has been updated by \citet{2008PASA...25..184G}  
to agree with more extensive pulsar data, where now the WIM distribution has a 
larger scale-height perpendicular to the Galactic plane: 2~kpc 
instead of 1~kpc in NE2001.
Therefore, we use the WIM $z$-distribution given by \citet{2008PASA...25..184G}.
The narrow plane component provides a small contribution to the 
overall emission, but is included in 
our modelling using a simplified form based on NE2001.

\subsubsection{Dust as a Tracer of Gas}
\label{sec:dust}

The use of dust as a tracer of gas for \gray{} studies goes back to
\citet{1982A&A...115..404S} and 
\citet{1982A&A...105..159S}.  
Infrared emission from cold interstellar dust is an alternative to surveys 
of \hi{} and CO emission lines, 
which may not trace all of the neutral gas due to various 
reasons (cold/optically thick \hi, variations in \Xco, H$_2$ not traced
by CO).
An extensive study of this topic with application to EGRET data has 
been performed by \citet{Grenier:2005}, where the total hydrogen column 
density was derived for each pixel using the \ebv{} reddening maps 
given by \citet{1998ApJ...500..525S}.
This procedure significantly reduced the residual in the DGE 
modelling of EGRET data \citep{Grenier:2005}.
The addition of dust as a tracer of gas has also been used successfully in
analysis of \fermilat{} data \citep{2010ApJ...710..133A,2011ApJ...726...81A}.

\begin{figure*}
\plotone{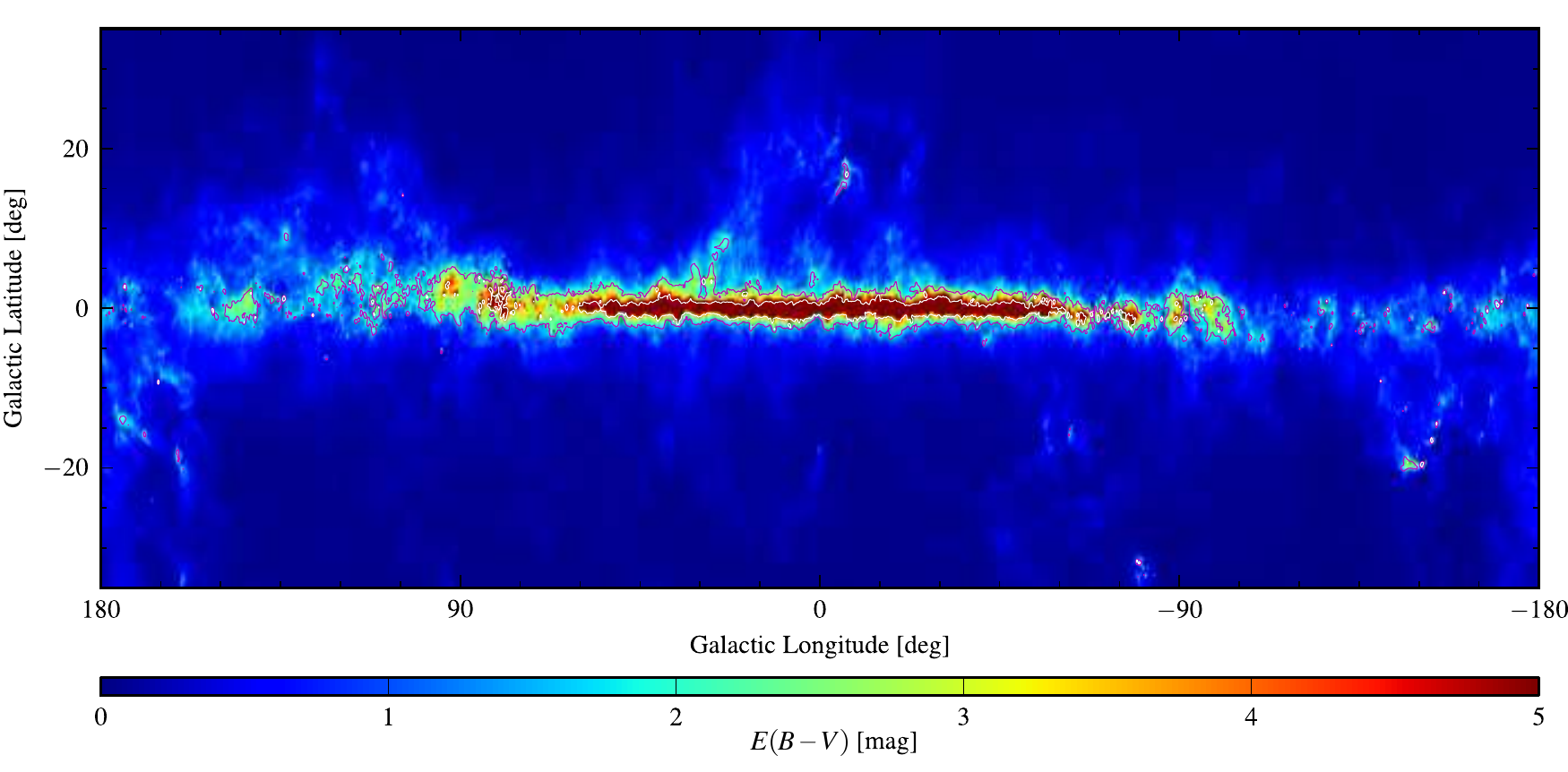}
\caption{The \ebv{} extinction map from \citet{1998ApJ...500..525S}.
Shown are contours for 2 mag (magenta) and 5 mag (white).  Note that the
latitude scale is stretched 2 times compared to the longitude scale for
clarity.  We also clip the scale for \ebv{} at 5 magnitudes.
\label{fig:magnitudeCutFilter}}
\end{figure*}

In this work, we apply a similar procedure as \citet{Grenier:2005} and create a
map of ``excess'' dust 
column density, \ebv$_{res}$. 
We obtain
a gas-to-dust ratio for both \hi{} and CO using a linear fit of the N(\hi) 
map and W(CO) map
to the \ebv{} reddening map of \citet{1998ApJ...500..525S}.  
For simplicity, we assume a constant gas-to-dust ratio throughout
the Galaxy.
To minimise errors in the fitting, we first determine the gas-to-dust ratio 
for \hi{} (\hi{} ratio) in regions where no CO is observed and then use that to 
determine the gas-to-dust ratio for CO (CO ratio) in regions rich in CO.
Because the quantity of dust traced by \ebv{} cannot be reliably determined in
regions with high extinction, we apply a 
magnitude cut to the \ebv{} map.
To gauge the uncertainty involved with this procedure, we consider 
two values: magnitude cuts at 2 and 5, respectively.
Figure~\ref{fig:magnitudeCutFilter} shows that the region affected by these
cuts is only a narrow strip around the Galactic plane for both values.
The gas-to-dust ratio obtained from our procedure depends on the 
assumed value of the 
spin temperature $T_S$ and the \ebv{} magnitude cut.
Our derived ratios are given in Table~\ref{tab:gas-to-dustRatio}.  
The \Xco{} factor in the table is
determined by assuming a constant proton-to-dust ratio as \Xco{} =
\hi{} ratio / ($2\times$ CO ratio).

For simplicity, we use the \hi{} gas-to-dust ratio to turn \ebv$_{res}$ 
into a column density map, N(\ebv$_{res}$).  
This should not cause a significant systematic effect because the \Xco\
values in Table~\ref{tab:gas-to-dustRatio} are similar to those found from the
\gray{} fits (see Section~\ref{sec:Xco}). 
But, the dust reddening map does not contain distance information.
Because the gas-related \gray{} emissivity varies throughout the 
Galaxy, and depends on the incident CR intensity together with the 
gas density, correct placement of the residual 
gas that is traced by the reddening map is essential.
While the \ebv$_{res}$ map corrects for 
uncertainties in N(\hi) and W(CO) and 
its density distribution along each LOS should be similar to 
N(\hi) and W(CO), 
unique assignment from dust to N(\hi) or N(H$_2$) is not possible.  
It is difficult to use the density distribution of W(CO) for this placement
for two reasons: the sky coverage of W(CO) is smaller than \ebv$_{res}$, and 
the \Xco{} factor is susceptible to variations.
N(\hi) is not limited in these ways.
Therefore, we choose to distribute
N(\ebv$_{res}$) proportionally to the density distribution of N(\hi) 
along each LOS.

N(\hi) in the optically thin limit provides a robust lower limit on the 
\hi{} column density.
To account for spurious negative residuals in the reddening map we limit 
the residual such that the
sum of \ebv$_{res}$ and the equivalent reddening of
N(\hi) and W(CO) is never less than the equivalent
reddening of N(\hi) in the
optically thin limit
The equivalent reddening of W(CO) and
N(\hi) is evaluated using the
determined gas-to-dust ratios, implicitly using the \Xco{} ratio
given in Table~\ref{tab:gas-to-dustRatio}.
W(CO) is included in the sum to account for possible variations in the
\Xco{} ratio in the Galaxy, i.e., N(\hi)$-$N(\ebv$_{res}$) might be less
than N(\hi) in the optically thin limit because we overestimate \Xco. 
We further limit the absolute value of the negative residuals to be less 
than the \hi{} column
density for each LOS so no pixels in the reddening 
corrected annular column density maps are negative.  
This last requirement is needed because our
method for calculating the expected model counts assumes no negative pixels.
The numbers of pixels affected by these two cuts is a small fraction of
the total and does not affect our results significantly.

Note that this method effectively replaces the N(\hi)
estimate with N(\hi)+N(\ebv$_{res}$) in the regions not affected by the \ebv\
magnitude cut (see Figure~\ref{fig:magnitudeCutFilter}).  
As described earlier, this changes the
meaning of $T_S$ because it now acts only as a proxy for the 
gas-to-dust ratio for
a large part of the sky.

\subsection{Interstellar Radiation Field}
\label{sec:ISRF}

The Galactic ISRF is the result of emission by stars, and the scattering, 
absorption, and re-emission of absorbed starlight by dust in the ISM. 
Because the ISM is not optically thin for the stellar emission 
due to the interstellar
dust, a radiation transport code must be used to model the 
distribution of low-energy photons throughout the Galaxy.
We calculate the ISRF using 
the {\em FRaNKIE\footnote[9]{Fast Radiation 
transport Numerical Kode for Interstellar Emission}} 
code \citep[][see Appendix~\ref{appendix:ISRF} for more details]{Porter:2008}.
The ISRF model we use in this 
paper \citep[the ``maximum metallicity gradient'' model from][]{Porter:2008} 
has an input bolometric stellar luminosity $\sim 4 \times 10^{10} L_\sun$.
This is distributed across the stellar components 
boxy bulge/thin disc/thick disc/halo with
fractions $\sim$0.1/0.7/0.1/0.1. 
Approximately 20\% of the input stellar luminosity is reprocessed by dust and 
emitted in the infrared.

A major uncertainty with the ISRF model is the overall input stellar 
luminosity and
how it is distributed amongst the components of the model.
Higher input stellar luminosities for a particular component, e.g., the bulge, 
will increase the CR electron/positron 
losses via IC scattering and hence the overall output in \gray{s} 
approximately over the spatial region where the stellar model component 
dominates.
Estimates available in the literature illustrating the 
range for the 
total Galactic stellar luminosity are, e.g., 
$6.7\times10^{10}$ $L_\odot$ \citep{Kent1991} and
$2.3\times10^{10}$ $L_\odot$ \citep{Freudenreich1998}, with different 
distributions of the total luminosity across the stellar components used 
in the models of these authors.
Also, the metallicity gradient is important for determining the distribution
of interstellar dust \citep[see][for the variation due to the range of 
Galactic metallicity gradients]{Porter:2008}.

Because of these details, the uncertainty in the ISRF can be considerable in 
regions like the inner Galaxy.
A full exploration of the model parameters for the ISRF is beyond the
scope of the current work, so we account for the uncertainty in the ISRF by
allowing freedom in the IC emission associated with the 
optical and infrared (IR) components.
This is done by separately calculating with \GP{} the contributions 
to the IC intensity by optical, IR, and cosmic microwave background photons.
Because the optical and IR are physically related, we use a common 
scaling parameter for both components.

\subsection{Comparison with \fermilat{} Data}

Once the parameters of the propagation model have been determined, the 
predicted \gray{} maps are compared to the \fermilat{} data.
The comparison is non-trivial due to the uncertainties in some of the DGE 
parameters described above, along with other \gray{} sources
emitting in the \fermilat{} energy range.
To account for the uncertainties we perform a maximum likelihood fit to the
data using the \gardian{} tool described in Appendix~\ref{appendix:gardian}
including in the model the detected point sources and an isotropic component
described below.

\subsubsection{Detected Sources}

Gamma-ray point sources from the \fermilat{} 1FGL are 
included explicitly in the model. 
This list contains 1451 sources and gives, amongst other information, their 
location and spectra.
In general, high-significance ($TS \gtrsim 200$) \gray{} point sources in
the Galactic plane and those outside of the Galactic plane, 
even down to the formal $TS > 25$ criterion, are 
relatively unaffected by changes in the assumed DGE model. 
However, there is some dependence on the DGE model for lower significance 
point sources in the Galactic plane, with the strongest effect at 
low energies. 
The relatively wide \fermilat\ 
PSF combines with the spatial and spectral structure of the DGE for our 
models in the vicinity of sources in the plane, which can give considerable
variations in the fluxes and spectra even for detected point 
sources significantly above the formal detection threshold.
The time range used in this analysis is also different from that used for the 
1FGL analysis, consequently 
the average spectra of variable sources might be different 
for the data set used in this paper.
Therefore, we use 
the spectral information given in the 1FGL catalogue only for the initial 
prescription in the fit.
Then, the flux of each point source in the list is determined for every energy 
bin independently.
Because a simultaneous fit is very computer intensive, we use an 
iterative method, fitting point sources 100 at a time starting with 
the brightest.
At each step, the fluxes and spectra of sources that have not been fitted are 
included but fixed at the 1FGL catalogue values.
However, our method has been shown to give results compatible with the 1FGL 
catalogue
when the data selection and background model are the same.


\subsubsection{Instrumental and Extragalactic Backgrounds}

The \fermilat{} data have a residual instrumental background (RIB) due to CR 
interactions in the instrument and spacecraft and also CR events 
misclassified as photons.
The CR background depends on geomagnetic latitude, but is considered 
isotropic in this paper because we average over many orbits.
The extragalactic diffuse \gray{} background (EGB), assumed to be isotropic, is 
also present. 
These must be taken into account when comparing any astrophysical model 
with data.
For a recent determination of the EGB and RIB components by the \fermilat{} 
team see \citet{EGBpaper}.

As is shown by \citet{EGBpaper}, the measured spectrum of the EGB 
is dependent on the assumed foreground DGE model, while the RIB is 
determined from the instrument Monte Carlo modelling. 
However, for the present work the distinction between EGB and RIB is not 
important.
We therefore determine the total ``isotropic'' background for each model 
where the flux in each energy bin is fitted independently.
The results for the combined EGB + RIB obtained for each model 
are compared to the total of these components derived by \citet{EGBpaper} 
as a consistency check.

\subsubsection{Fitting Region Subdivision}
\label{sec:Regions}

\begin{figure*}
\plotone{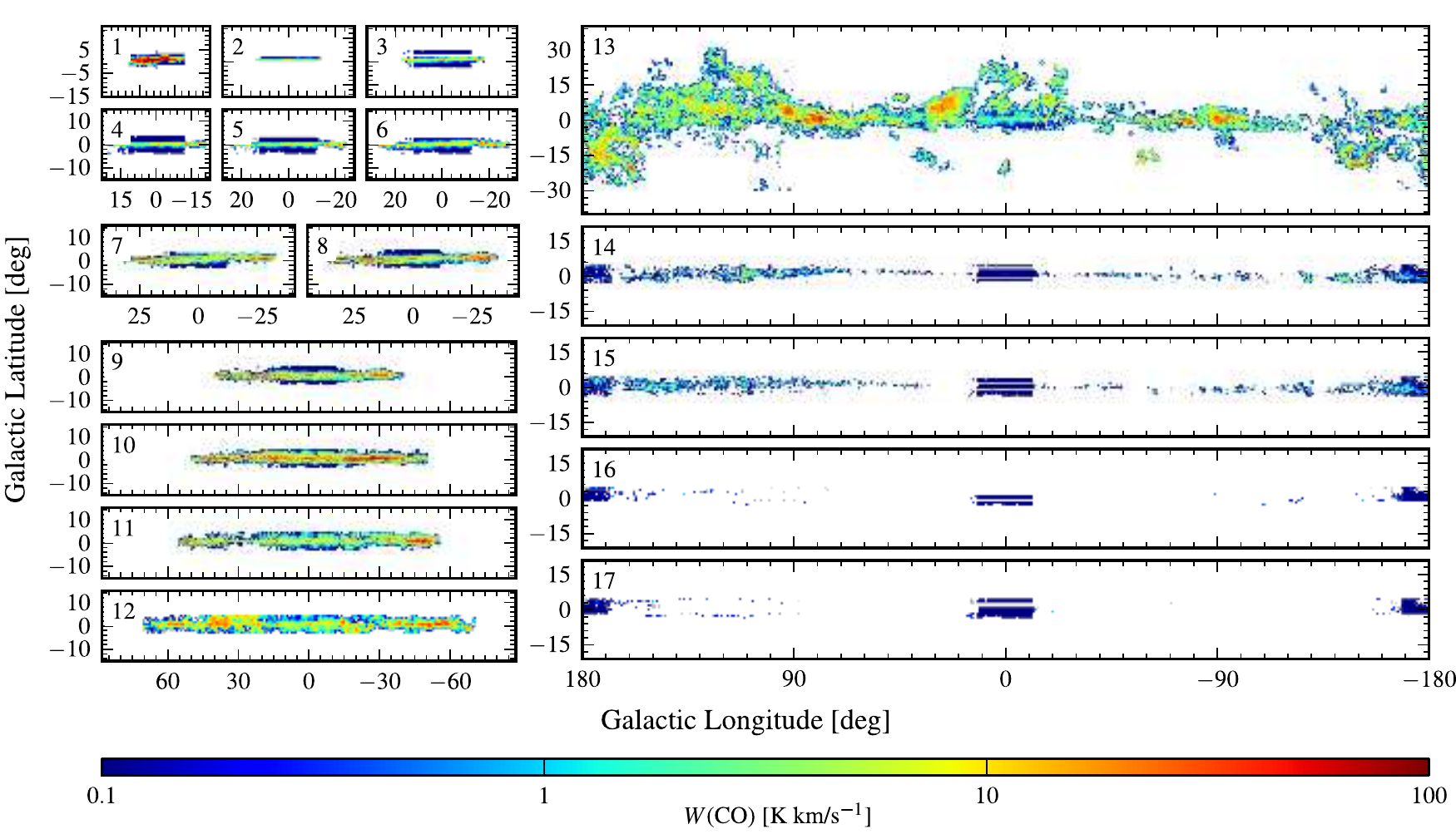}
\caption{Integrated line intensity of CO as a function of 
Galactocentric radius.
The logarithmic color scale is clipped at a value of 100~K~km~s$^{-1}$.
The actual scale reaches over a 1000~K~km~s$^{-1}$ in annulus 1.
The numbers in the top left corner of each panel label the annuli whose
boundaries are given in Table~\ref{tab:AnnuliBoundaries}.
Note that there is very little CO outside of 16.5 kpc (annuli 16 and 17).
The interpolation regions around the Galactic centre and anti-centre are clearly
visible as low density (blue) bands.  
They are a significant
contributions to the line intensity of CO in the outer Galaxy annuli (14
through 17).
For details on the creation of these maps see Appendix~\ref{appendix:gasRings}.
\label{fig:COrings}}
\end{figure*}

Figure~\ref{fig:COrings} shows the CO annuli used in this analysis.
There is very little CO emission in the outer Galaxy.
To minimise the effects that the bright and complex inner Galaxy has on the 
determination of the CO scaling parameters in the outer Galaxy, we split 
the maximum 
likelihood fit into regions, separating the inner and outer Galaxy.
In addition, we also minimise the effect of the bright Galactic plane
when determining the isotropic background by fitting low and
high-latitude regions separately.
This subdivision results in 3 regions: $|b| >
8^\circ$ (local), $|b| \le 8^\circ$ and $80^\circ < l < 280^\circ$ (outer
Galaxy), and $|b| \le 8^\circ$ and $l < 80^\circ$ or $l > 280^\circ$ 
(inner Galaxy).
A latitude cut of $b = 8^\circ$ was 
chosen because all CO with $|b| > 4^\circ$ is
considered to be in the local annulus, with the extra $4^\circ$ accounting 
for the extension of the \fermilat{} PSF, and also to reduce effects of the 
bright plane for the determination of the isotropic spectrum.
We first fit in the local region and determine the scaling parameter for the 
local CO annulus (8 kpc -- 10 kpc) and the spectrum of the isotropic emission.
We also allow freedom in the ISRF scaling parameter because there is 
significant IC emission
in this region, a fraction of which originates in the inner Galaxy
where the ISRF is most uncertain.
These parameters are then fixed and the fit is performed for the outer Galaxy
region to determine the 
CO scaling parameters there. 
Finally, we fit the remaining CO scaling parameters in the inner Galaxy 
region and 
allow the ISRF scaling parameter to be free in the fit because the 
fraction of IC 
emission originating from the
inner Galaxy is much higher in this region than the local region.
The fluxes and spectra of point sources in each region are fit as 
described above.

To account for the overlap between regions caused by the \fermilat{} PSF 
we create a model of the whole sky for each fit, setting non-fitted scaling
parameters to their nominal values of 1 and the spectra of the point sources to
their values in the 1FGL catalogue. 
This does not affect our results significantly, because the overlapping 
area is a
small fraction of the total area for each region, the scaling parameters do 
not vary significantly from 1, 
and the point source fluxes and spectra are close to the 1FGL catalogue values.

\subsection{Iterating the procedure}

To account for the effect of the radial variation of \Xco{} on the 
CR propagation and the LOS integration when generating
the \gray{} sky maps with \GP, the 
above process is iterated using the \Xco{} distribution found from 
the \gray{} fit back into the propagation parameter determination/transport 
calculation.
The iteration is done in two steps.
First, we calculate for each annulus the average \Xco{} value, weighted 
with both the parameterised CO gas distribution used in \GP{} and the 
integrated CO intensity from the annular maps.
These values are then scaled with the values found from the \gray{} fit.
To have a smoothly varying \Xco$(R)$, we use power-law interpolation 
between the scaled values.

We use \Xco$(R) = 2 \cdot 10^{19+0.1 R/(1 kpc)}$ cm$^2$(K km/s)$^{-1}$ for the
initial radial variation of \Xco, 
compatible with the results from \citet{2004A&A...422L..47S}.
There is no formal criterion for stopping the process, but we have 
found that it converges after a few iterations.
The results we report in this paper are obtained after 4 iterations.

\section{Results}
\label{sec:results}

For brevity, we use the short hand notation
\model{X}{z_h}{R_h}{T_S}{c} where X is the first letter of the CR source
distribution\footnote[10]{S: SNR, L: Lorimer, Y: Yusifov, O: OB stars, see
Figure~\ref{fig:CRsources}}, $z_h$ and $R_h$ are given in units of kpc,
$T_S$ in units of K\footnote[11]{We use $T_S=\infty$ for the optically thin
case}, and $c$ is the \ebv{} magnitude cut. 
In addition, for figures comparing the entire set of 
models, each set of model parameters is given a number.
The number is a binary encoding of the input parameters and the mapping is
given in Table~\ref{tab:numberMapping}.  
As an example, the model with a
Yusifov CR source distribution, $z_h=10$~kpc, $R_h=30$~kpc, $T_S=150$~K,
and \ebv{} magnitude cut of 2~mag gets the number 1011100+1 = 93.

Due to the limited freedom in the DGE model, the
parameters determined from the \gray-fit can be biased if some
important
component is not included in the model or because of some 
systematic uncertainty in the DGE model.  
However, determining the parameters from the data is appropriate because their 
values are known {\it a priori} only with some error.
Note that this is a general limitation of any parameter determination from a
maximum likelihood fit where the model does not perfectly parameterise the data.

\subsection{Statistical evaluation of models using \gray{} data}

The best-fit DGE models to the \gray{} data are determined by comparing their 
maximum likelihoods (see Appendix~\ref{appendix:gardian}) where higher values 
are a qualitatively better fit.
Figure~\ref{fig:gammaLikelihood} shows the logarithm of the 
maximum likelihoods of all the models for the 3 different fit regions: 
local, outer Galaxy, and inner Galaxy.
No single model stands out as providing the best-fit 
in all three regions simultaneously. 
The largest difference between models occurs in the outer Galaxy. 
Because the difference is about 3 times larger in the outer Galaxy than other
regions, the outer Galaxy would dominate in an all-sky likelihood ratio test.

While none of the models provides a best-fit for all three regions
simultaneously, there are patterns in the likelihood results that are 
similar between regions.
The most general trend is that increasing $z_h$ improves the likelihood in all
regions, though the effect is strongest for the outer Galaxy.
It is also in the outer Galaxy that the difference between models 
employing different CR source distributions is most pronounced, 
with the flat SNR distribution favoured over the
distributions of pulsars and OB stars, which are more peaked in the 
inner Galaxy.
However, this is strongly dependent on $z_h$ and the effect nearly
disappears for $z_h=10$~kpc.
The outer Galaxy also shows an increase in likelihood with 
larger values of $R_h$, especially combined with high values of $z_h$.
The models giving the highest CR flux in the outer Galaxy therefore give the
largest likelihood. 
This need for an increased flux in the outer Galaxy compared to standard
propagation models has been shown in other \fermilat{} analyses
\citep{2010ApJ...710..133A,2011ApJ...726...81A}.

The value of $T_S$ also has a significant 
impact on the likelihood values of the models, although the effect 
differs from region to region.
A value of $T_S=150$~K is preferred in the outer Galaxy, 
which is consistent with requiring an increased flux in this region. 
Lower values of $T_S$ give higher column densities of \hi{} that increase the
\gray{} intensity of the models.
The effect of $T_S$ is different in the local region, where 
the optically thin assumption for \hi{} is preferred.
As discussed in Section~\ref{sec:dust}, the \hi{} column density is replaced by
that estimated from dust and $T_S$ becomes a proxy for a certain gas-to-dust
ratio given in Table~\ref{tab:gas-to-dustRatio}.
A similar consideration applies in the inner Galaxy region, where optically 
thin \hi{} gives both the maximum and minimum likelihood depending on the 
value of the adopted \ebv{} cut.
The higher cut of 5 magnitudes gives the best-fit and thus 
the \ebv{} column density estimator seems to be preferred even in the 
inner Galaxy region.
The lower gas-to-dust ratio from the optically thin \hi{} assumption is also
preferred while the large difference in the likelihood for different cuts
of \ebv{} indicate that the optically thin assumption for \hi{} is not
appropriate in the Galacic plane as 
is generally known \citep[see e.g.,][]{CGPS}.
An \ebv{} cut of 5 magnitudes is also preferred in the outer Galaxy for both
values of $T_S$, showing that \ebv{} is a better total column density tracer
than \hi{} and CO combined in the Galactic plane.  

While the likelihood ratio test allows comparison between different models, 
it is not an absolute measure.
As we show in Section~\ref{sec:visuals}, there are large scale residuals 
remaining after model subtraction, which indicate missing components in the 
models that might bias the comparison.
However, because the
residuals exhibit a spatial structure that is different from the DGE, we do not
think there is a strong bias.

\subsection{Comparison with Spectra, Longitudinal and Latitude Profiles, and Residual Maps\label{sec:visuals}}

\begin{figure}
\epsscale{0.53}
\plotone{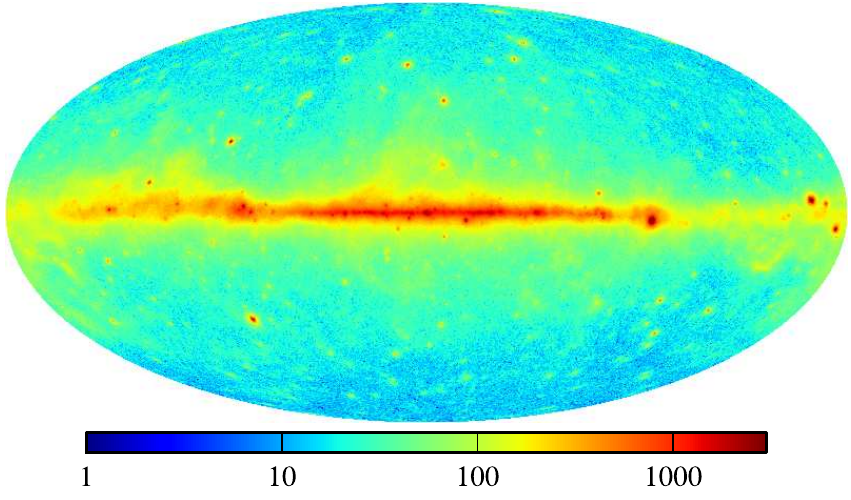}\\
\plotone{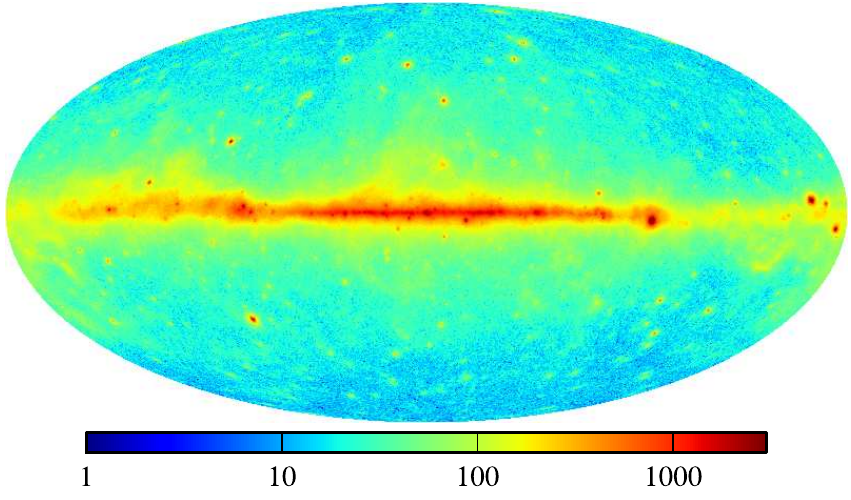}
\caption{Upper panel: observed \fermilat{} counts in 
the energy range 200 MeV to 100 GeV used in this paper.  
Lower panel: predicted counts for model \model{S}{4}{20}{150}{5}
in the same energy range.
To improve contrast we have used a logarithmic scale and clipped the 
counts/pixel scale at 3000.
The maps are in Galactic coordinates in Mollweide projection with longitudes
increasing to the left and the Galactic centre in the middle.
\label{fig:countsExample}}
\end{figure}

While the likelihood ratio test is effective for comparing different 
models, it is not able to describe the accuracy of each model separately.
Examining residual maps and spectra for different sky regions, along with 
the longitude and latitude profiles, is a direct method for comparison of 
models with data. 
Figure~\ref{fig:countsExample} shows the counts observed with 
the \fermilat{} in the 
energy range 200~MeV to 100~GeV considered in this paper and also the 
predicted counts from model \model{S}{4}{20}{150}{5}, which we take 
as our reference model (the use of this as the reference model is not arbitrary 
because its parameters are similar to the ``conventional'' 
model employed in earlier work \citep{2004ApJ...613..962S}).
This illustrates the general good agreement across the sky between model and 
data.
However, looking in detail reveals discrepancies in particular regions.
We discuss these in the following sections.
Due to space constraints, we will not show figures for all of the models
considered in the paper.  
A few models are chosen for
display, selected to show the range of results, emphasising the
differences between the models.  
The figures for all of the models are
available in the online supplementary material.
Note, the comparison models incorporate the factors found from the
fit to the \gray{} data so directly comparing the \GP{} output using
the GALDEF files provided in the online supplementary material will not
give identical results.

\subsubsection{Residual Sky Maps \label{sec:residualMaps}}

\begin{figure}
\epsscale{0.53}
\plotone{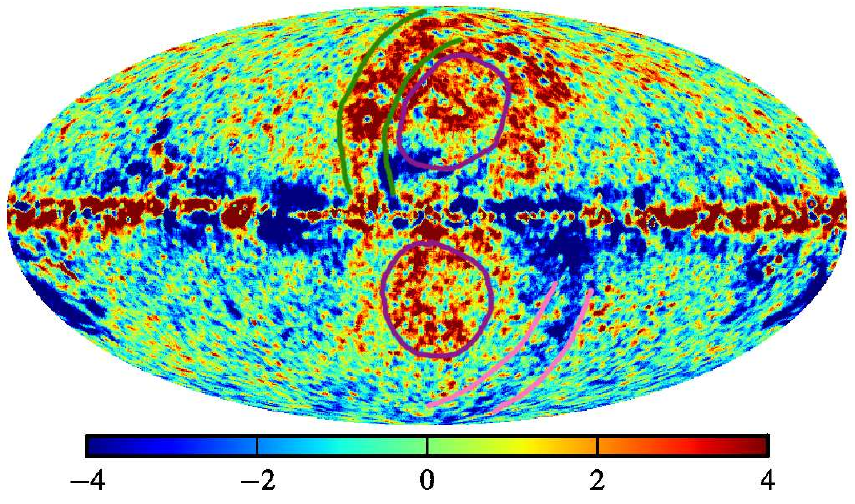}\\
\plotone{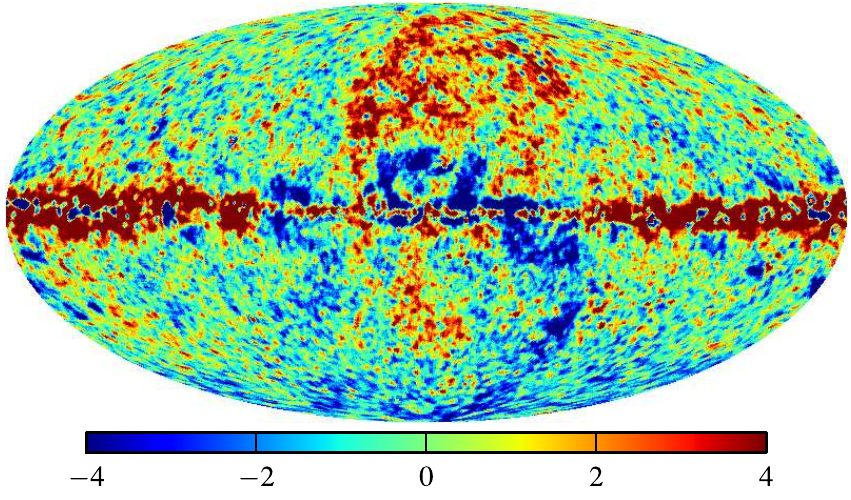}
\caption{
Residual maps in units of standard deviation in the energy range 200 MeV -- 100 GeV.
Shown are residuals for model \model{S}{4}{20}{150}{5} (top) and
model \model{L}{6}{20}{\infty}{5} (bottom).
The top map shows in addition a sketch of a few identified large scale residuals,
Loop I (green), Magellanic stream (pink), and features coincident with
those identified
by \citet{2010ApJ...724.1044S} and \citet{2010ApJ...717..825D} (magenta).
The maps have been smoothed with a $0.5^\circ$ hard-edge kernel.  
The kernel
is inclusive so that every pixel intersecting the kernel is taken into
account.
\label{fig:sigmaResiduals}}
\end{figure}

\begin{figure}
\epsscale{0.53}
\plotone{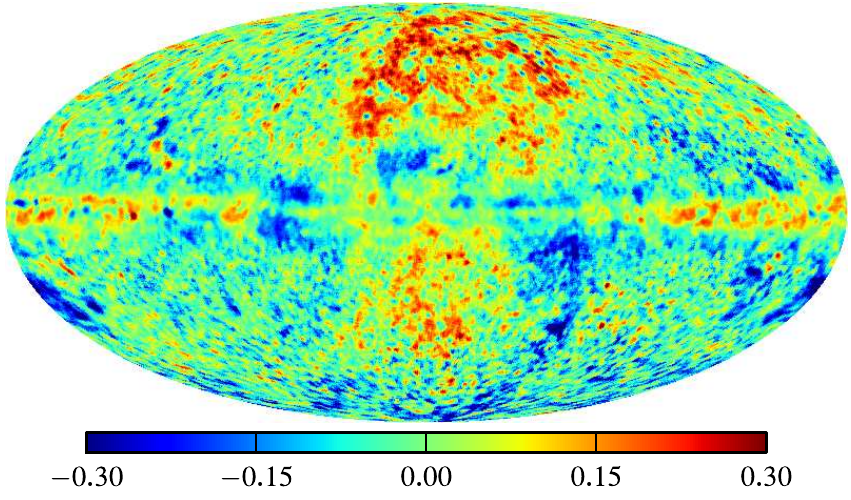}\\
\plotone{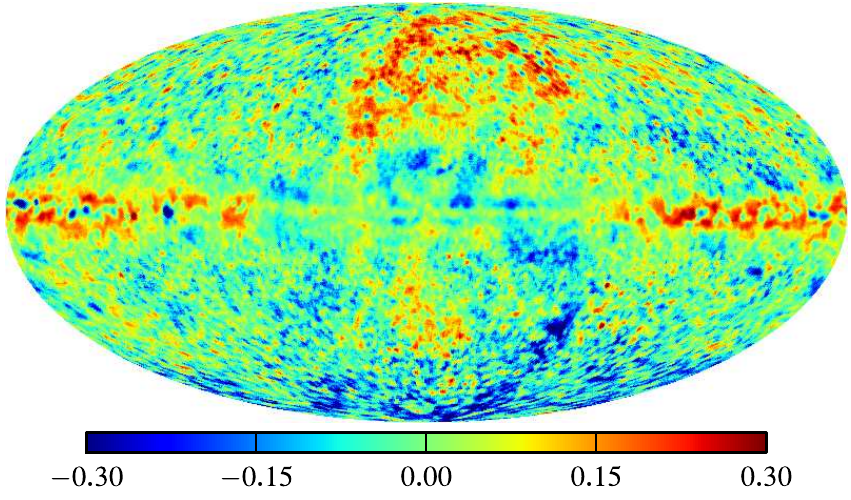}
\caption{
Fractional residual maps, $(model-data)/data$, in the energy range 200 MeV -- 100 GeV.
Shown are residuals for model \model{S}{4}{20}{150}{5} (top) and
model \model{L}{6}{20}{\infty}{5} (bottom).
The maps have been smoothed with a $0.5^\circ$ hard-edge kernel, see
Figure~\ref{fig:sigmaResiduals}.
\label{fig:fractionalResiduals}}
\end{figure}

Figure~\ref{fig:sigmaResiduals} shows the residual sky maps in units of 
standard deviations\footnote[12]{Calculated as ${\rm sign}(\Delta) *
\sqrt{2(data*\log(data/model) - \Delta)}$ with $\Delta = data - model$.} 
for models \model{S}{4}{20}{150}{5} and \model{L}{6}{20}{\infty}{5}.
All models display large-scale residuals with similar, but not identical,
features.
A more physical way of comparing the models to the data are fractional 
residual maps, $(data-model)/data$, shown in
Figure~\ref{fig:fractionalResiduals} for the same models.
The Galactic plane shows significant (greater than $4\sigma$) positive and 
negative structure in the inner Galaxy, but mainly positive in the outer Galaxy.
While the residuals are statistically significant,
Figure~\ref{fig:fractionalResiduals} shows that the fractional difference in
the inner Galactic plane is less than 10\%.

All of the models considered have large positive residuals at intermediate 
and high 
latitudes about the Galactic centre, most notably features
coincident with those
described by \citet{2010ApJ...724.1044S} and \citet{2010ApJ...717..825D}, 
and a feature that is similar to the radio-detected 
Loop~I \citep{2009arXiv0912.3478C}.
The negative residual of the Magellanic stream is also visible 
in the southern hemisphere.
It was not subtracted from the \hi{} annular column density 
maps because its contribution to the column density was incorrectly assumed 
to be negligible.  
However, this does not affect our model comparison because the
models all include this same extra column density.
Due to the limited freedom in our fits of the DGE to the \gray{} sky, no
attempt will be made here to characterise these residual structures but we do
note that their shapes depend on the assumed DGE model.

Point sources are also evident in the large-scale residuals, indicating that 
the point-source fluxes 
determined by the fit are biased in these areas.
However, their PSF-like spatial extent prevents them from affecting the 
DGE modelling significantly.
Only in areas with many overlapping point sources, such as in the Galactic 
ridge, can they mimic the structure of the DGE.  
Our tests have shown that inaccurate source modelling causes less than
20\% variations in the derived \Xco{} factors, less
than the variation caused by the CR source distribution and gas
properties (see Section~\ref{sec:Xco}).

The track of the Sun along the ecliptic can also be seen (particularly 
in the north), although it is not very prominent.
The quiet Sun is a source of high-energy \gray{s} from CR nuclei interacting
in its atmosphere \citep{Seckel1991} and CR electrons and positrons IC 
scattering of the 
heliospheric photon field \citep{SolarIC1,SolarIC4,SolarIC2,SolarIC3}.
However, when averaged over a year the overall intensity of this component
is very small, being less than 5\% of the isotropic background 
over most of the sky \citep{EGBpaper}, and does not affect the large-scale
DGE modelling significantly. 
The Moon also contributes to the emission from the ecliptic, being nearly as
bright as the Sun around 100~MeV.
But, the equivalent diffuse intensity from the moving Moon is less
due
to the inclination of its orbit relative to the ecliptic.
The \gray{} spectrum from the Moon is also steeper than that of the 
Sun \citep{MoonAlbedo} and does not contribute at a detectable level above 
10~GeV.

\begin{figure}
\epsscale{0.43}
\plotone{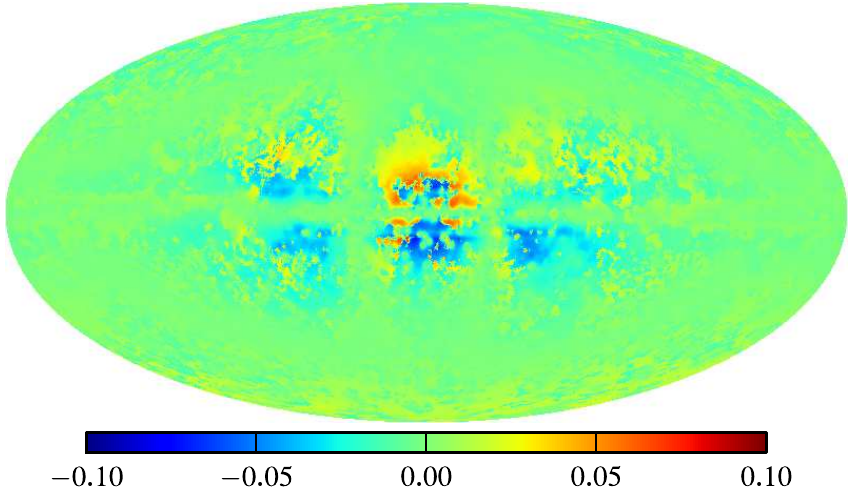}\\
\plotone{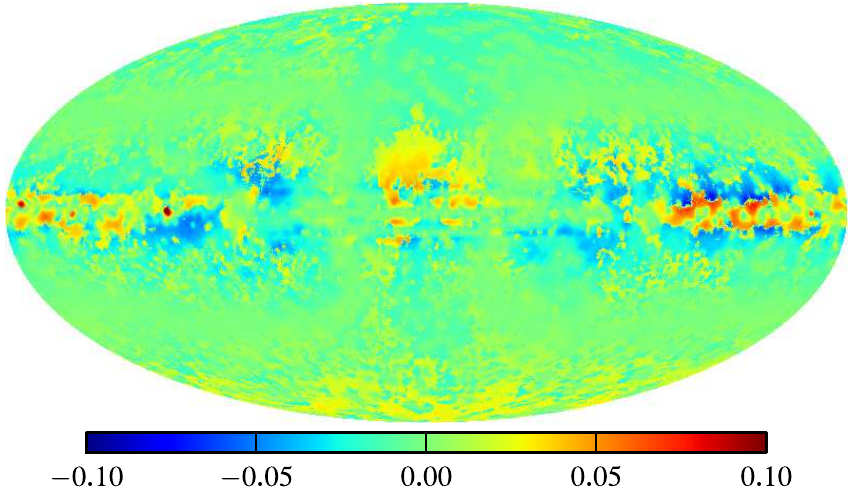}\\
\plotone{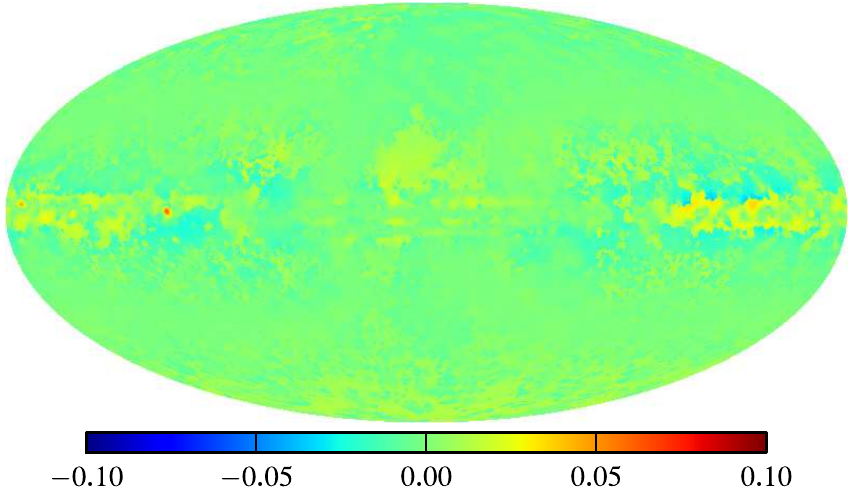}
\caption{
The difference between the absolute values of the fractional residuals of models where only the CR source distribution is changed.
Top: model \model{L}{10}{30}{150}{5} minus model \model{O}{10}{30}{150}{5}, middle: model \model{L}{10}{30}{150}{5} minus model
\model{S}{10}{30}{150}{5}, bottom: model \model{L}{10}{30}{150}{5} minus model
\model{Y}{10}{30}{150}{5}.
Negative pixels represent a better fit with the first mentioned model.
The maps have been smoothed with a $0.5^\circ$ hard-edge kernel, see
Figure~\ref{fig:sigmaResiduals}.
\label{fig:distDiffFractResiduals}}
\end{figure}

\begin{figure}
\epsscale{0.53}
\plotone{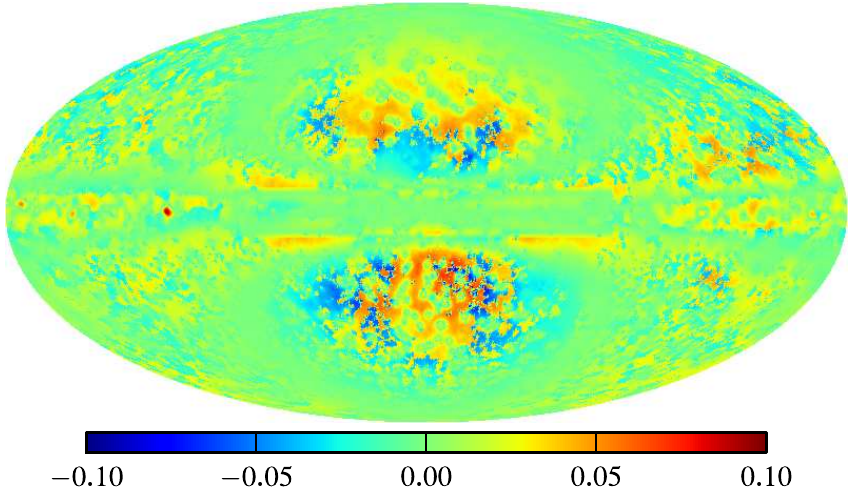}\\
\plotone{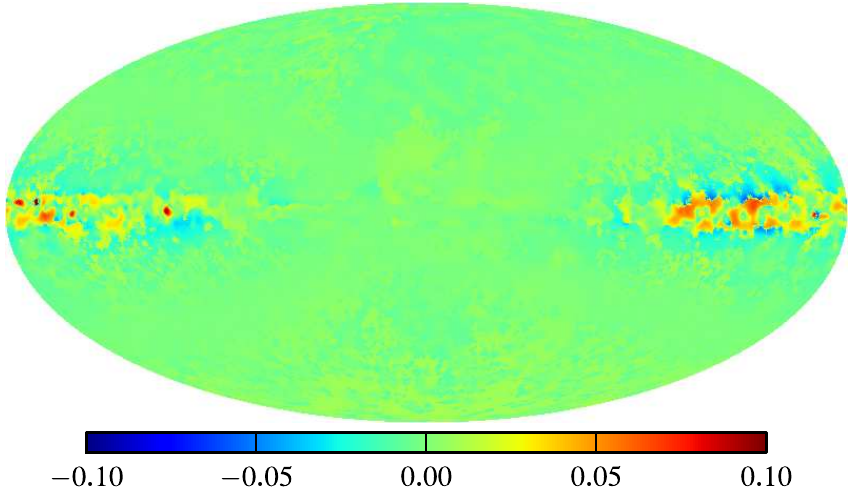}
\caption{
The difference between the absolute values of the fractional residuals of models where only the halo size is changed.
Top: model \model{S}{4}{20}{150}{5} minus model
\model{S}{10}{20}{150}{5}, bottom: model
\model{Y}{10}{20}{150}{2} minus model \model{Y}{10}{30}{150}{2}.
Negative pixels represent a better fit with the first mentioned model.
The maps have been smoothed with a $0.5^\circ$ hard-edge kernel, see
Figure~\ref{fig:sigmaResiduals}.
\label{fig:haloDiffFractResiduals}}
\end{figure}

\begin{figure}
\epsscale{0.43}
\plotone{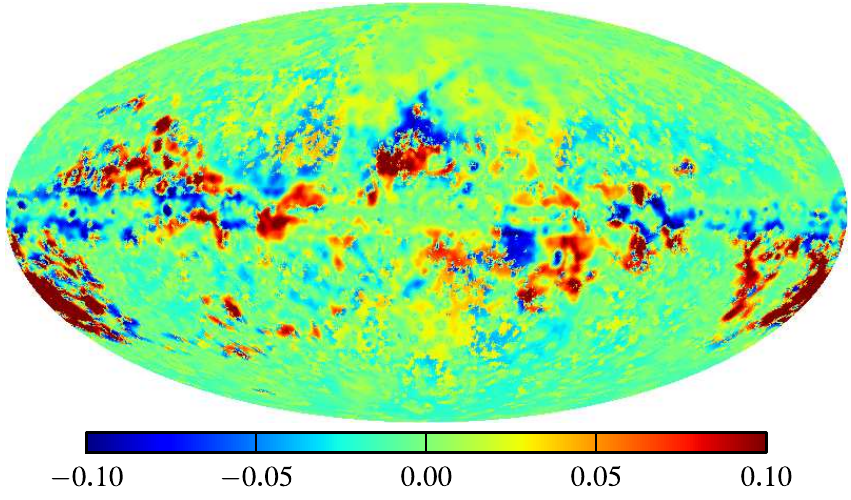}\\
\plotone{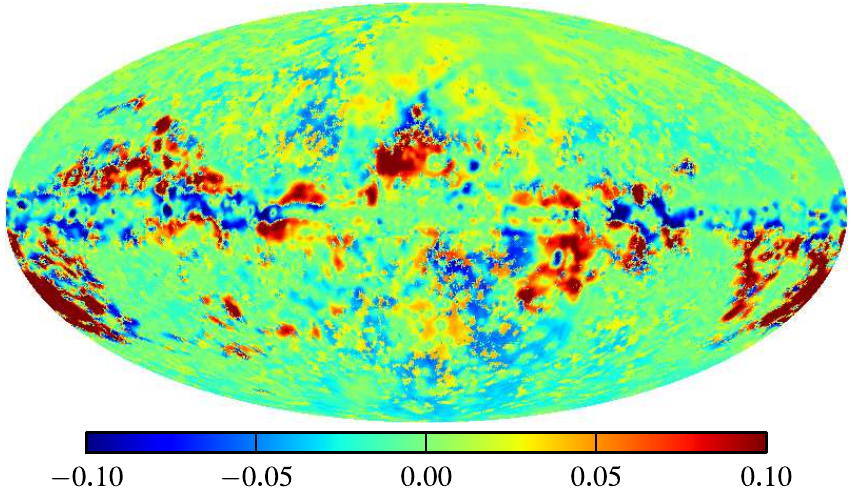}\\
\plotone{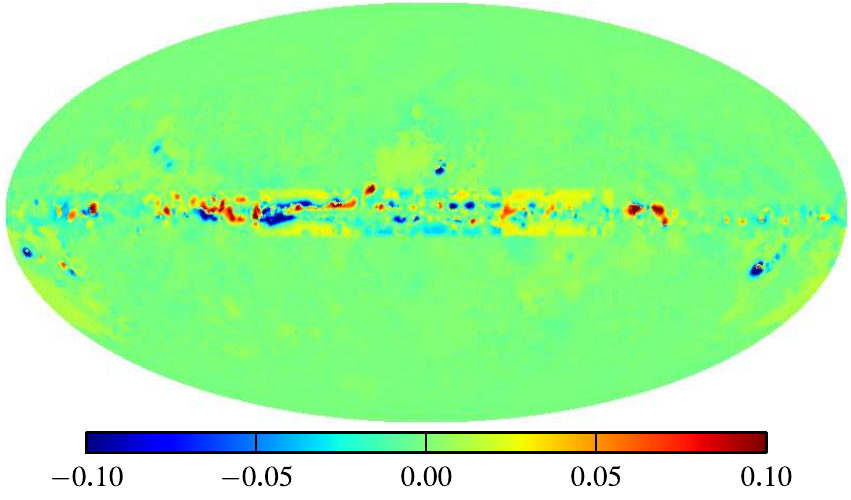}
\caption{
The difference between the absolute values of the fractional residuals of models where only the properties of the gas distribution is changed.
Top: model \model{S}{4}{20}{150}{5} minus model
\model{S}{4}{20}{\infty}{5}, middle:
model \model{Y}{10}{30}{150}{2} minus
model \model{Y}{10}{30}{\infty}{2},
and bottom: model \model{S}{4}{20}{150}{2} minus model
\model{S}{4}{20}{150}{5}.
Negative pixels represent a better fit with the first mentioned model.
The maps have been smoothed with a $0.5^\circ$ hard-edge kernel, see
Figure~\ref{fig:sigmaResiduals}.
\label{fig:gasDiffFractResiduals}}
\end{figure}

For comparisons between models, we calculate the differences 
between the absolute values of the fractional residual maps for the models.
These maps show directly which model better fits the data while the 
difference between the models might be larger than shown in these plots.
To study the effects of individual parameters, we 
compare models where only a single model grid 
parameter is varied.
In Figure~\ref{fig:distDiffFractResiduals} we show the difference 
residual maps for variation of only the CR source 
distribution, changes in the size of the
CR confinement volume in Figure~\ref{fig:haloDiffFractResiduals}, and 
variations of the gas properties in Figure~\ref{fig:gasDiffFractResiduals}.
While only a single model grid parameter is changed between the models, 
there are
related changes in the propagation parameters, CR source injection spectrum, 
ISRF scale factor, \Xco-factors, isotropic spectrum, and
point source spectra resulting from the CR and \gray{} fits that also 
affect the results.
The variation of the gas-related parameters has the largest and most 
distributed effect across the sky.
Varying the \ebv{} magnitude cut produces differences that are mostly 
confined to
the Galactic plane, but the small change in the gas-to-dust ratio
(see Table~\ref{tab:gas-to-dustRatio}) has an effect at higher latitudes.
Changing $T_S$ gives large positive and negative 
differences over the sky.  
The most striking feature is toward the outer Galaxy, where changing 
from $T_S = 150$~K to the optically thin approximation mostly improves the
agreement at intermediate latitudes, but generally worsens it in the outer
Galaxy plane.   The improvement at intermediate latitudes seems to
correlate at least somewhat with the distribution of CO at intermediate
latitudes seen in Figure~\ref{fig:COrings}, indicating that the
gamma-ray signal is sensitive to the ratio of the gas-to-dust ratios for
\hi{} and CO.  Another explanation might be a Galacto-centric gradient
of the gas-to-dust ratios, it being higher in the outer Galaxy in
agreement with the increased metallicity in the outer Galaxy.
Other variations at high latitude are not as strong but still significant.
Due to the diffusive propagation of CRs throughout the Galaxy, the 
steady state CR
distribution should not show
strong variations on the scales that are needed to account for the differences
shown in the figure.
This indicates rather that
the assumption of a single $T_S$ value, and hence gas-to-dust ratio, should
be reconsidered in subsequent work.  

While variation of the assumed CR source distribution does not show as 
strong as an effect as for the gas properties, significant differences are
still seen over the sky.
No single CR source distribution is best for all regions of the sky.
A strong asymmetric feature in the direction of the inner Galaxy can be seen
in the top panel of Figure~\ref{fig:distDiffFractResiduals}, having opposite 
signs above and below the plane, indicating a missing asymmetry in the model, 
either in terms of gas properties or CR flux.
The outer Galaxy shows similar features, where the intermediate latitudes and
the plane have opposite signs.  This is most easily seen in the middle panel of
Figure~\ref{fig:distDiffFractResiduals} where we have an improvement in 
the fit in the plane but worsening at intermediate latitudes.
Because the SNR distribution provides more CR flux in the outer Galaxy, this 
indicates that
the gas distribution could be more closely confined to
the plane in the outer Galaxy than estimated in our modelling.
This possibility has been studied by \citet{2007A&A...469..511K} who found 
a reduced extension in $z$ of the gas distribution in the outer Galaxy 
when assuming the gas in the halo rotated more slowly than gas in the plane.

Variation of the halo size parameters produces low-level residuals, both 
positive and negative, in different regions of the sky.
The halo size, $z_h$, has the strongest influence in 
the outer Galaxy and in the region above and below the Galactic plane in
the direction of the Galactic centre.  
The former can be explained by
increased CR flux in the outer Galaxy when $z_h$ is increased, while 
the latter is due to
increased IC emission in the direction of the Galactic centre caused by
a longer integration path length along the LOS.
The increase in IC emission is suppressed somewhat because the 
normalisation of the ISRF 
is anti-correlated with $z_h$ (see Section~\ref{sec:ISRFscale}). 
Increasing $R_h$ affects only the outer Galaxy significantly, 
where the models with larger $R_h$ better agree with the data.

The effect of varying a single parameter on the derived residuals 
can be strongly interdependent on the other adopted input parameters.
This is illustrated in Figure~\ref{fig:gasDiffFractResiduals} where the 
difference in residuals when varying $T_S$ for two different sets of the 
other input parameters is shown.
The changes are clearly different depending on the input parameters and
the resulting parameters found from the CR and \gray{} fits.

\begin{figure}
\epsscale{0.43}
\plotone{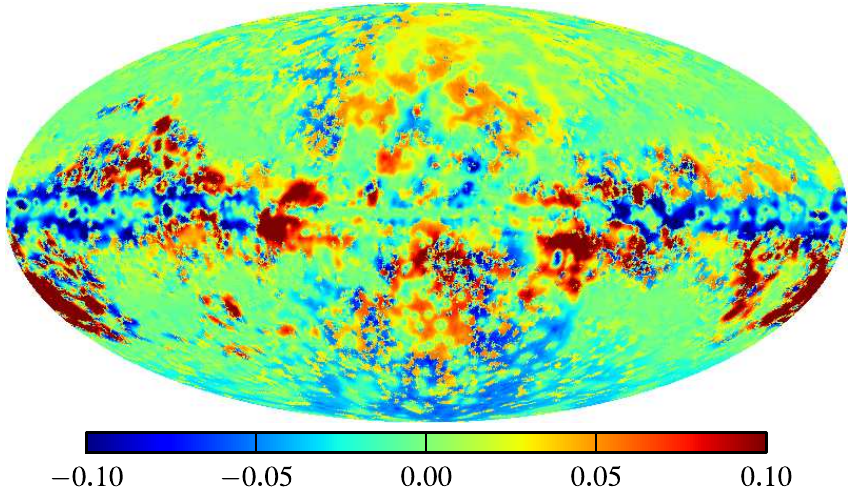}\\
\plotone{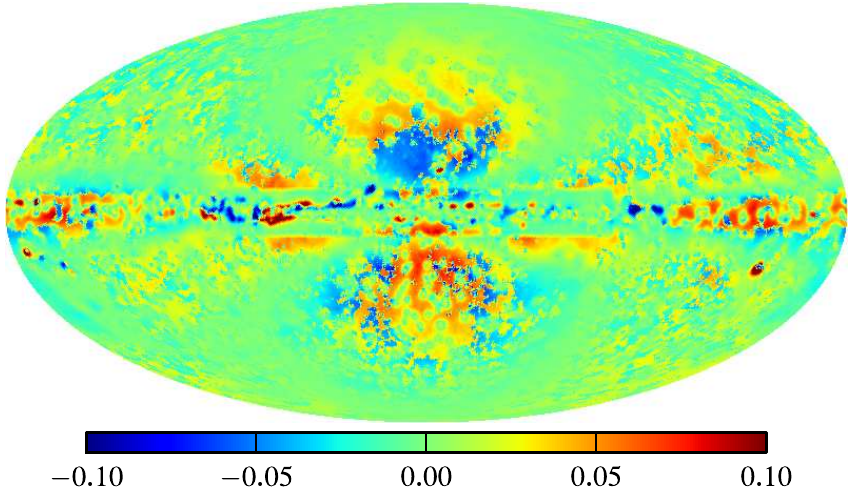}\\
\plotone{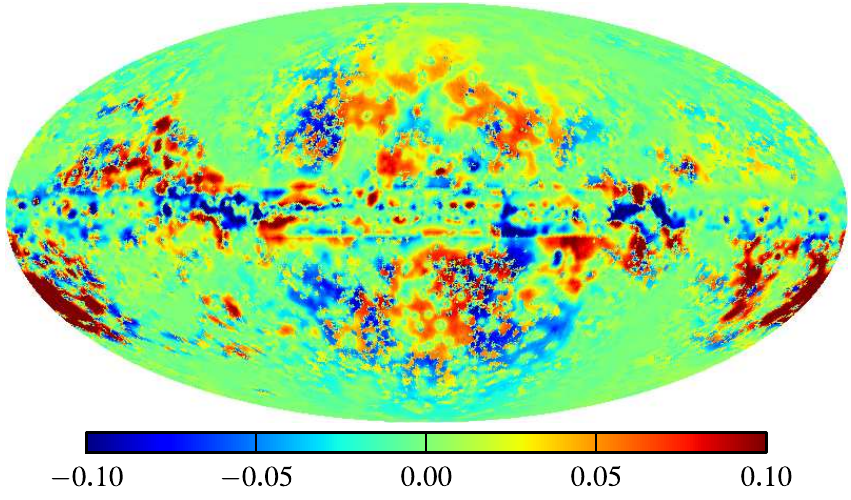}
\caption{
The difference between the absolute values of the fractional residuals of
model \model{S}{4}{20}{150}{5} and model \model{L}{6}{20}{\infty}{5} (top), model \model{S}{4}{20}{150}{5} and model \model{Y}{10}{30}{150}{2} (middle), and model \model{S}{4}{20}{150}{5} and
model \model{O}{8}{30}{\infty}{2} (bottom).
Negative pixels represent a better fit with model \model{S}{4}{20}{150}{5}, while positive pixels are better fit with the other models.
The maps have been smoothed with a $0.5^\circ$ hard-edge kernel, see
Figure~\ref{fig:sigmaResiduals}.
\label{fig:referenceDiffFractResiduals}}
\end{figure}

Finally, to illustrate the differences between models where more than
one parameter is changed, we compare in
Figure~\ref{fig:referenceDiffFractResiduals} models \model{L}{6}{20}{\infty}{5},
\model{Y}{10}{30}{150}{2}, and \model{O}{8}{30}{\infty}{2} to the reference 
model \model{S}{4}{20}{150}{5}.  
The models were chosen to illustrate the range of our parameter scan.
While some models fare better than others in the likelihood ratio tests
(see Figure~\ref{fig:gammaLikelihood}), there is no model that is 
uniformly better than the other models.
There are large-scale positive and negative differences between the models 
that are distributed over the entire sky, although the greatest differences
are at low latitudes.
We emphasise that differences between models can be non-linear, and 
caution should therefore be exercised when interpreting low-level residual
structures because of subtle interplay between the different model parameters.

\subsubsection{Plots of Spectra \label{sec:spectra}}

\begin{figure}
\epsscale{0.43}
\plotone{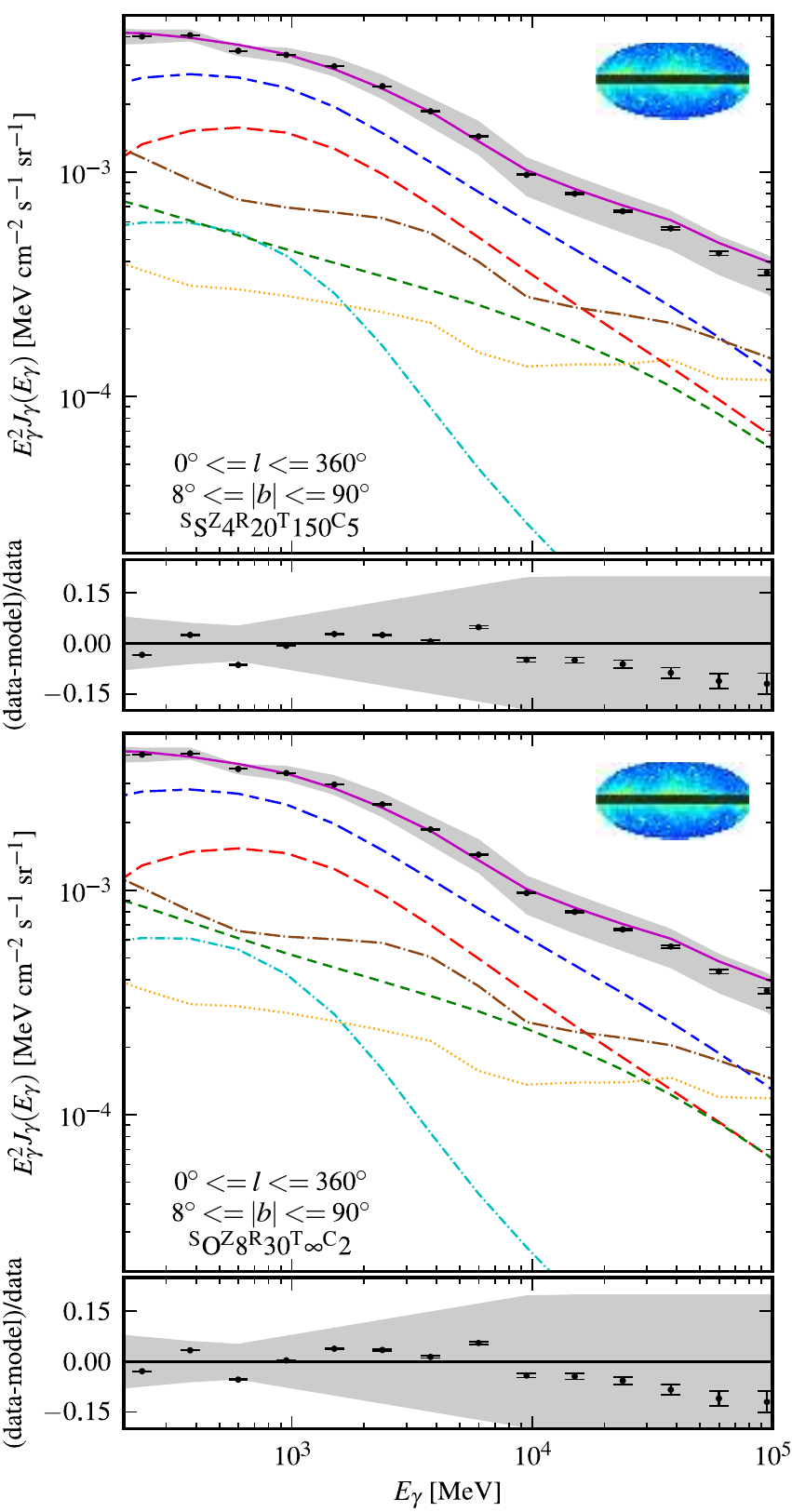}
\caption{Spectra extracted from the local region for model \model{S}{4}{20}{150}{5} 
 (top) and model \model{O}{8}{30}{\infty}{2} (bottom) along with the 
isotropic background (brown, long-dash-dotted) and the detected
sources (orange, dotted).
The models are split into the three basic emission components: 
$\pi^0$-decay (red, long-dashed), IC (green, dashed), and 
bremsstrahlung (cyan, dash-dotted).
All components have been scaled with parameters found from the \gray-fits.
Also shown is the total DGE (blue, long-dash-dashed) 
and total emission
including detected sources and isotropic background (magenta, solid).
The \fermilat{} data are shown as points and the error bars represent 
the statistical errors only that are in many cases smaller than the point size.
The gray region represents the systematic error in the \fermilat{} effective 
area.
The inset skymap in the top right corner shows the \fermilat{} counts in the 
region plotted.
Bottom panel shows the fractional residual $(data-model)/data$.
\label{fig:localRegion}}
\end{figure}

\begin{figure}
\epsscale{0.43}
\plotone{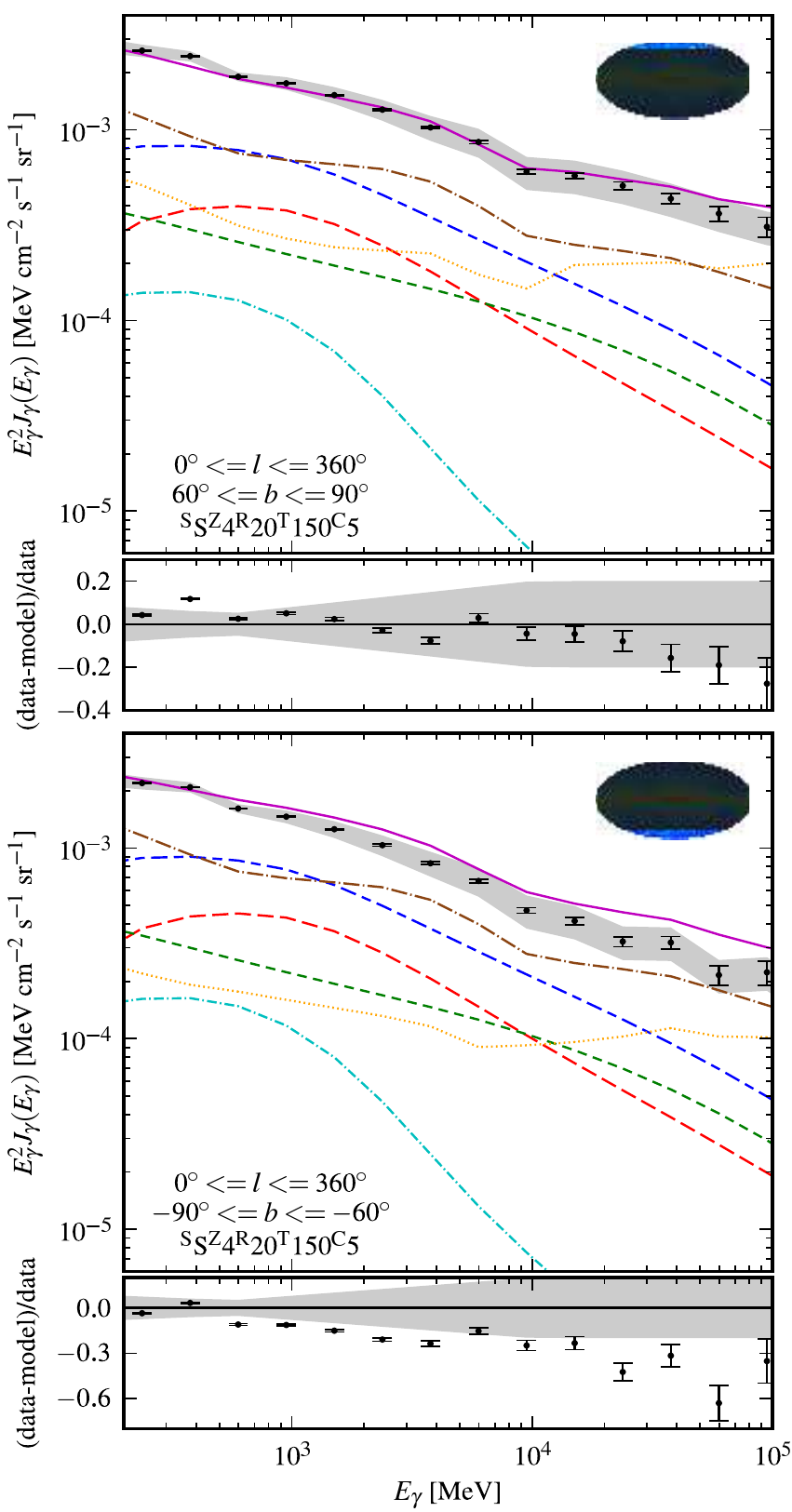}
\caption{Spectra extracted from the polar cap regions, north (top) and south
(bottom) for model \model{S}{4}{20}{150}{5}.  
See Figure~\ref{fig:localRegion} for legend.  
Note that the model shows a north-south asymmetry in the residuals that is most prominent at 
high energies but can be seen over the entire spectral range.
\label{fig:polarCapRegion}}
\end{figure}

\begin{figure}
\epsscale{0.43}
\plotone{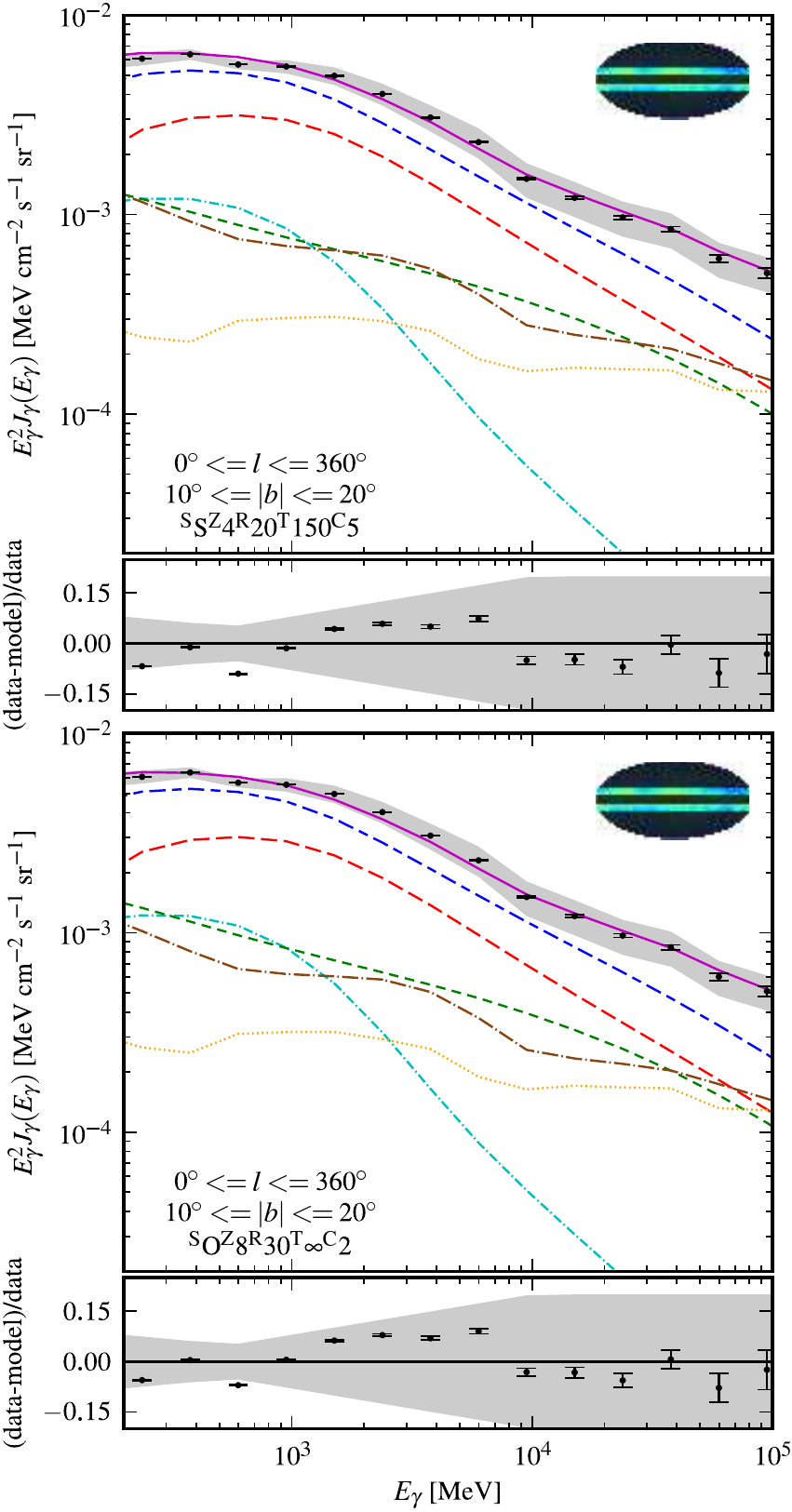}
\caption{Spectra of the low intermediate laditude region for model \model{S}{4}{20}{150}{5} (top) and model \model{O}{8}{30}{\infty}{2} (bottom).  
This region was also used by \citet{LAT:GeVExcess}.
See Figure~\ref{fig:localRegion} for legend.  
\label{fig:lowIntermLatRegion}}
\end{figure}

We plot the spectra of models and data for several selected regions.
It is evident from Figure~\ref{fig:localRegion} that the models give 
on-average a good description of the \fermilat{} 
data at high and intermediate latitudes.
Even though the likelihoods of the models differ significantly
the model predictions for the total intensity fall within the systematic error
of the \fermilat{} effective area, deviating less than 10\% from the data over 
the entire energy range.
This is partly due to the freedom we have when fitting for the isotropic
background (see Section~\ref{sec:isotropic}).
Figure~\ref{fig:polarCapRegion} shows that we over-predict the data in
the south polar cap, an indication that the isotropic component is too
large and compensating for inaccuracies in the DGE models.
But, even for the intermediate latitude region 
shown in Figure~\ref{fig:lowIntermLatRegion}, where the 
DGE dominates the isotropic component, the agreement is very good.

The models in the current paper agree better with the
intermediate latitude data
(Figure~\ref{fig:lowIntermLatRegion}) than the model presented in
\citet{LAT:GeVExcess} for two main reasons.
First, we use dust as an additional tracer for gas densities that has been
shown to give better results than using 
only \hi{} and CO tracers \citep{Grenier:2005}.
This is especially true for intermediate latitudes in the direction
toward the inner Galaxy, which is the brightest part of the low intermediate 
latitude region.
Second, we allow for freedom in both the ISRF scale factor and \Xco{} to tune
the model to the data, which is well motivated given the uncertainty in those
input parameters.

\begin{figure*}
\epsscale{0.85}
\plotone{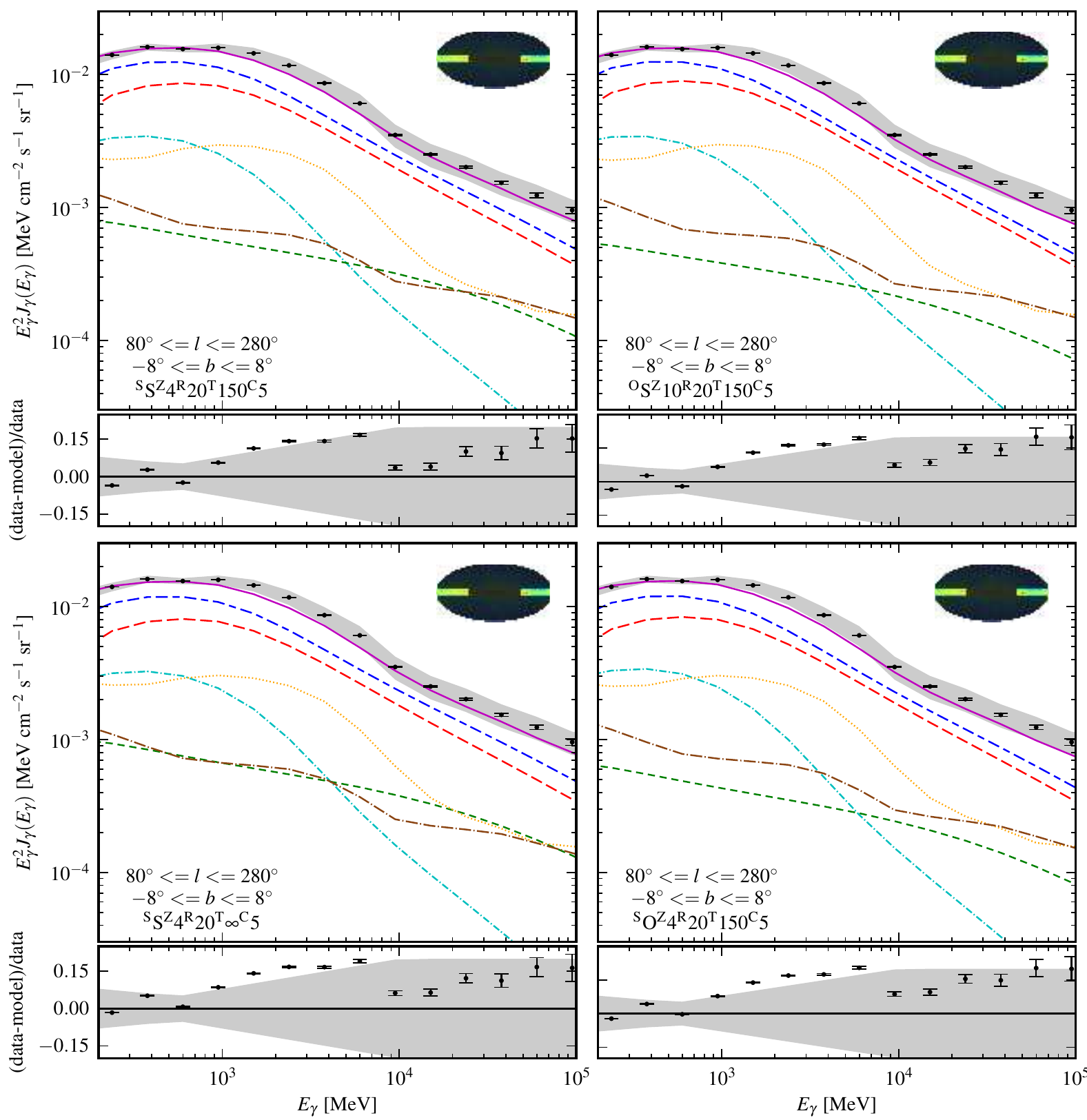}
\caption{Spectra extracted from the outer Galaxy region for model
\model{S}{4}{20}{150}{5} (top left), \model{O}{10}{20}{150}{5} (top
right), \model{S}{4}{20}{\infty}{5} (bottom left), and \model{O}{4}{20}{150}{5} (bottom right).
See Figure~\ref{fig:localRegion} for legend.  
\label{fig:outerRegion}}
\end{figure*}

The models in general do not fare as well in the Galactic plane where
they systematically under-predict the data above a few GeV but
over-predict it at energies below a GeV.
This is most pronounced in the inner Galaxy
(Figure~\ref{fig:innerRegion}), but can also be seen in the outer 
Galaxy (Figure~\ref{fig:outerRegion}), with even a small 
excess at intermediate latitudes (Figure~\ref{fig:lowIntermLatRegion}).
Possible explanations for this discrepancy are deferred to the
discussion section.
We note that the dip in the data visible between
10 and 20 GeV is due to 
the IRFs used in the present analysis.
Figure~\ref{fig:Pass7Spectra} shows a 
comparison of model \model{S}{4}{20}{150}{5} to the data
in the outer Galaxy using the Pass 7 clean photons. 
The dip between 10 and 20~GeV is greatly reduced because of the improved
effective area of the new photon class.
Because our results do not depend strongly on the exact shapes of the
spectra of the data, these improvements in the effective area do not 
affect our conclusions. 

The maximum likelihood trend of preferring models 
with larger $z_h$, lower $T_S$, and
flatter CR source distribution (see Figure~\ref{fig:gammaLikelihood}) is 
illustrated in Figure~\ref{fig:outerRegion}.  
Going from the SNR distribution to the OB star distribution has very similar
effects as changing from $T_S=150$~K to the optically thin approximation and
also increasing $z_h$ from 4~kpc to 10~kpc.  We note that changing $z_h$ with
the SNR distribution has a much smaller effect.
It is also evident
that all of the models under-predict the data in this region 
above $\sim 800$ MeV, and some even for the entire energy range.

\subsubsection{Longitude and Latitude Profile plots \label{sec:profiles}}

We compare longitude and latitude profiles of representative models and 
data for selected
regions.
For the profile plots, we use 3 energy bands (200~MeV--1.6~GeV, 1.6--13~GeV, 
and 13--100~GeV) to increase the 
statistics in the profiles.
Our discussion below is mainly focussed on the lowest energy band, because this
has the highest statistics and even though the PSF is broader than at higher
energies the profiles are wide enough to be relatively unaffected.
In general, the models agree well with the data, deviating 
less than $\sim 10$\% from the data over
a large fraction of the sky while covering almost 
2 decades of dynamic range in the latitude profiles.
From the profile figures, the component associated with CRs interacting 
with the \hi{} dominates the DGE in most sky 
regions and for most of the energy range of the \fermilat.
The IC component approaches a similar intensity to the \hi{} 
for high latitudes, and dominates only in the $13 - 100$ GeV energy band.
The H$_2$ component extends only a few degrees from the Galactic plane 
and is dominant only in the inner Galaxy.

\begin{figure*}
\epsscale{0.85}
\plotone{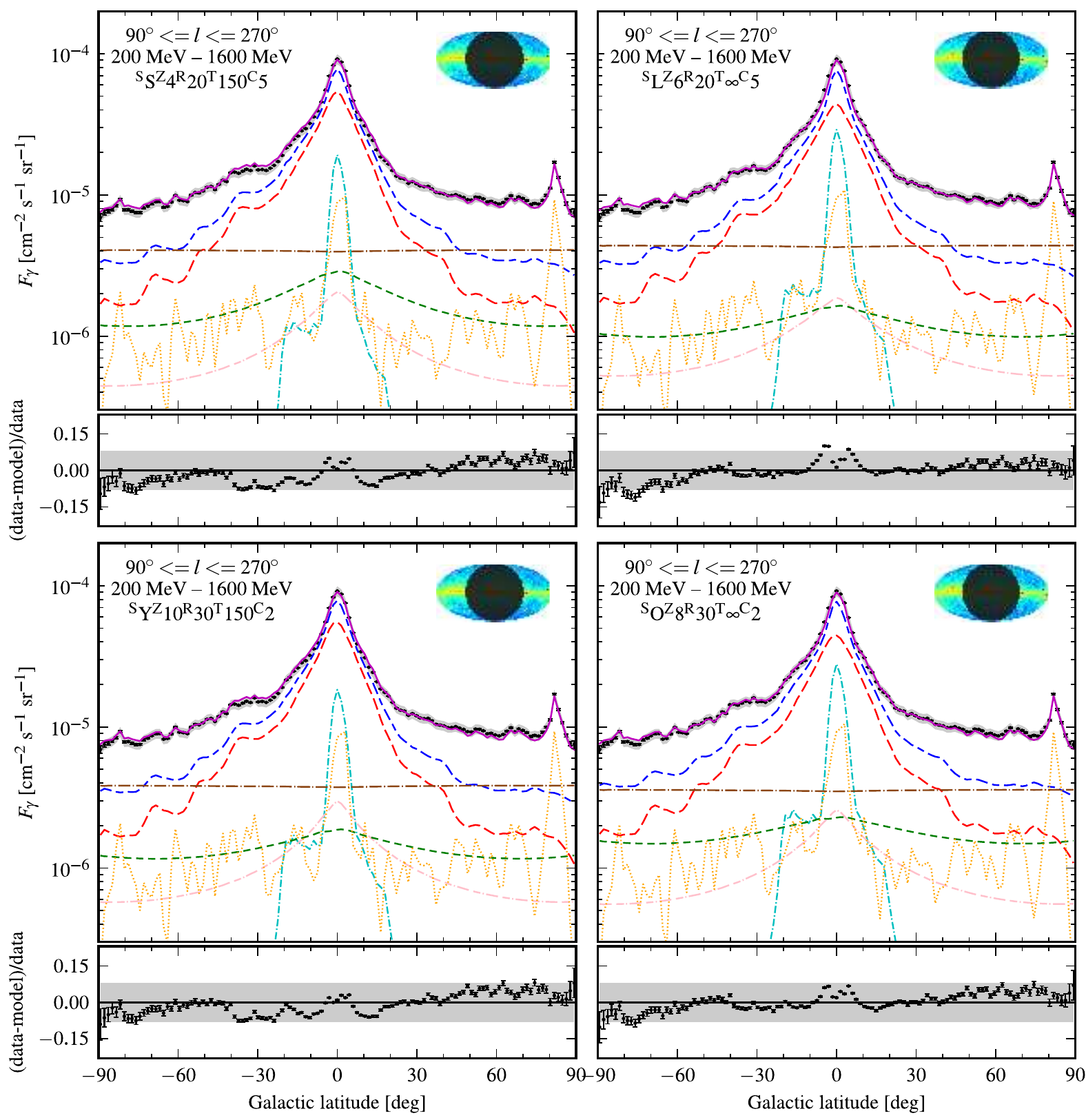}
\caption{Latitude profile showing the outer Galaxy in the energy range 
200~MeV -- 1.6~GeV.
Shown are profiles for models \model{S}{4}{20}{150}{5} (top left),
\model{L}{6}{20}{\infty}{5} (top right), \model{Y}{10}{30}{150}{2} (bottom left), and \model{O}{8}{30}{\infty}{2} (bottom right).
The DGE model is split into the three different gas components: \hi{} (red, long-dashed), H$_2$ (cyan, dash-dotted), and \hii{} (pink, long-dash-dash-dotted) and also IC (green, dashed).
Also shown are the isotropic component (brown, long-dash-dotted), the detected sources (orange, dotted), total DGE (blue, long-dash-dashed) and total model (magenta, solid).
\fermilat{} data are shown as points with statistical error bars and the systematic uncertainty in the effective area is shown as a grey band.
Due to the evenness of the sky exposure of the \fermilat, the systematic error is not expected to be position dependent, only global normalisation for the profile.
The inset skymap in the top right corner shows the \fermilat{} counts in the region plotted.
The bottom panel shows fractional residuals $(data-model)/data$.
\label{fig:latProfileOuterGalaxy200}}
\end{figure*}

\begin{figure}
\epsscale{0.85}
\plotone{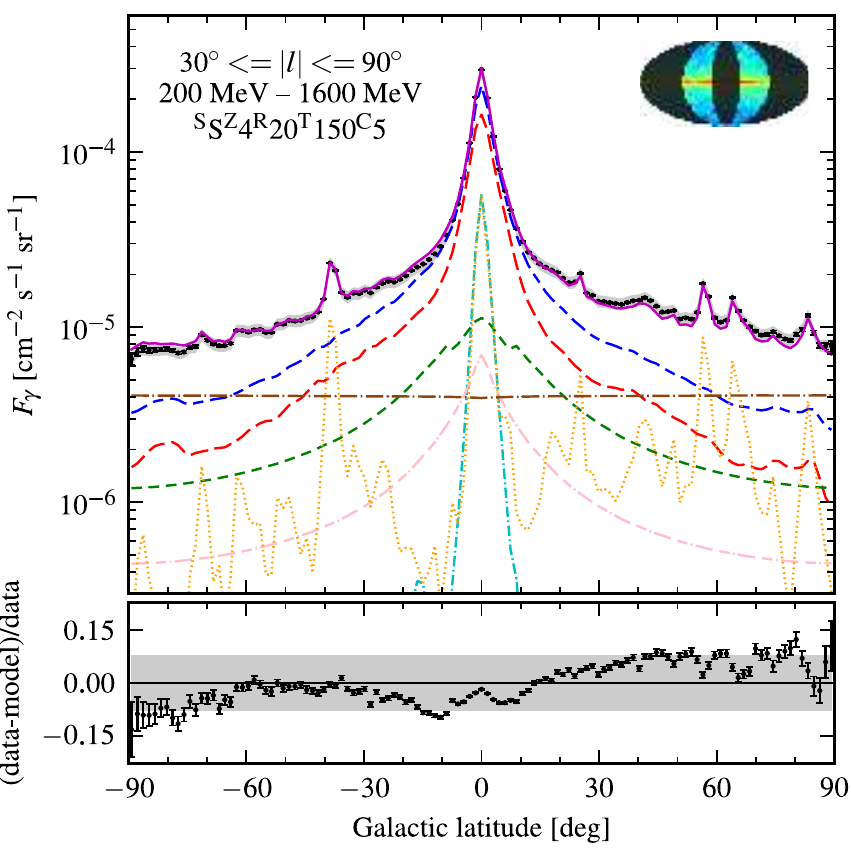}
\caption{Latitude profile for model \model{S}{4}{20}{150}{5} showing the 
inner Galaxy without the inner $\pm 30^\circ$ about the Galactic centre for 200 MeV -- 1.6 GeV.
See Figure~\ref{fig:latProfileOuterGalaxy200} for legend.
\label{fig:latProfileEastWestGalaxy200}}
\end{figure}

\begin{figure}
\epsscale{0.33}
\plotone{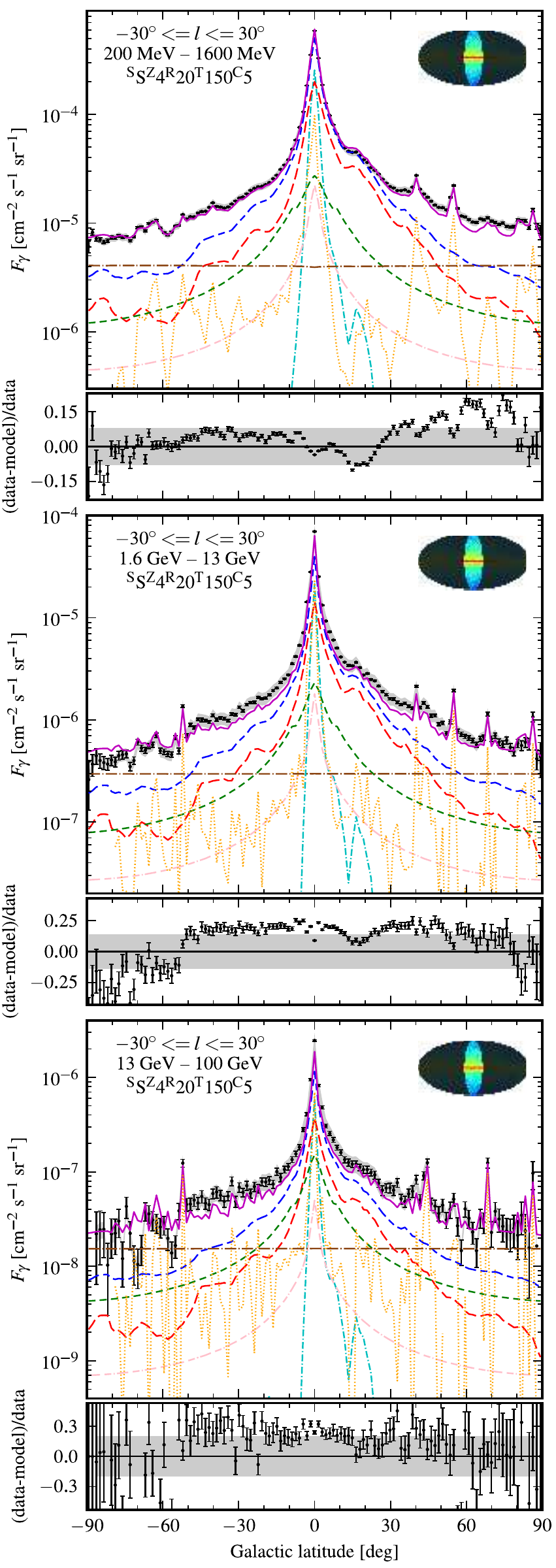}
\caption{Latitude profile for model \model{S}{4}{20}{150}{5} showing the innermost $l \pm 30^\circ$ about 
Galactic centre for 200 MeV -- 1.6 GeV (top), 1.6 GeV -- 13 GeV (middle), and
13 GeV -- 1000 GeV (bottom).
See Figure~\ref{fig:latProfileOuterGalaxy200} for legend.
\label{fig:latProfileGalacticCenter}}
\end{figure}

Despite the overall good agreement, the profile residuals do show 
structure on scales from few degrees to tens of degrees.
For the latitude profile in the outer Galaxy shown in
Figure~\ref{fig:latProfileOuterGalaxy200}, it is evident that 
the models under-predict the data in the Galactic plane, but over-predict 
it at intermediate latitudes.
The exact shape and magnitude of this residual depends on the model. 
The under-prediction in the plane is mostly dependent on the CR flux in 
the outer Galaxy (CR source distribution and halo height), while 
the over-prediction at intermediate latitudes depends mostly on the 
assumed $T_S$ value and therefore gas-to-dust ratio 
(see Section~\ref{sec:dust}). 
These effects can be seen also toward the inner Galaxy 
(Figure~\ref{fig:latProfileEastWestGalaxy200}), 
but the effect is mostly absent toward the Galactic 
centre (Figure~\ref{fig:latProfileGalacticCenter}).
The residual map differences in Figures~\ref{fig:distDiffFractResiduals}
and~\ref{fig:gasDiffFractResiduals} also illustrate this.

\begin{figure}
\epsscale{0.63}
\plotone{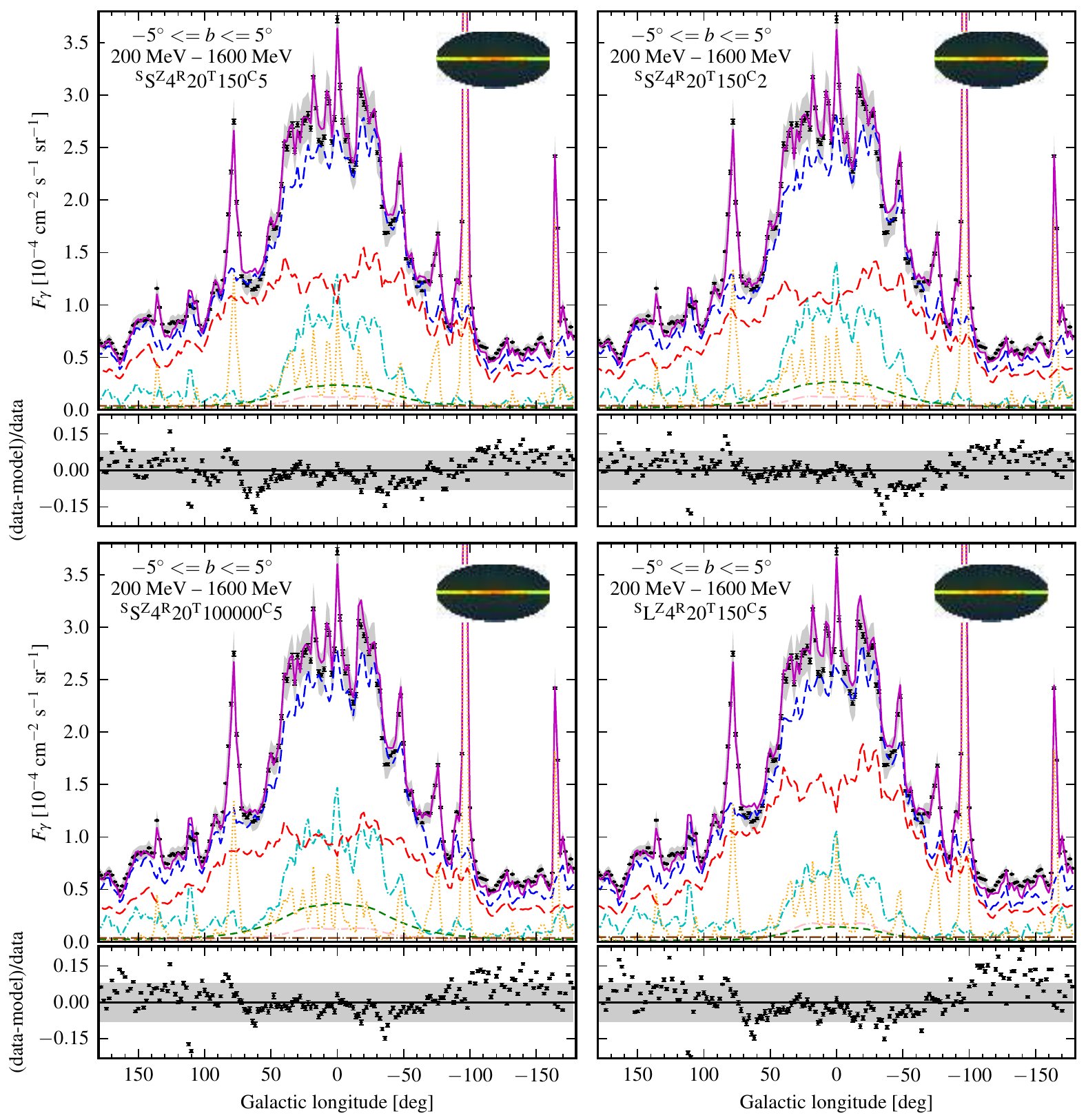}
\caption{Longitude profile showing the Galactic plane in the energy range 
200 MeV -- 1.6 GeV.
Shown are profiles for model \model{S}{4}{20}{150}{5} (top left), \model{S}{4}{20}{150}{2} (top right), \model{S}{4}{20}{\infty}{5} (bottom left), and \model{L}{4}{20}{150}{5} (bottom right).
See Figure~\ref{fig:latProfileOuterGalaxy200} for legend.
\label{fig:lonProfileGalacticPlane200}}
\end{figure}

The dip around the Galactic plane in the residual in
Figure~\ref{fig:latProfileOuterGalaxy200} is caused by unreasonably 
large \Xco{} factors found
from the fits (see Section~\ref{sec:Xco}), artificially increasing 
the H$_2$ component.
A residual structure coincident with the H$_2$ component is not seen 
in any of the other latitude profiles.
The under-prediction in the outer Galaxy can also be seen in the
longitude profiles in the Galactic plane
(Figure~\ref{fig:lonProfileGalacticPlane200}) where peaks in the H$_2$
component corresponds with dips in the residual.
The contribution from detected point sources is also strongest in the 
plane with a similar profile as the H$_2$ component, which
can also compensate for a lack of freedom in the DGE model 
during the fitting procedure.
The longitude profile in the Galactic plane does not show a correlation
of peaks in the source intensity and dips in the residual indicating
that sources from the 1FGL catalogue are not able to compensate for large scale
inaccuracies in the diffuse emission.

All of the latitude profiles display a north-south asymmetry
in the residuals, as was shown in the spectra of the polar cap regions in
Figure~\ref{fig:polarCapRegion}.  
The effect is most noticeable in 
Figure~\ref{fig:latProfileEastWestGalaxy200}, which is 
caused mostly by the gas from the Magellanic 
stream \citep{1974ApJ...190..291M} that was not removed from the
\hi{} annular column density maps as mentioned earlier.
As the north-south asymmetry is also visible in the
outer Galaxy profile where the Magellanic stream has very little effect,
there must be some underlying asymmetry.
The origin of this asymmetry is not currently known. 
It is more likely associated with an
asymmetry in the CR flux rather than the ISM, because the ISM 
is more observationally constrained. 

\begin{figure}
\epsscale{0.53}
\plotone{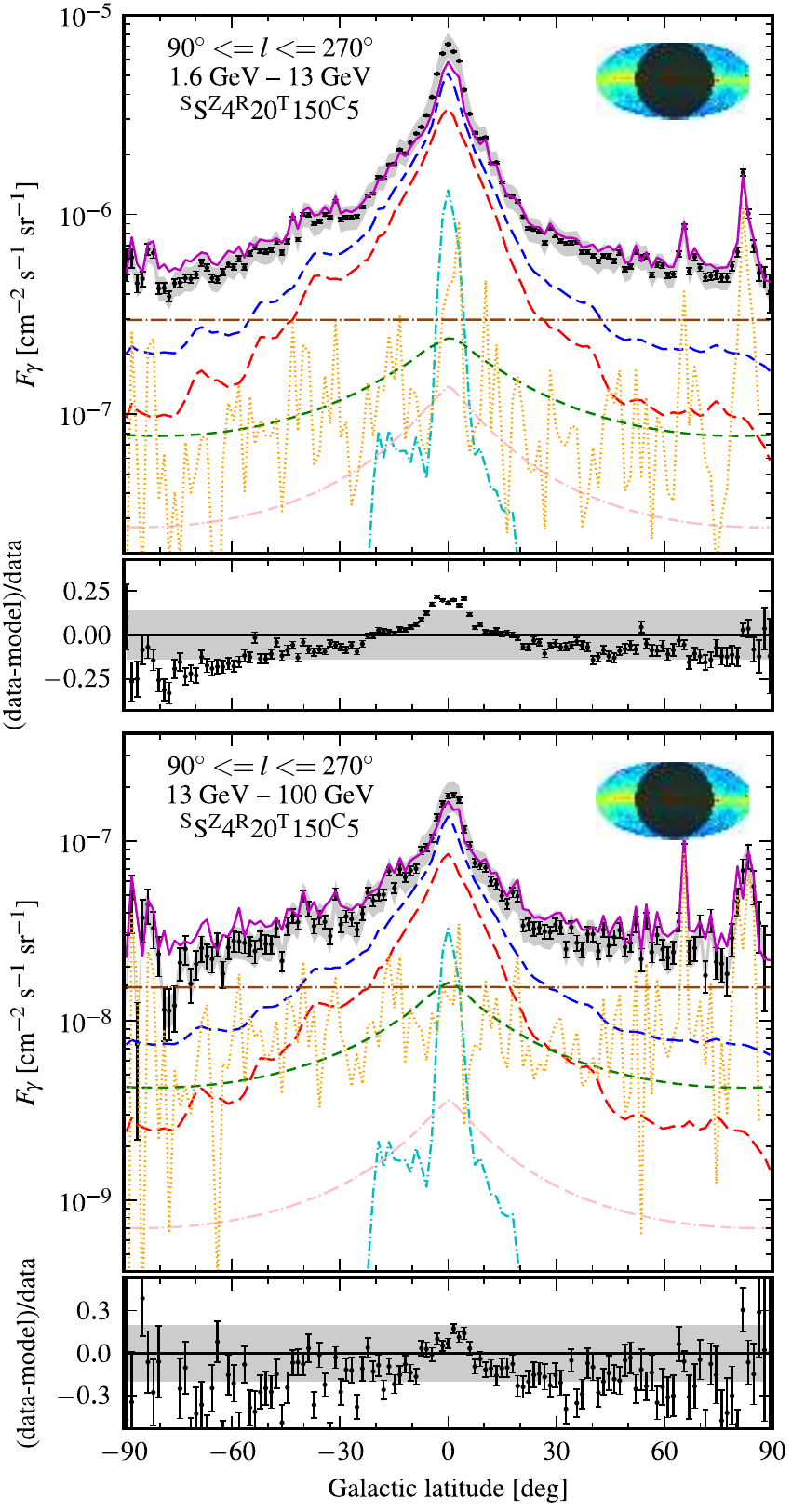}
\caption{Latitude profile showing the outer Galaxy in the 
$1.6 - 13$ GeV (top) and $13 - 100$ GeV (bottom) energy range 
for model \model{S}{4}{20}{150}{5}.
See Figure~\ref{fig:latProfileOuterGalaxy200} for legend.
\label{fig:latProfileOuterGalaxyOther}}
\end{figure}

The model under-prediction above a few GeV seen in 
Figure~\ref{fig:innerRegion} and
Figure~\ref{fig:outerRegion} is confined to the Galactic plane, as can be
seen in Figure~\ref{fig:latProfileOuterGalaxyOther}.
The model systematically under-predicts the data in the plane in the 
$1.6 - 13$ GeV and $13 - 100$ GeV energy bands, but very 
little excess emission is seen at higher latitudes.
This is not seen as clearly in the Galactic centre profile
(Figure~\ref{fig:latProfileGalacticCenter}) because that 
region also includes other large scale residuals, most notably due to features 
coincident with those described by 
\citet{2010ApJ...724.1044S} and \citet{2010ApJ...717..825D}.  
Note that while these are prominent above 1.6 GeV, 
they can also be seen at lower energies, but the details of the residual 
features depend on the DGE model.

\begin{figure}
\epsscale{0.53}
\plotone{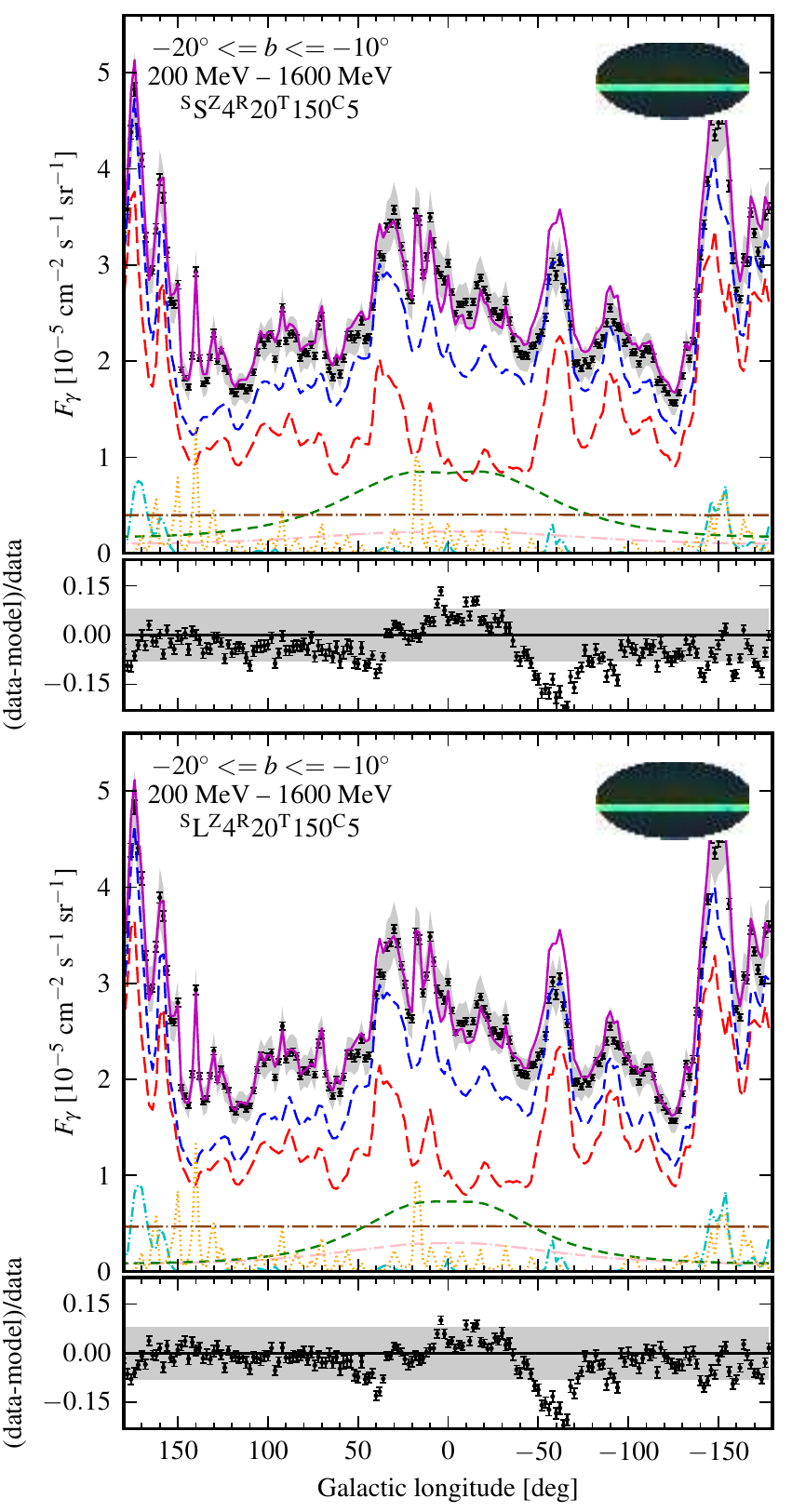}
\caption{Longitude profile for models \model{S}{4}{20}{150}{5} (top) and \model{L}{4}{20}{150}{5} (bottom) showing south intermediate latitudes in the energy 
range 200 MeV -- 1.6 GeV.
See Figure~\ref{fig:latProfileOuterGalaxy200} for legend.
\label{fig:lonProfileSouthIntermediateLat200}}
\end{figure}

Figure~\ref{fig:lonProfileGalacticPlane200} shows the longitude profile 
about the Galactic plane for a few different models.
It shows how the \hi{} component is affected by 
different assumptions for $T_S$, the magnitude cut in the dust map, and 
the different CR source distributions.
The difference in the CR source distribution is also seen in the IC component
that is more peaked for the Lorimer source distribution than the SNR
distribution.  
This can be better seen at intermediate latitudes in
Figure~\ref{fig:lonProfileSouthIntermediateLat200}.
The effect is noticeable both at intermediate latitudes as well as in the outer
Galaxy where CO from the local annulus dominates.

\begin{figure}
\epsscale{0.53}
\plotone{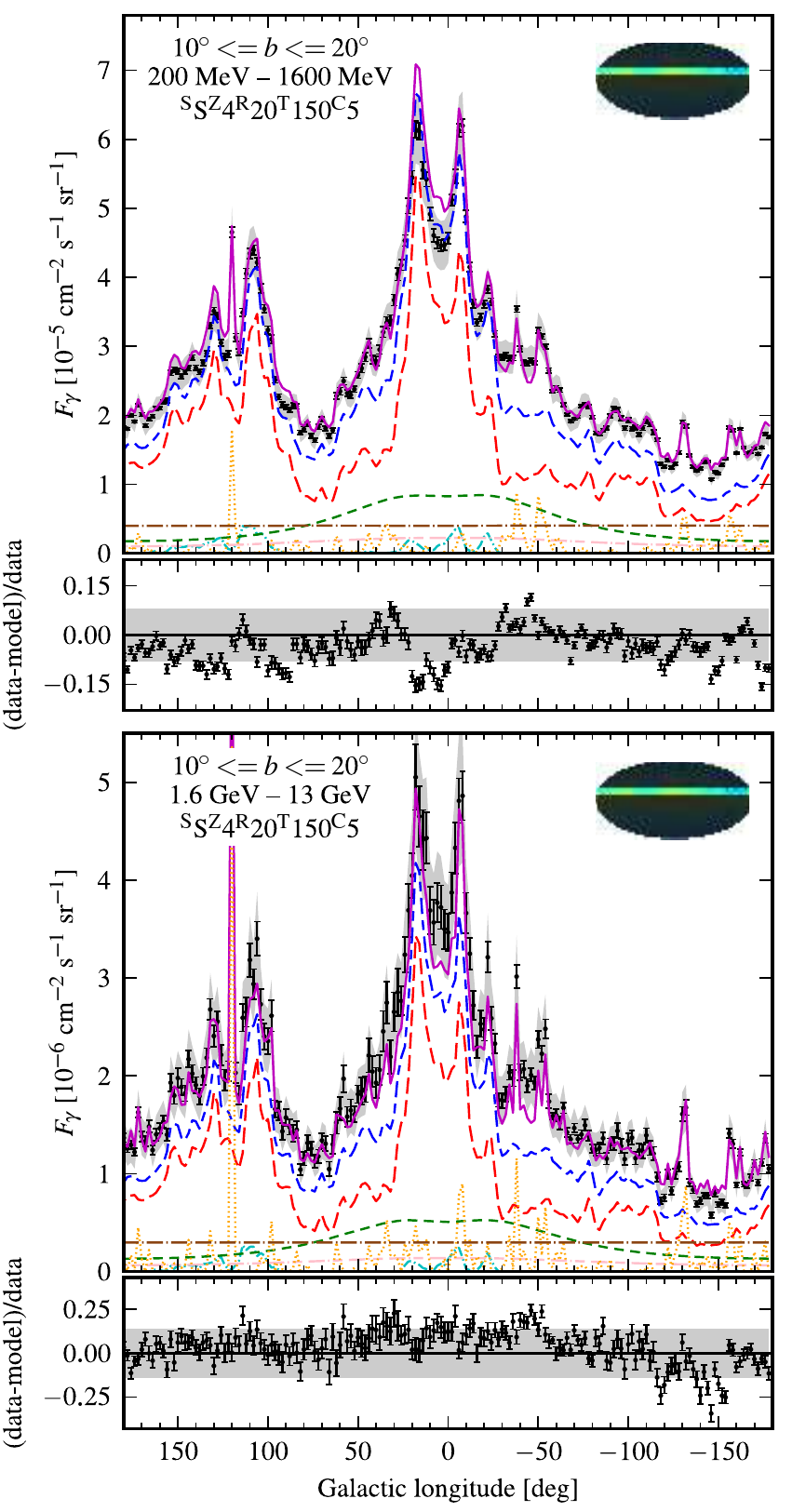}
\caption{Longitude profile for model \model{S}{4}{20}{150}{5} showing north intermediate latitudes in the 
energy range 200 MeV -- 1.6 GeV (top) and 1.6 GeV -- 13 GeV (bottom).
See Figure~\ref{fig:latProfileOuterGalaxy200} for legend.
\label{fig:lonProfileNorthIntermediateLat}}
\end{figure}

The residuals in the plane show signs of small-scale features that are not
compatible with statistical fluctuations.
Similar residual structure is also seen at intermediate latitudes in
Figure~\ref{fig:lonProfileSouthIntermediateLat200}
and Figure~\ref{fig:lonProfileNorthIntermediateLat}, 
where the most significant
structures in the residuals are correlated with peaks in the \hi\
distribution.  
Note that some peaks in the \hi{} distribution are not
associated with residual structure. 
It is unlikely that the small angular scale fluctuations are due to 
small-scale CR intensity variations because the bulk of the CR
nuclei producing 
the DGE for the energy range shown are smoothly distributed. 
The variations are then mostly caused
by features in the annular gas maps that introduce artifacts 
on small angular scales.
This suggests that the gas-to-dust ratio is not constant over the sky and 
can fluctuate by at least 10\%. 
However, comparing the 
panels in Figure~\ref{fig:lonProfileNorthIntermediateLat},
the residual structure can be seen to be energy dependent.
The largest variation is towards the inner Galaxy that can be associated
with structure coincident with 
those identified by \citet{2010ApJ...724.1044S} and
\citet{2010ApJ...717..825D} but smaller variations around $l=100\degr$ 
indicate spectral variations in the CR flux.
See, e.g., \citet{Bykov:1992} for how OB-associations and super-bubbles
might have an effect on the CR flux on smaller spatial scales.

\subsection{Radial dependence of \Xco}
\label{sec:Xco}

\begin{figure}
\epsscale{0.93}
\plotone{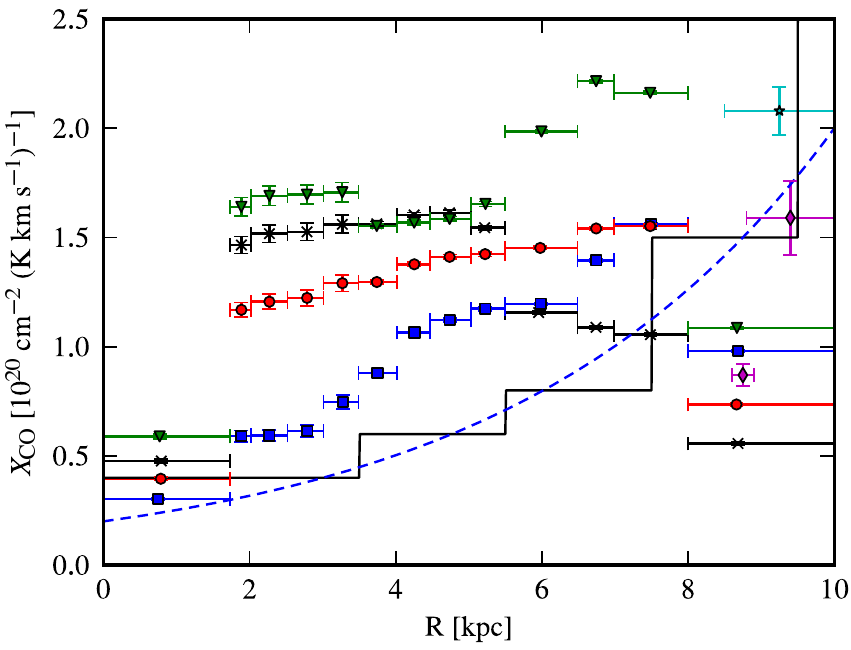}
\caption{Radial distribution of \Xco{} for model \model{S}{4}{20}{150}{5} (black X), \model{L}{6}{20}{\infty}{5} (blue
squares), \model{Y}{10}{30}{150}{2} (red circles), and \model{O}{8}{30}{\infty}{2} (green triangles).
We do not show the \Xco{} values in the outer Galaxy because they are strongly
biased by the lack of \gray{} intensity in the outer Galaxy in our models.
For comparison, we also show data from \citet{2010ApJ...710..133A} 
(purple diamonds), \citet{2011ApJ...726...81A} (cyan stars), and
\citet{2004A&A...422L..47S} (solid curve).
The blue dashed curve shows the initial value we used in our iterative
procedure.
\label{fig:XCO_radial}}
\end{figure}

Figure~\ref{fig:XCO_radial} shows the radial dependence of \Xco{} for a 
few selected models.
\Xco{} for all models can be found in the online supplementary material.
Our analysis finds that \Xco$(R)$ depends both on the assumed CR source
distribution and the gas properties.  
This is illustrated in
Figure~\ref{fig:XCO_annuli}, which shows \Xco{} derived
for the local annulus for all models.  
The local \Xco{} varies by
up to a factor of 2 depending on the value of $T_S$ and the \ebv\
magnitude cut but is nearly independent of the other input parameters.
Because the emissivity of the local gas is well determined by observations of
CRs, this shows that an accurate determination of the \hi{} column
density is important for determining \Xco{} values from \gray{} data.

The scatter in the \Xco{} is not surprising.
The limited freedom in
the \gray{} fit can bias the derived values as has already been mentioned.
For \Xco{} the bias can be twofold.  
For an accurate determination of \Xco{} we need to know the \gray{} intensity 
associated with CO as well as the emissivity per H$_2$ atom.  
For our case, we
calculate the emissivity assuming a CR distribution and estimate the intensity
from a fit to the \gray{} data.  
If the emissivity is incorrectly estimated, the
intensity associated with CO will be biased in the opposite direction, 
resulting in the twofold bias. 
Methods that determine the emissivity of
the gas simultaneously with the intensity associated with CO are independent
of this bias, as long as the emissivity is accurately determined from the
data \citep[e.g.,][]{2010ApJ...710..133A,2011ApJ...726...81A}  
Note that the effect of variations in N(\hi) applies in all cases.
However, the scatter in the \Xco{} we determine does not significantly 
affect our comparison between the models except
possibly in the inner Galaxy where the molecular gas content is greatest.

Despite the large variations of the \Xco{} factors between the models 
there are several features that are consistent.
The \Xco{} factors in the inner Galaxy are systematically higher than
the estimate from \citet{2004A&A...422L..47S}, even when using the 
same CR source distribution.
Only in the innermost annulus is there agreement between our
\Xco{} values and those of \citet{2004A&A...422L..47S}.  
The strong decrease of \Xco{} in the innermost
annulus is consistent for all our models as has also
been seen in other analyses \citep[see e.g.,][]{2007A&A...467..611F}.
This has been attributed to the breakdown of \Xco{} as a tracer of H$_2$ 
because the $^{12}$CO line becomes less optically thick in the 
Galactic centre region \citep{Dahmen:1998}.
There also seems to be a dip in \Xco{} for the local annulus that was
not in \citet{2004A&A...422L..47S}.
Our values for the local region \Xco{} agree very well with the value
found for the nearby Gould belt by \citet{2010ApJ...710..133A}.
Because the \Xco{} values for this annulus are determined from fits to the
local region, the value is associated with high latitude clouds.
This indicates that molecular clouds in the vicinity of the solar system may 
have
different properties than clouds at a similar Galactocentric distance.
High latitude translucent clouds have also been shown before to have 
lower \Xco{} values \citep{deVries:1987, Heithausen:1990} but more recent work 
based on other tracers of molecular hydrogen show that CO intensities 
might not be linearly related to H$_2$ column densities in those 
clouds \citep{Magnani:2003}.

\begin{figure}
\epsscale{0.93}
\plotone{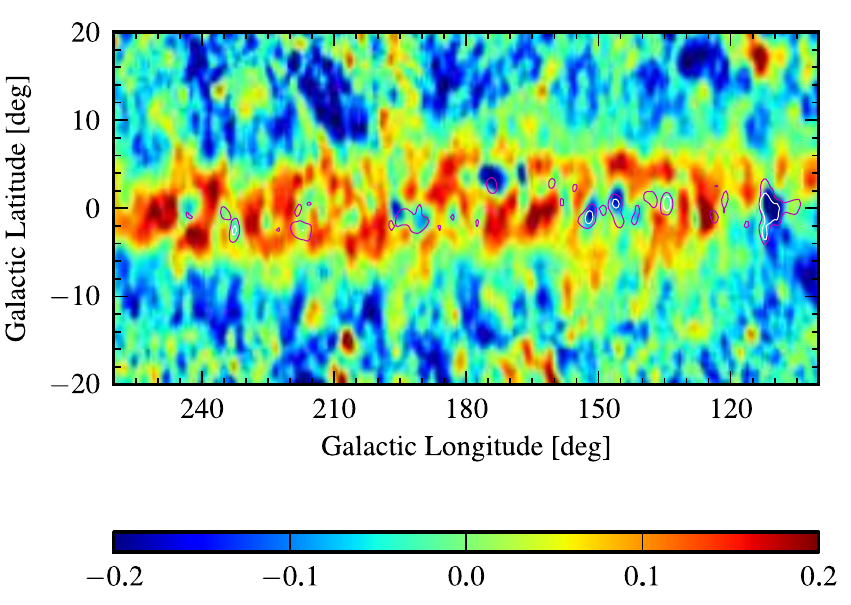}
\caption{Relative residual $(data-model)/data$ for model \model{S}{4}{20}{150}{5}.
Overlaid are contours for the integrated CO emission in the outer Galaxy ($R > 10$~kpc) at
2~K~km~s$^{-1}$ (magenta) and 5~K~km~s$^{-1}$ (white).
The CO is clearly correlated with negative residuals in the map indicating
that our \Xco{} factors in the outer Galaxy are overestimated.
Note that the latitude scale is stretched 2 times compared to the longitude scale for clarity.
\label{fig:COouterGalaxy}}
\end{figure}

There is an exponential increase in the outer Galaxy that is
strongly dependent on the assumed CR source
distribution, halo size, and $T_S$.
Figure~\ref{fig:COouterGalaxy} shows the fractional residual (see
Section~\ref{sec:visuals}) for model \model{S}{4}{20}{150}{5} with the 
integrated CO emission in the outer Galaxy overlaid.
A correlation between the CO emission and negative residuals is evident.
On this basis, we conclude that the \Xco{} values in the outer Galaxy 
derived in our analysis are strongly biased
and we do not show them in Figure~\ref{fig:XCO_radial}.
This bias is caused by the under-prediction of \gray{} intensity in the outer
Galaxy by all of the models considered here. 
Because there is very little CO in the outer Galaxy (see
Figure~\ref{fig:COrings}) this bias will not strongly affect our
results, only slightly reduce the scatter when comparing the models in
the outer Galaxy.

\subsection{ISRF scaling factors \label{sec:ISRFscale}}
The scaling factor of the ISRF is shown
in Figure~\ref{fig:ISRFscale} for each model in both the local and inner region.
The derived ISRF scaling factor is model dependent and 
varies with CR source distribution, gas densities, and halo size.
In general, the scaling factor is smaller for the pulsar CR source 
distributions that are more peaked towards the inner Galaxy than both 
the OB stars and SNR distributions.
More CRs will be injected in the inner Galaxy from the pulsar-like 
source distributions.
This produces a corresponding increase in the IC emission in this region 
because of the larger number of CR electrons and positrons for these source
distributions.

The ISRF scaling factor is also dependent on $T_S$ and the \ebv{} magnitude
cut, indicating that the normalisation of the ISRF (and of the IC intensities) 
serves to compensate for different gas densities in the fits.
This is despite the IC component being both spectrally and also 
spatially different from the \hi{} component.
The latitude dependence of the IC component is similar enough to the \hi\
component (see Figure~\ref{fig:latProfileGalacticCenter}) to allow for the
correlation between the \hi{} and IC components in the fit.
Coupling that with the trend between increased likelihood in the inner region
for the optically thin case seems to indicate that a greater intensity of IC 
is needed for all models, either
from an increased ISRF, more CR electron sources, or a larger $z_h$.
The longer confinement time for the CR electrons/positrons for the 
larger halo sizes, together with the approximately $1/z$ decrease 
of the ISRF perpendicular to the Galactic plane \citep{Porter:2008}, 
produces more IC emission for cases with 
larger $z_h$ \citep[see also][]{LuminosityPaper}.
This effect can be seen as a decrease in the scaling factor in the local 
region with
increasing $z_h$, although the magnitude of the decrease depends on the 
assumed CR source distribution and gas densities.
The ISRF scaling factors follow a trend that they are larger for the 
inner Galaxy region for larger $z_h$. 
This indicates that further enhancement is required in addition to that 
provided by the
large $z_h$, which could be due to additional CR electrons/positrons, 
or an increase in the ISRF in the inner Galaxy region.
Disentangling these effects is difficult.
One way could be to consider the IC from CMB photons along with the 
bremsstrahlung emission.
Unfortunately, their overall intensity is not high enough
to allow them to be used to constrain the CR electron component.
The ISRF scaling factors therefore only allow us to infer that 
the IC modelling requires 
modifications but a detailed investigation is needed to unravel their origin. 

\subsection{Isotropic background \label{sec:isotropic}}

\begin{figure}
\epsscale{0.93}
\plotone{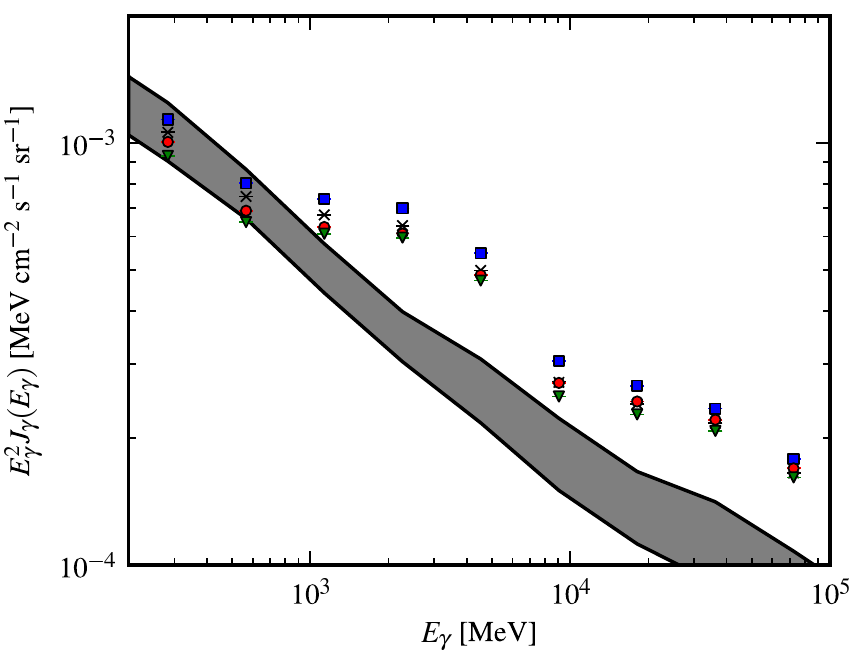}
\caption{The spectra of the isotropic component determined from the local
region fit for model \model{S}{4}{20}{150}{5} (black X), \model{L}{6}{20}{\infty}{5} (blue squares), \model{Y}{10}{30}{150}{2} (red circles), and \model{O}{8}{30}{\infty}{2} (green triangles).
The grey shaded area is the isotropic component from \citet{EGBpaper}
combining their EGB and residual component, including the 
effective area systematic uncertainty. 
Other systematic 
uncertainties estimated by \citet{EGBpaper} are not included in the figure.
Their magnitude could be comparable to the effective area systematic uncertainty. 
\label{fig:isotropic}}
\end{figure}

The spectrum of the derived isotropic background is shown in 
Figure~\ref{fig:isotropic} for selected models.
Isotropic 
spectra for all models can be found in the online supplementary material.
As for the ISRF normalisation and \Xco{} values, the isotropic background 
is model dependent.
However, the variation in the overall normalisation is not very 
large because it is spatially very different from the other 
fitted components.
The derived spectral shape is also very similar amongst the various models.
Both of these indicate that the isotropic background 
is not strongly biased by variations between the different models.

Figure~\ref{fig:isotropic} also shows a comparison with the isotropic 
spectrum from \citet{EGBpaper} after combining the derived EGB and RIB
components from that work.
Our estimates of the isotropic background are consistent with \citet{EGBpaper} 
below 1~GeV, 
and systematically higher above this energy.
Despite our efforts to minimise the contribution of the Galactic ridge
to the determination of the isotropic background by fitting for the
local region, Figure~\ref{fig:lowIntermLatRegion} shows that the model 
slightly underestimates the data at low intermediate latitudes above 1~GeV 
while still
being within the systematic uncertainty of the \fermilat{} data.  
The
fit will compensate for this with the isotropic component as can be seen
in Figure~\ref{fig:polarCapRegion} where the model 
overestimates the data in the polar cap regions above 1~GeV.
This is especially true in the south polar cap.
The difference in the two estimates of the isotropic background can 
therefore be attributed to the
additional freedom allowed in the diffuse Galactic emission model in the
analysis of \citet{EGBpaper} where the intensity spectrum of the local
\hi{} annulus and the IC component was allowed to vary freely.  
The motivations for this additional freedom were 
the uncertainties associated with the observed CR intensities that are 
around few percent at energies above 100 GeV reaching more than ten 
percent below 10 GeV caused by the uncertainty in solar modulation, and 
the size of the CR halo.
The combination of these can produce variations both in the \hi-related and 
IC emission.
Because our models
predict the \fermilat{} data  within the systematic error we do not
try to account for this uncertainty in this analysis.
The isotropic spectrum from \citet{EGBpaper} is therefore a better measurement.
This does not affect the comparison between the models considered in this 
paper because they are all treated identically.

\subsection{CR Propagation Parameters}

Because the main purpose of the fit to CR data is to obtain a propagation 
model consistent with CR observations, we defer most of the discussion 
of the results to Appendix~\ref{appendix:CR} and only summarise the 
few key points here.
We emphasise that all the models give a good
representation of the CR data as can be seen in
Figure~\ref{fig:CRspectrumPrimary} and Figure~\ref{fig:CRspectrumSecondary}. 
This has been shown earlier for similar diffusive reacceleration 
models~\citep{1998ApJ...509..212S}.
But, models with $z_h = 10$ kpc are at the limit of consistency with
the observed $^{10}$Be/$^9$Be ratio and therefore considering 
larger values for $z_h$ is not warranted. 

The propagation parameters from the fits shown in
Figure~\ref{fig:CRparametersNuc} and Figure~\ref{fig:CRparametersNuc2} 
are generally in agreement with values found
from similar analyses~\citep{Strong:2004,Trotta2010}.
The values are also stable between the different models, with notable 
exceptions for $D_0$ and $\valf$.
These parameters are strongly correlated with $z_h$ and the properties of the
gas distribution, that is the \Xco{} profile\footnote[13]{The analytical \hi\
distribution in \GP{} used for the propagation calculation is not 
corrected for the assumed $T_S$ or dust
correction. 
Therefore, it is only the \Xco{} profile that differs between models.  
This simplification does not change our results qualitatively.}.  
The values of $D_0$ and $\valf$ are also slightly affected by the assumed CR 
source distribution.
Therefore, the values for these parameters should be interpreted in the context
of the entire model that they were determined from because they depend not 
only on the propagation set up, but also the distributions of gas and 
assumed CR sources.

\section{Discussion}

\begin{figure}
\plotone{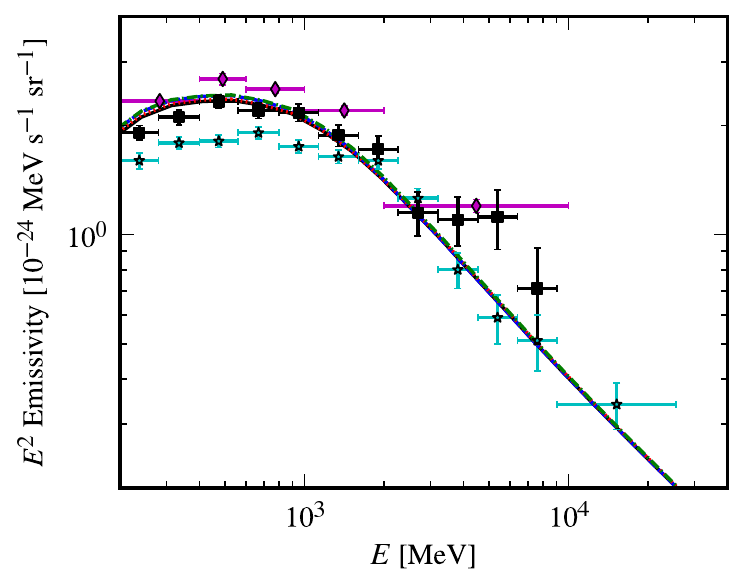}
\caption{
The average emissivity of the local annulus for model
\model{S}{4}{20}{150}{5} (solid black), \model{L}{6}{20}{\infty}{5}
(blue dashed), \model{Y}{10}{30}{150}{2} (red dotted), and
\model{O}{8}{30}{\infty}{2} (green dash-dotted).  Shown for comparison
are emissivities derived from \fermilat{} data using a template fitting
approach.  Cyan stars are from \cite{2011ApJ...726...81A}, magenta
diamonds are from \cite{2010ApJ...710..133A}, and black squares are
from \cite{LAT:HIEmissivity}.
\label{fig:localEmissivity}}
\end{figure}

The agreement between all our models and the \fermilat{} data is, on average, 
good despite minimal fitting.
The models generally agree within $\approx 15$\% of the data over most of the 
sky except for regions discussed below. 
In Figure~\ref{fig:localEmissivity} we show the predicted
emissivities of the local annulus for a few example models and compare
them with emissivities derived from the \fermilat{} data using different methods
\citep{2010ApJ...710..133A,2011ApJ...726...81A, LAT:HIEmissivity}.
The agreement is in line with our results from the local region; the
models predict the local emissivities within the scatter of the observations.
The scatter between different \fermilat{} results is most likely caused
by uncertainties in the column densities of gas used in the template
fitting.

The parameter scan was deliberately limited due to computational constraints.
However, it does provide insight into the effects associated with the 
variations of different parameters.
For a given assumption on the CR propagation model, 
variations of the gas-related parameters give the largest variations 
in log-likelihoods for the \gray{} fit over the entire sky. 
The CR source distribution and halo size, however, show such large effects 
on the likelihood ratio in the outer Galaxy only.
Similarly, the gas-related parameters have a large
effect on the residual sky maps at all angular scales, which contrasts 
with the much smoother structures caused by changes in the assumed 
CR source distribution and size of the confinement volume.
Understanding the Galactic gas distribution and its different phases is 
essential for accurate modelling of the DGE.

The most important parameter in our scan is the \hi{} spin 
temperature used for the optical depth correction for deriving N(\hi).
For our analysis, we considered two values, $T_S$ = 150 K and $T_S$ very 
large (i.e., the optically thin approximation).  
This is a highly simplified approach because observations
show $T_S$ varies from 10s~to 1000s~of~K 
\citep[e.g.,][]{2009ApJ...693.1250D}.  
A preliminary study showed that alternative
approximations for the value of $T_S$ can significantly improve the 
agreement between
DGE models and \fermilat\ data \citep{TsFermiSymposium2010}.  
Further improvements
include taking into account \hi{} self-absorption \citep{Gibson2002, Gibson2010}
and uncertainties in the rotation curve for the \hi{} distribution, which
will be explored in future work.

We confirm the need for augmenting the use of CO and \hi{} line observations
as tracers of the interstellar gas with infrared observations of interstellar 
dust \citep{Grenier:2005,2010ApJ...710..133A,2011ApJ...726...81A,2011arXiv1101.2029P}.
Dust reveals some molecular hydrogen that is not traced by CO which can 
compensate to some extent for errors in or variations of the spin temperature 
of the interstellar hydrogen.
The dust column
density is represented in this work as the equivalent interstellar
reddening \ebv{} and limitations of the color correction method used to
derive the reddening maps make them less accurate near the Galactic plane.
But our analysis shows that the \gray{} fit improves with the inclusion of
dust near the Galactic plane, up to a reddening magnitude of 5.
The disadvantage of \ebv{} as a tracer of interstellar gas is that it
provides no distance information analogous to the Doppler shifts of the CO
and \hi{} lines due to differential rotation of the Milky Way, and for
this work we distributed the ``excess'' \ebv-associated column densities
like the inferred distribution of \hi.
This distribution of the ``excess'' is reasonable in regions with little or 
no CO emission but is suspect near large molecular clouds, where 
the ``excess'' should be mostly low density H$_2$ gas not traced by CO 
\citep[e.g.,][]{2011MNRAS.412..337G}.

\begin{figure}
\epsscale{0.85}
\plotone{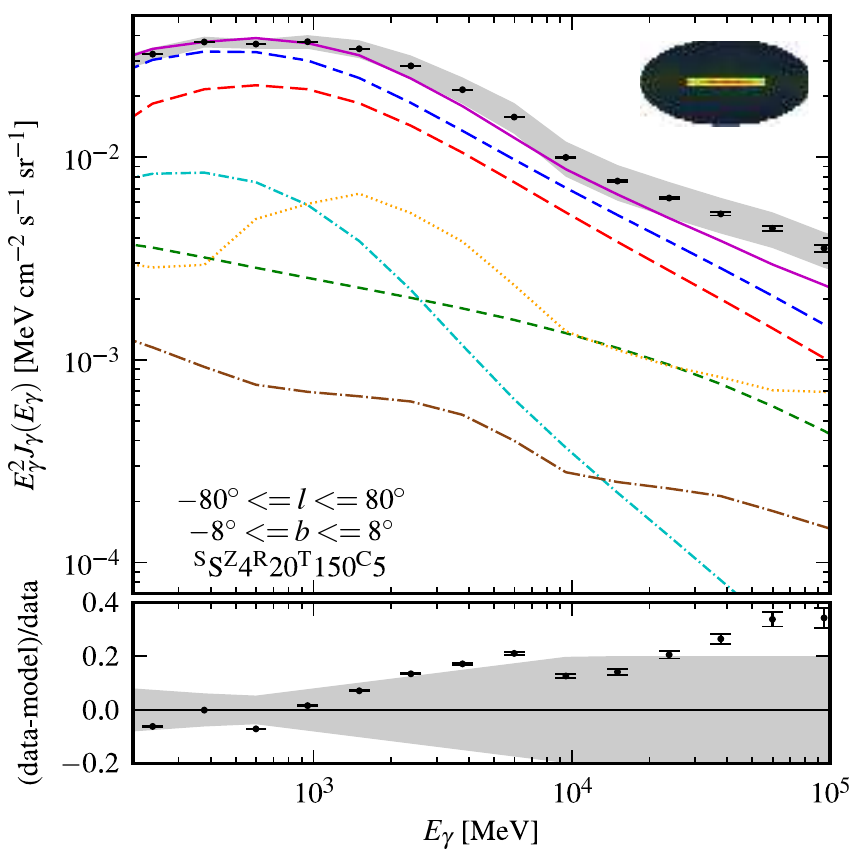}
\caption{Spectra extracted from the inner Galaxy region  for model \model{S}{4}{20}{150}{5}.
See Figure~\ref{fig:localRegion} for legend.  
\label{fig:innerRegion}}
\end{figure}

\begin{figure}
\epsscale{0.85}
\plotone{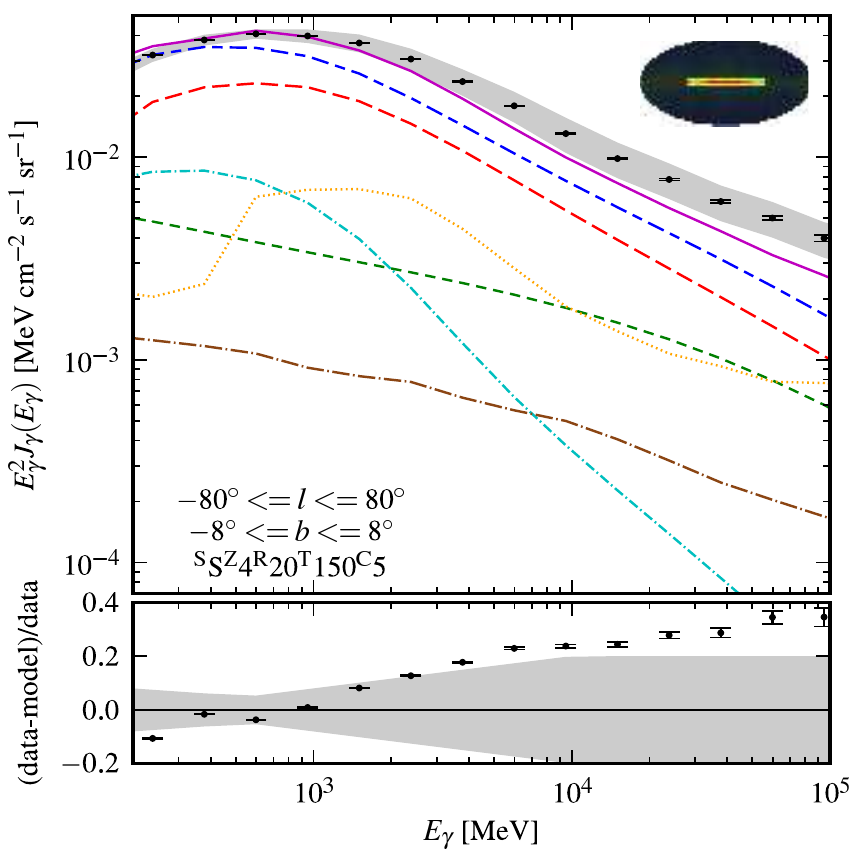}
\caption{Spectra extracted from the inner Galaxy region for model
\model{S}{4}{20}{150}{5} using Pass 7 clean photons.
The dip between 10 and 20 GeV is greatly reduced
compared to Figure~\ref{fig:innerRegion}.
See Figure~\ref{fig:localRegion} for legend.  
\label{fig:Pass7Spectra}}
\end{figure}

The models all under-predict the data in the Galactic 
plane (Figure~\ref{fig:outerRegion}) above a few
GeV and the difference is most pronounced in the inner 
Galaxy (Figure~\ref{fig:innerRegion}).
The magnitude of the difference is much less than the 
so-called EGRET ``GeV excess'' and is also confined to the plane.
A possible explanation for this is a contribution by point
sources 
such as pulsars, SNRs, and pulsar wind nebulae (PWN).  
Only a fraction of these are
actually detected individually by the \fermilat. 
The contribution by source populations below
the detection threshold is currently undetermined and it may have a 
diffuse-like spatial distribution.  
A study of this topic based on EGRET data,
with predictions for the \fermilat, was made by \citet{2007Ap&SS.309...35S}, 
giving estimates of a $\sim10$\% contribution by sources below
the \fermilat{} detection threshold.  
Pulsars are the largest contributor
to detected Galactic sources in the 2FGL catalogue 
\citep{FermiCatalogue2}, being far more numerous than SNRs and PWN.  
This class 
might dominate the contribution by undetected 
sources, but due to their 
spectral cutoffs above $\sim 10$~GeV they cannot completely explain the
spectral difference that we find at higher energies.
This will be addressed in a subsequent paper, employing a 
population-synthesis approach and comparison with the 2FGL catalogue
to constrain possible source contributions.

Another possibility is that the inner Milky Way contains fresh CR sources 
with a harder spectrum in addition to the presumed steady-state CR population
that has undergone propagation.
Signs of freshly accelerated CRs have been recently observed in the
Cygnus region \citep{CygnusCRs} and more
regions are likely to be discovered in the future.
Observations with \hess{} of the Galactic plane 
at TeV energies may partly support this explanation.
The \hess{} observations showed \gray{} emission associated with molecular 
clouds 
that have harder spectra than expected from extrapolation of lower-energy 
spectra \citep{HessGC}.  
But, 
freshly accelerated CRs are an unlikely explanation for the excess emission in
the outer Galaxy.  
Another alternative is that local CR measurements do not sample 
the average CR spectra in the
Galaxy \citep[e.g.,][]{Porter1997,2004ApJ...613..962S}.  
As discussed in
Section~\ref{sec:isotropic}, the uncertainties in the observed CR intensities 
have not
been propagated to the DGE models.
A 10\% decrease in \gray{} intensity below 1 GeV and a few percent
increase above 1 GeV with additionally 
an overall increase in the total CR intensity
would make the model agree with the data.  
The current analysis
cannot distinguish between these alternatives, and most likely 
a combination will contribute to providing the required additional \gray{} 
emission.

Contrasting with the under-prediction at $\gtrsim$ few~GeV, the models
over-predict the data at lower energies with the largest 
residual in the $200-400$~MeV bin.  
While being most prominent in the inner Galaxy 
(Figure~\ref{fig:innerRegion}),
this can also be seen to a lesser extent at low intermediate latitudes 
(Figure~\ref{fig:lowIntermLatRegion}).  
The over-prediction at low energies is
fractionally smaller than the under-prediction at higher energies and 
is partly contained within
the systematical uncertainty of the effective area of the \fermilat{}.  
We
note, however, that the discrepancy is still visible using the updated
instrument response in the
Pass 7 event selection (Figure~\ref{fig:Pass7Spectra}) suggesting that the
effect is not entirely instrumental.
The effect is strongest in the plane indicating
that the $\pi^0$-decay spectrum is primarily responsible for the mismatch. 
The \gray{s} below $\lesssim 1$~GeV are produced mainly by CR protons with 
energies $\lesssim 10$ GeV. 
The locally measured CR proton spectrum 
(Figure~\ref{fig:CRspectrumPrimary}) for these energies 
is affected by the solar modulation and is therefore 
subject to any errors we make in correcting for this effect.
While the force-field approximation used in this
paper is a useful parameterisation, realistic
models of the heliospheric CR transport 
\citep[e.g.,][]{Florinski2010,Ngobeni2010} could allow better 
determination of the low-energy interstellar CR nuclei spectra that
are relevant for the calculation of the $\pi^0$-decay \gray{} emission in 
our DGE models.  
The force-field approximation is however sufficient for the
analysis done in this paper as we are mainly concerned with comparing the
models with each other.
Finally, we note that reducing the spectrum below 1~GeV and
using a higher overall normalisation for the CR nuclei results in an 
increase in the $\pi^0$-decay DGE spectrum above $\gtrsim 1$~GeV.  
Therefore, the excess above a few GeV can also be partly due
to uncertainties in how the solar modulation is handled when fitting to
the observed CR spectra.

\begin{figure}
\epsscale{0.33}
\plotone{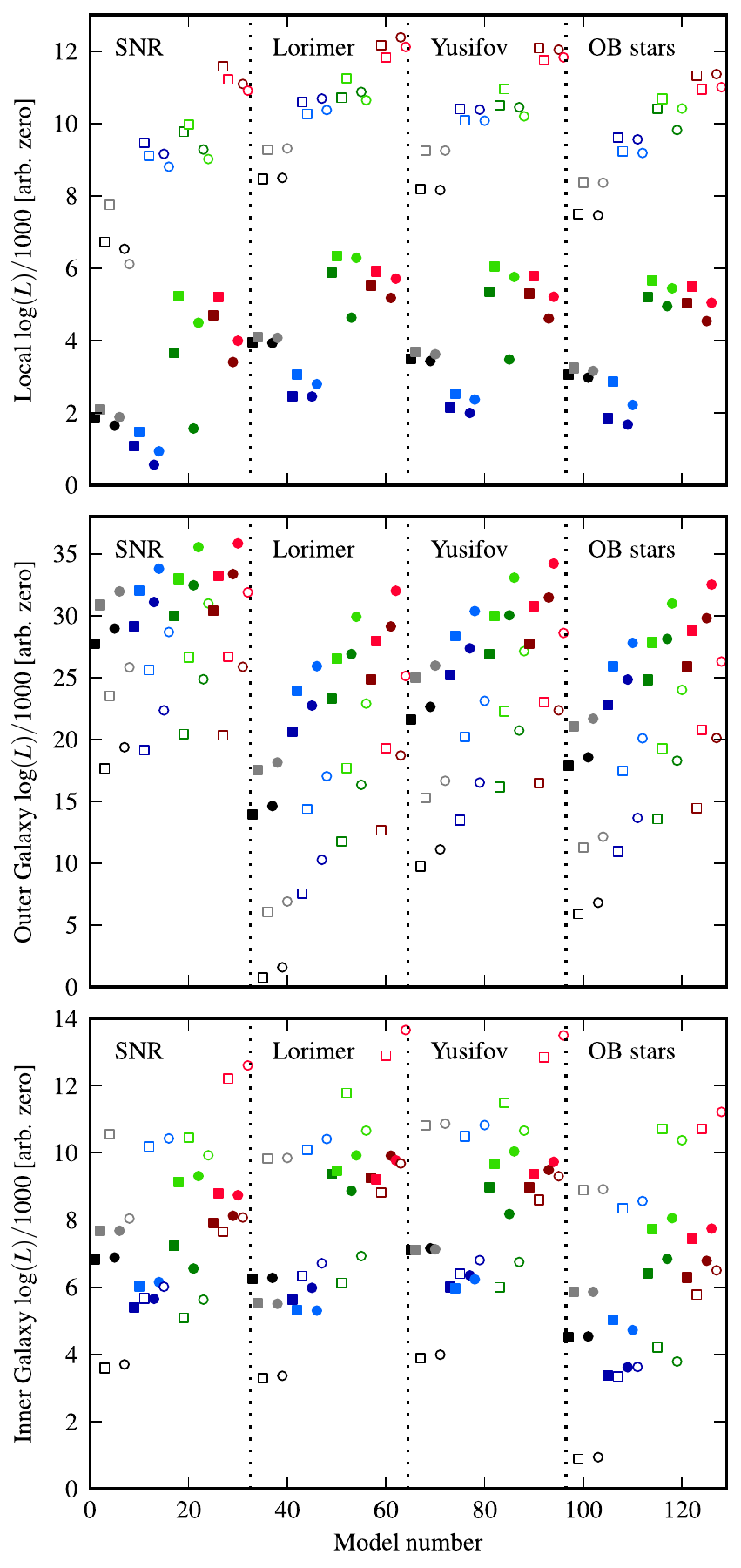}
\caption{The log-likelihood values found from the separate fits for the local 
region (top), the outer Galaxy region (middle), and the inner Galaxy region (bottom).
The zero level of the log-likelihood values is arbitrary but the 
difference between two models within a region gives their likelihood ratio 
for that region and a sum of differences in all regions gives the all-sky 
likelihood ratio.
The model number is a binary encoding of the input parameters 
(see Section~\ref{sec:results}).
The values of $z_h$ are color coded: 4~kpc is black, 6~kpc is blue, 
8~kpc is green, and 10~kpc is red.
Light colors represent a \ebv{} magnitude cut of 5 while dark 
have a magnitude cut of 2.
Filled symbols have $T_S=150$ K while open symbols use the optically 
thin assumption. 
Squares have $R_h=20$ kpc while circles have $R_h=30$ kpc.
The dotted vertical lines delineate the results for the different 
CR source distributions.
\label{fig:gammaLikelihood}}
\end{figure}

\begin{figure}
\epsscale{0.93}
\plotone{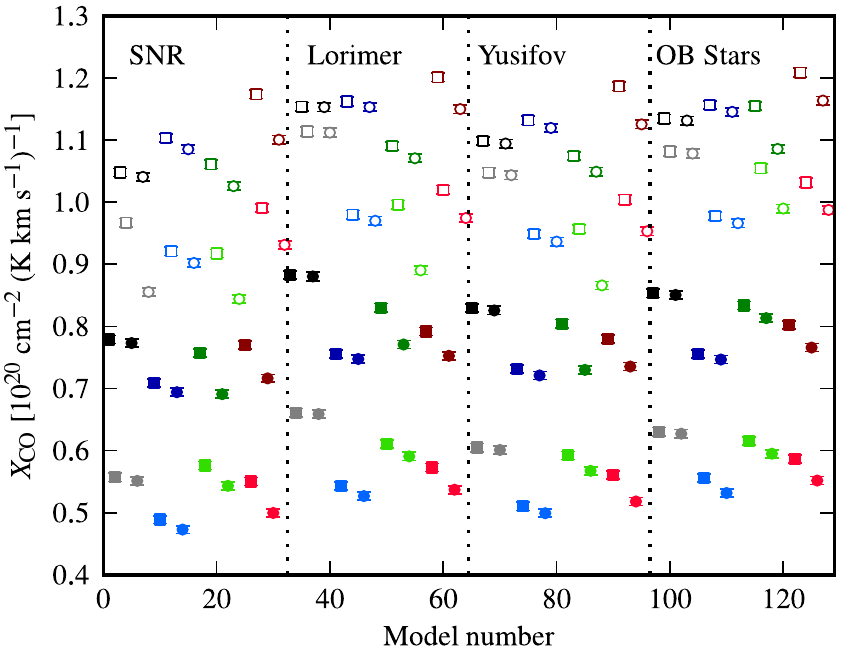}
\caption{The determined values of \Xco{} associated with the local annulus for
all models.
See Figure~\ref{fig:gammaLikelihood} for legend.
\label{fig:XCO_annuli}}
\end{figure}

\begin{figure}
\epsscale{0.53}
\plotone{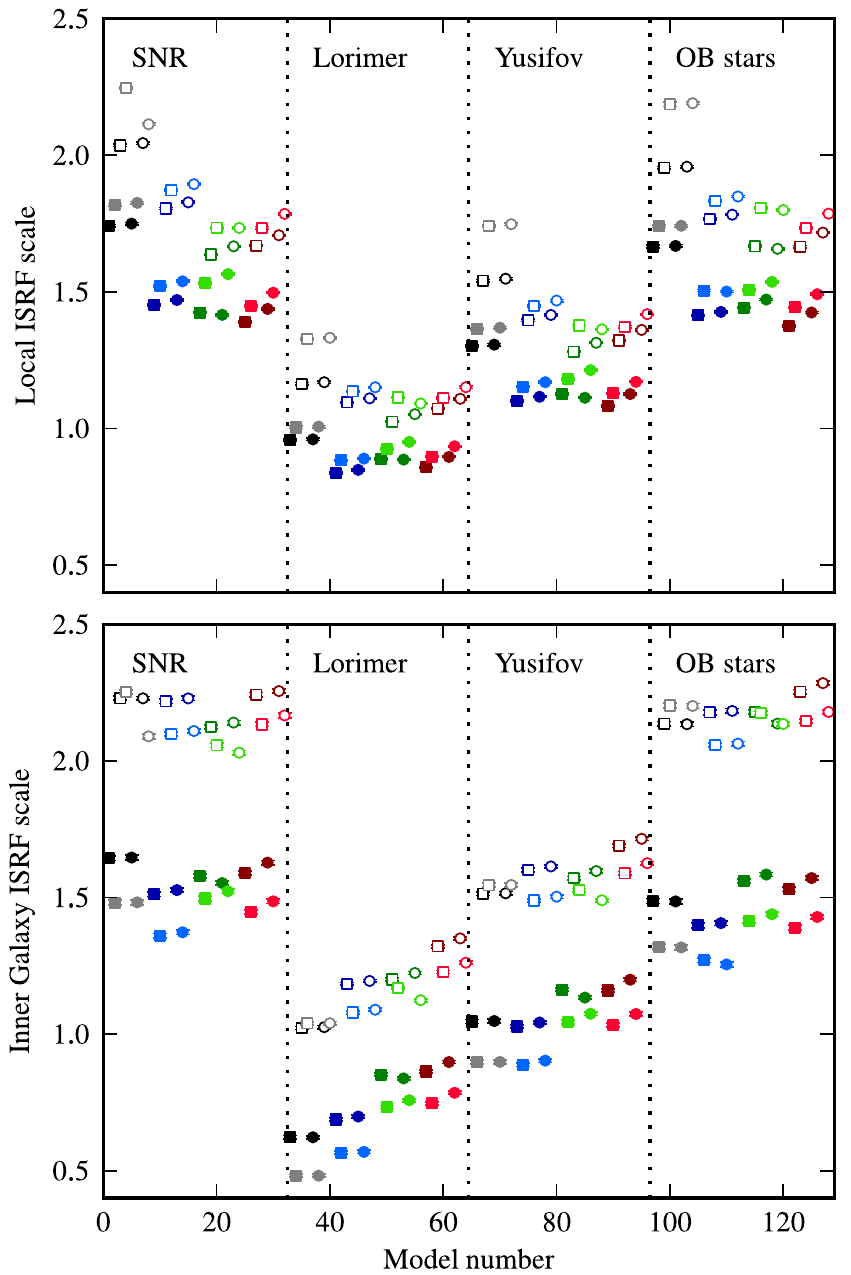}
\caption{Scaling factor of the ISRF
resulting from the fits in the local region (upper panel) and inner 
region (lower panel).
See Figure~\ref{fig:gammaLikelihood} for legend.
\label{fig:ISRFscale}}
\end{figure}

Despite the good agreement between model and data on average, there are 
structures seen on both small and large scales in the residual 
sky maps (Figure~\ref{fig:fractionalResiduals}).
Small-scale discrepancies are inevitable for any large-scale \GP-based model 
because simplifications have to be made to treat the CR transport together 
with the CR source, gas and ISRF distributions.  In particular, the assumption
of axisymmetry for the CR source distribution and the ISRF is not a realistic
model for the 3D distribution.
Apart from Loop~I, the Magellanic Stream, and the structures 
coincident with the features found by 
\citet{2010ApJ...724.1044S} and \citet{2010ApJ...717..825D},
the most prominent of the
large-scale residuals is the excess in the outer Galaxy.
Our analysis shows the observed intensities in the outer Galaxy are greater 
than predicted using conventional choices for the values of the parameters
that we studied.
Despite differing by more than a factor of 3 in CR density outside of 10~kpc, 
all of the CR source distributions considered in this paper gives a 
model that under-predicts the observed \gray{} emissivity in the outer Galaxy.
The models all show an increased likelihood for 
larger $z_h$ (Figure~\ref{fig:gammaLikelihood}) that has been 
shown to increase the CR flux in the outer 
Galaxy \citep{2010ApJ...710..133A, 2011ApJ...726...81A} but the large 
values of $z_h$ are approaching the bounds for consistency with the 
observed $^{10}$Be/$^9$Be ratio (Figure~\ref{fig:CRspectrumSecondary}).
Another possibility is to increase the density of CR sources in the 
outer Galaxy.
This would mean that SNRs, or CR source classes 
having a similar spatial distribution as the proxies 
considered in this paper, are not the only source of CRs \citep{Butt2009Nature}.
Modifications of the CR propagation mechanism have also been proposed in
the literature as an explanation for increased CR flux in the outer
Galaxy \citep{Breitschwerdt:2002,Shibata:2007}.
Alternatively, large amounts of gas not traced by 21-cm-emitting \hi{} and 
CO (as the tracer of H$_2$) surveys may be present in the outer 
Galaxy \citep{2002ApJ...579..270P}, which would also increase the \gray{}
emission.
Breaking this degeneracy and deriving the correct 
CR source distribution and 
propagation model requires using all available data for CRs and their 
interactions in the ISM.
Such a determination would inevitably have to take into account uncertainties 
involved in the astrophysical input to the models discussed in this paper.

The increase in log-likelihood for larger values of $z_h$ is also seen in
the local region (Figure~\ref{fig:gammaLikelihood}).  
An increase in $z_h$ is accompanied by a decrease in the ISRF
scaling factor (Figure~\ref{fig:ISRFscale}), indicating that 
additional IC emission at high latitudes is needed compared
to the previously assumed $z_h = 4$~kpc propagation models 
\citep[e.g.,][]{Strong:2004}.  
The longer confinement time for the CR
electrons/positrons in the larger halo sized models results in more IC emission
for these cases. 
From our analysis, 
the derived normalisation of the IC component and its intensity 
varies considerably between CR source distributions.
Because the spectral shape of the normalised 
IC emission is similar in all cases, the spatial distribution of this
component determines the inferred IC contribution
to the DGE for each model.
This emphasises that accurate modelling of the spatial distribution of 
the IC emission is essential to properly assess its intensity and spectrum 
from \gray{} data.

We have also explored how the uncertainties affect the 
CR propagation (Figure~\ref{fig:CRparametersNuc}), 
aiming for a self-consistent model by incorporating the \Xco{}
values found from the \gray{} fit into the propagation parameter 
determination and transport calculations.
Self-consistency, as used in this paper, is intended to ensure that 
the CR secondary-to-primary ratios and other direct measurements are 
consistent with the assumed $T_S$ and fitted \Xco, which affect the 
gas density and hence CR secondary production.
For an assumed set of input parameters, this 
is obtained by adjusting the spatial and momentum-space 
diffusion coefficients via $D_0$ and the Alfv\'{e}n speed $\valf$, respectively.
For the CR protons and He the propagated CR intensities and spectra 
are 
determined mostly by the assumed CR source distribution and boundary conditions
because their energy loss time scales for the energies of interest in this 
paper are long compared to the propagation time scale.
The CR electrons and positrons are more strongly affected by changes in the
diffusion coefficient and halo size because their energy losses are much faster.
The modelled CR intensities and spectra are also 
constrained by their normalisation
to the locally measured data, so the self-consistency requirement does not 
significantly change the \gray{} models and results for these models in this
paper.
Nevertheless, it is an important criterion to ensure that the origin 
of systematic effects from the assumed input parameters can be properly 
attributed.
It is important to also emphasise that 
uncertainties in the input parameters can also 
affect the determination of the propagation parameters.
Simply assuming that one set of propagation parameters applies equally to
all variations of, e.g., assumed CR source distributions, is incorrect. 
Note, that even though we have assumed a diffusive-reacceleration model 
for the CR propagation, this applies to the 
other variants such as pure diffusion models, models with convection, and 
so forth.

\section{Summary}

This paper presents a systematic study of several basic parameters for
global models of the DGE using \fermilat{} data.  
The parameters, all inputs to the \GP{} CR propagation code, are
related to the distributions of interstellar gas and of CRs, and
for each combination of parameters considered the models were calculated
self consistently and were constrained to be consistent with local
measurements of CRs.   
The evaluation of the models with respect
to the data, taking into account the point sources in the 1FGL catalogue,
was made with the \gardian{} software package, which was developed for
studying diffuse emission in the \fermilat{} data.

We find that augmenting the CO and \hi{} column density estimate with
column density estimates from dust improves agreement between 
model and \gray{} data. 
Our analysis finds this to be true even near the Galactic plane, where the
dust column density estimated from the equivalent interstellar reddening
\ebv{} is less accurate due to limitations of the color correction
method used to derive the reddening maps.

The DGE in the outer Galaxy is better fit by models
with larger-than-expected CR halo sizes.   
There are other possiblities for the
models to predict large enough intensities in the outer Galaxy.  
These include modifications of the assumed distributions of CR sources or of
propagation of CRs in the outer Galaxy, or even the presence of much
greater amounts of interstellar gas than currently assumed.  

From our \gray{} fits in the region with $|b| >
8^\circ$ we show that larger IC intensities provide a better
fit to the data for most models.  
In our approach, the 
single normalisation factor for the optical and IR components 
of the ISRF that is shown to decrease with larger halo size provides this
information. 
In addition to accounting for uncertainties in the ISRF, this 
normalisation factor also encapsulates uncertainties
in the distribution of CR electrons in the Galaxy.

All of the models considered in the paper under-predict the data above a
few GeV in the Galactic plane.  
The magnitude of the difference is much
less than the ``GeV excess'' observed by EGRET and mostly confined to the
Galactic plane.  
Two possible explanations were discussed, contribution from undetected sources
and variations in the CR spectra.  
Further analysis is needed to estimate the contribution from each.

We derived the radial distribution of \Xco{} for all models and
confirmed the systematically lower values of \Xco{} for the inner-most 
annulus (0 -- 1.5 kpc).  
The \Xco{} determined for the local annulus (8 -- 10 kpc)
above $8^\circ$ latitude was also systematically lower than for the rest of
the Galaxy, implying that the local high latitude clouds have
different properties than the Galactic average.  
Our \Xco{} values for
other radial ranges show that accurate determination of the CR gradient
and the column density of the \hi{} gas distribution are essential in
determining the \Xco{} gradient from \gray{} observations.

Our fits to the \gray{} data reveal the difficulties in
accurately determining the properties of the ISM or CR propagation with
DGE modelling.  
The derived \Xco{} values and ISRF scaling
parameters depend strongly on the assumed input parameters, both on the
propagation setup and also on the properties of the \hi{} gas distribution.
Past studies using \gray{} data to determine these properties
were thus susceptible to unexplored systematic uncertainties. 
The measured properties of Galactic plane sources even well above the 
detection limit can
be affected by the assumed DGE model, especially for energy
ranges where the scale of the PSF is comparable to the scale of the structure
in the DGE.
Accounting for systematic uncertainties of the
astrophysical input needed for a DGE model is a
necessary step in accurate analysis of \gray{} data when the observed
signal is comparable or less than the DGE.

\acknowledgments

The \fermilat{} Collaboration acknowledges generous ongoing support
from a number of agencies and institutes that have supported both the
development and the operation of the \fermilat{} as well as scientific data analysis.
These include the National Aeronautics and Space Administration and the
Department of Energy in the United States, the Commissariat \`a l'Energie 
Atomique
and the Centre National de la Recherche Scientifique / Institut National 
de Physique
Nucl\'eaire et de Physique des Particules in France, the Agenzia Spaziale 
Italiana
and the Istituto Nazionale di Fisica Nucleare in Italy, the Ministry of 
Education,
Culture, Sports, Science and Technology (MEXT), High Energy Accelerator Research
Organization (KEK) and Japan Aerospace Exploration Agency (JAXA) in Japan, and
the K.~A.~Wallenberg Foundation, the Swedish Research Council and the
Swedish National Space Board in Sweden.

Additional support for science analysis during the operations phase is 
gratefully
acknowledged from the Istituto Nazionale di Astrofisica in Italy and the 
Centre National d'\'Etudes Spatiales in France.

\GP{} development is partially funded via NASA grants NNX09AC15G and 
NNX10AE78G.

Some of the results in this paper have been derived using the HEALPix 
\citep{Gorski:2005} package.


\appendix

\section{\gardian{} Package}
\label{appendix:gardian}

The {\bf Ga}mma {\bf R}ay {\bf Di}ffuse {\bf An}alysis (\gardian) package 
is a full sky binned maximum-likelihood analysis tool.
It was designed for fitting DGE models to the \fermilat{} data, although 
it is general enough to accommodate other instruments.
While the framework is capable of fitting non-linear models, the main 
emphasis has been on DGE models consisting of a linear combination of 
template sky maps.
At this point, the \gardian{} package is not publicly available.

The photon data and model are spatially binned on a HEALPix grid
\citep{Gorski:2005} that is hierarchical, equal-area, and isolatitude 
allowing for fast spherical harmonics decomposition and nearest neighbor search.
As is appropriate for photon limited data we use a forward folding method 
for the analysis, turning the model into counts using the IRFs.
The GaRDiAn package requires knowledge of the exposure as a function of 
energy for each direction on the sky and a sky average representation of 
the PSF as a function of energy.
This information needs to be provided in tabulated form, which is standard 
for the \fermilat{} IRFs\footnote[14]{http://fermi.gsfc.nasa.gov/ssc/data/analysis/documentation/Cicerone/Cicerone\_LAT\_IRFs}.

Denoting the model flux by $f(\mathbf{\theta}, E)$ and the exposure 
as $Exp(\mathbf{\theta}, E)$, we can write the model counts for 
energy bin $i$ as

\begin{equation}
	F_i(\mathbf{\theta}) = \int_{E_{{\rm min},i}}^{E_{{\rm max},i}} {\rm d}E\, f(\mathbf{\theta}, E) \, Exp(\mathbf{\theta}, E).
	\label{eq:exposureCorrection}
\end{equation}

\noindent
Here, $\mathbf{\theta}$ is a position in the sky, $E$ is the photon energy, and 
we have assumed the photon data has been binned in energy bins defined 
by $E_{{\rm min},i} \le E \le E_{{\rm max},i}$.
To account for the energy dependence of the PSF, $\psi(\alpha, E)$, we 
calculate an effective PSF for each energy bin as the spectrally-weighted average with the 
spectra of the model

\begin{equation}
	\Psi_i(\alpha) = \frac{\int_{E_{{\rm min},i}}^{E_{{\rm max},i}} {\rm d}E\, \psi(\alpha, E)\, <F>(E)}{\int_{E_{{\rm min},i}}^{E_{{\rm max},i}} {\rm d}E\, <F>(E)}.
	\label{eq:weightedPSF}
\end{equation}

\noindent
Here, $\alpha$ is the angle between the true photon direction and the 
reconstructed direction and 
$<F>(E) = \int {\rm d}\Omega\,F(\mathbf{\theta},E) / 4\pi$ is the sky 
average of the model counts as a function of energy.
Using the spherical harmonics $Y_{lm}(\mathbf{\theta})$, we can write

\begin{align}
	F_i(\mathbf{\theta}) &= \sum_{l=0}^{l_{\rm max}} \sum_{m=-l}^l a_{lm,i} Y_{lm}(\mathbf{\theta}), \\
	\Psi_i(\alpha) &= \sum_{l=0}^{l_{\rm max}} c_{l0,i} Y_{l0}(\alpha).
	\label{eq:harmonicsDecomposition}
\end{align}

\noindent
When using the spherical harmonic decomposition of the PSF, we 
assume $\alpha$ is the angle between a point in the sky and the north pole.  
We also assume the PSF is azimuthally symmetric so 
all $c_{lm} = 0$ for $|m| \ge 1$.
The convolved model is now calculated as
\begin{equation}
	\tilde{F_i}(\mathbf{\theta}) = \sum_{l=0}^{l_{\rm max}} \sum_{m=-l}^l \sqrt{\frac{4\pi}{2l+1}} c_{l0,i} a_{lm,i} Y_{lm}(\mathbf{\theta}).
	\label{eq:convolution}
\end{equation}

To allow for arbitrary energy binning of the photon data while still 
handling strong energy dependence of the IRFs, we integrate 
equations~\eqref{eq:exposureCorrection} and~\eqref{eq:weightedPSF} 
semi-analytically.
We use power-law interpolation of the tabulated input values of the IRFs 
and model.  
For the PSF weighting in Eq.~\eqref{eq:weightedPSF}, we use a single 
effective power-law index for the entire bin because fine structure within the 
energy bin is lost in the conversion to counts.

While using the spherical harmonics decomposition for convolving the 
model sky maps with the PSF is extremely efficient it has limitations.
We are limited to using an azimuthally-symmetric PSF and must assume the 
PSF is the same over the entire sky.  
Fortunately, the tabulated \fermilat{} PSF is azimuthally symmetric and
its variations over the sky are minimal due to both the
uniform exposure of the \fermilat{} in its nominal survey mode operation, 
and the small variations of the PSF with incident
angle\footnote[15]{\url{http://www-glast.slac.stanford.edu/software/IS/glast\_lat\_performance.htm}}.

Having the model converted to counts and properly convolved, we calculate 
the likelihood using
\begin{equation}
	L(\mathbf{X}) = \sum_{i,j} D_i(\mathbf{\theta_j})\log(F_i(\mathbf{\theta_j}, X)) - F_i(\mathbf{\theta_j}, X) - \log(D_i(\mathbf{\theta_j})!),
	\label{eq:likelihood}
\end{equation}

\noindent
where $D_i(\mathbf{\theta_j})$ are the binned photons for energy bin $i$ and 
HEALPix pixel $j$, and $X$ are the parameters of the model.
The best-fit parameters are found by maximising the likelihood using 
Minuit2\footnote[16]{http://seal.web.cern.ch/seal/MathLibs/Minuit2/html/}.

\section{Generation of \hi{} and CO Gas Annuli} 
\label{appendix:gasRings}

Under the assumption of uniform circular motion around the Galactic
centre with rotation curve $V(R)$, the velocity with respect to the
local standard of rest of a region with Galactocentric distance $R$ viewed  
toward direction $l$, $b$ (in Galactic coordinates) is
\begin{equation}
	v_{\rm LSR} = R_\odot \left (\frac{V(R)}{R} -
	\frac{V_\odot}{R_\odot}\right) \sin(l) \cos(b).
	\label{eq:LSRVelocityRadiusRelation}
\end{equation}
This relation provides a one-to-one  
relationship between $v_{\rm LSR}$ and $R$ for any given LOS.
We use the parametrised rotation curve 
of \citet{1985ApJ...295..422C} 
using the IAU-recommended values $R_\odot = 8.5$ kpc for the distance 
from the Galactic centre to the Sun
and $V_\odot = 220$ km s$^{-1}$ for the velocity of the Sun around the Galactic
centre \citep{1986MNRAS.221.1023K}. 
\footnote[17]{Use of more recent rotation curves and LSR 
\citep{2009PASJ...61..227S,2009NewA...14..615F} 
would not significantly affect our analysis.}   
We applied this relation to the 21-cm  
Leiden-Argentine-Bonn (LAB) survey of \hi{} \citep{2005A&A...440..775K} 
and the 115~GHz Center for  
Astrophysics survey of CO \citep{2001ApJ...547..792D} 
to transform the spectral measurements  
into maps of the emission for a range of Galactocentric annuli.  
The  
boundaries of the annuli are given in Table~\ref{tab:AnnuliBoundaries}.
The $\sim1$ kpc width of the annuli is set by the  
finite non-circular (random and systematic) motions of the gas traced  
by these surveys as well as internal velocity dispersions of molecular  
clouds.  
These non-circular and internal motions limit the practical  
linear resolution of the velocity-to-distance relation.  
The outer  
annuli are broader because the gradient of $v_{\rm LSR}$ with Galactocentric  
distance decreases approximately as $1/R$ beyond the solar circle.

Due to non-circular motion of gas in the Galaxy, a small
fraction of the emission has forbidden velocities. 
This can be due to the $v_{\rm LSR}$ being greater than the terminal
velocity or having an incorrect sign.
In our procedure, for the former case the emission is assigned to 
the tangent point
annulus, while for the latter the gas is assigned to the local annulus
(i.e., the one that spans $R_\odot = 8.5$ kpc).
In addition, if the gas is placed above a certain height
above the Galactic plane, it is assumed to be local.  
The height differs between the gas distributions and was chosen to be
1~kpc for \hi{} and $0.2$~kpc for CO.  
These
values were chosen to be significantly larger than the scale heights of
the gas distributions \citep[e.g.,][]{2003PASJ...55..191N,2006PASJ...58..847N}.

The kinematic resolution of the method vanishes for directions near
the Galactic centre and Galactic anti-centre.
Therefore, we linearly  
interpolate each annulus independently across the ranges
$|l| < 10^\circ$
and $|180-l| < 10^\circ$ to get an estimate of the radial profile of the
gas.
To estimate N(\hi) or 
W(CO) at the edge of the region, we calculate the
average over a longitude range $\Delta l = 5^\circ$ on each side of the
boundary.
The interpolated values are then scaled to match
the total N(\hi) or W(CO) along each LOS in the regions that
were interpolated.

Note that the innermost annulus is entirely enclosed within the
interpolated region, necessitating an alternate method to estimate 
its column density.
For \hi{} this is accomplished by assuming the innermost annulus contains
60\% more gas than its neighbouring annulus.  
This is a conservative
number considering that observations have shown that there is gas
depletion in the radial range $\sim 1.5 - 3$ kpc \citep[see][for a
review]{2007A&A...467..611F} 
For CO, we assign all high velocity emission in the innermost
annulus.  
Here, high velocity means
\begin{equation}
	v_{\rm LSR} < (-50 + 3l)\;{\rm km\,s^{-1}},
	\label{eq:COHighVelocity}
\end{equation}
and
\begin{equation}
	v_{\rm LSR} >
	\begin{cases}
		25\; {\rm km\,s^{-1}} &\qquad l < 0\\
		(10 + 3l)\; {\rm km\,s^{-1}}&\qquad l >= 0
	\end{cases}
	\label{eq:COHighVelocity2}
\end{equation}
\noindent
These values were found after visual inspection of the CO data.
The specific distribution in the innermost 1.5~kpc does not alter the
results of this paper in a significant way.

The CO data are from the 115~GHz composite survey 
of \citet{2001ApJ...547..792D} 
covering the latitude range $|b| < 30^\circ$.  
The coverage is not
complete for that range but it is believed that no significant emission
is missing. 
To increase the signal to noise in the data the CO data have been filtered with
the moment masking technique \citep{2001ApJ...547..792D} 
applied to each component
of the survey independently to accurately account for varying noise
levels.
The sampling grid spacing of the component surveys varies from $0.125^\circ$
to $0.25^\circ$, but we rebin to a resolution of 0.25$^\circ$ for the annuli.
This degradation of angular resolution does not affect the DGE analysis 
significantly for two main reasons.
First, the angular resolution of \fermilat{} below 5 GeV where the majority of 
photons are
detected is larger than 0.25$^\circ$.
Second, the N(\hi) annuli are anyway limited
to 0.5$^\circ$ sampling (see below) 
limiting any gains from better CO sampling to the inner Galactic ridge.

The \hi{} data are from the 21-cm composite LAB survey of
\citet{2005A&A...440..775K} 
covering the entire sky with a 0.5$^\circ$ sampling.  
Limited correction has been made for absorption against
bright background radio sources and pixels with large negative
brightness temperature are replaced with a linear interpolation in
longitude between neighboring pixels.
Emission from the Small Magellanic Cloud, Large Magellanic Cloud, and
Andromeda M31 is excluded from the annuli.
The observed brightness temperature, $T_B$, is converted to column 
density under
the assumption of a uniform spin temperature, $T_S$, 
using the equation
\begin{equation}
	N_{{\rm H{\sc I}}}(v,T_S) =
        -\log\left(1-\frac{T_B}{T_S-T_{bg}}\right) T_S Ci \Delta v,
	\label{eq:OpticalDepthCorrection}
\end{equation}

\noindent
where $T_{bg} \approx 2.66$ K is the brightness temperature of the 
microwave background at 21-cm and $C = 1.83\times10^{18}$ cm$^{-2}$.
In cases where $T_B > T_S - 5$ K, we truncate $T_B$ to $T_S - 5$ K.

\section{Interstellar radiation field}
\label{appendix:ISRF}

The Galactic ISRF is the result of emission
by stars, and the scattering, absorption, and re-emission of absorbed
starlight by dust in the ISM.
The first calculation considering the broadband (optical to far-IR) spectral
energy distribution (SED) as a function of Galactocentric distance 
was made by \citet{Mathis1983}.
Subsequently, \citet{Chi1991} extended the \citet{Mathis1983} 
calculation, and this work formed the basis for the ISRF model used 
in the EGRET-team DGE models \citep[e.g.,][]{Bertsch1993}.
A new calculation of the ISRF was made by \citet{SMR00}, using 
emissivities based on stellar populations from
COBE/DIRBE fits by \citet{Freudenreich1998} and the SKY model of 
\citet{Wainscoat1992} together with COBE/DIRBE derived 
infrared emissivities \citep{Sodroski1997,Dwek1997}.
However, no radiative transport was done for the 
stellar light in the \citet{SMR00} model, and hence there was no direct
coupling between the stellar emission and the output in the IR.
A full radiation transport calculation using ray tracing 
was done by \citet{PS2005}, which 
treated the scattering and absorption of the stellar light using a dust
model consistent with COBE/DIRBE data, for the first time 
directly relating the spatial variation of the ISRF SED throughout the Galaxy.
Subsequent work \citep{Porter:2008} extended the code to calculating the 
full angular distribution of the intensity of the ISRF from ultraviolet 
to far-IR 
wavelengths, which is essential for the calculation of the anisotropic IC
emission \citep{MS2000}.
Here, we describe further extension of the code, which has been rewritten
to use a parallel Monte Carlo radiative transfer method. 

In our model, we represent the stellar distribution by four 
spatial components: the thin and thick disc, the bulge, and the 
spheroidal halo. 
We follow \citet{Garwood1987} and \citet{Wainscoat1992}
and use a table of stellar spectral types comprising
main sequence stars, giants, and exotics to represent the luminosity 
function (LF) for each of the spatial components.
The spectral templates for each stellar type are taken from the 
semi-empirical library of \citet{Pickles1998}.
The normalisations per stellar type are obtained by adjusting the 
space densities to reproduce the observed LFs in the V- and K-band for
the thin disc.
The LFs for the other spatial components are obtained by adjusting 
weights per component for each of the stellar types relative to the 
normalisations obtained for the thin disc LF.
The spatial densities of the 
thin and thick disc are modelled as exponential disks.
For the thin disc available estimates for the radial scale length 
range from $\sim 2$~kpc to
$\sim 4$~kpc while for the thick disc estimates give
the range $\sim 3-4$ kpc \citep[e.g.,][]{1998A&A...330..136P,2001ApJ...556..181D,Juric2008,2010ApJ...714..663D,2011MNRAS.414.2446M}.
We use radial scale lengths of 2.5~kpc and 3.5~kpc for the thin and thick 
disc, respectively, in the present work. 
The thin disc has a hole interior of a Galactocentric radius of 
$\sim 1.7$ kpc, following \citet{Freudenreich1998}.
The scale height of the stellar classes in the thin disc follows
\citet{Wainscoat1992}, while the thick disc is characterised by a
single scale height of 0.75 kpc, which is approximately the middle
of values from the literature \citep[e.g.,][]{2010ApJ...714..663D}.
The local thick disc to thin disc 
normalisation, $\rho_{\rm thick} (R_\odot)/\rho_{\rm thin} (R_\odot)$, is 
assumed to be 5\%.
We take the
stellar halo as described by an oblate symmetrical spheroid with axial 
ratio $c/a = 0.7$ and power-law density profile $\rho_{halo} \propto r^{-2.8}$, 
intermediate between values found from Sloan 
data \citep{Juric2008,2010ApJ...714..663D}.
The local halo to thin disc 
normalisation, $\rho_{\rm halo}(R_\odot)/\rho_{\rm thin} (R_\odot)$, is 0.5 \%.
The bulge is assumed to be ``boxy'' following \citet{LopezCorredoira2005} 
with geometrical parameters taken from their paper.
For our nominal ISRF, the 
bulge input luminosity is normalised to the K-band 
luminosity of \citet{Freudenreich1998} for an assumed $R_\odot = 8.5$~kpc.

We use a dust model including graphite, polycyclic aromatic hydrocarbons
(PAHs), and silicate.
Dust grains in the model are spherical and the absorption and scattering 
efficiencies for graphite, PAHs, and silicate grains are taken from 
\citet{Li2001}.
The dust grain abundance and size distribution are taken from 
\citet{Weingartner2001} (their best-fit Galactic model).
We assume a purely neutral ISM.
The absorption and reemission by the dust is treated as described below in 
the Monte Carlo method.

Dust follows the Galactic gas distribution and we assume uniform 
mixing between the two in the ISM \citep{Bohlin1978}.
The dust-to-gas ratio scales with the Galactic metallicity gradient.
Estimates for the Galactic [O/H] gradient vary in the range $0.04-0.07$
dex kpc$^{-1}$ \citep[][and references therein]{2004A&A...422L..47S}.
The variation of the metallicity gradient influences
the scattering and redistribution of the mainly UV and blue 
component of the ISRF into
the infrared: increased metallicity implies more dust, and therefore increased
scattering and absorption of the star light.

The ISRF is calculated for a cylindrical geometry with a maximum radial
extent of 50 kpc and maximum height above the Galactic plane of 50 kpc.
We use a Monte Carlo method for the photon 
propagation through the ISM similar to those described by \citet{Gordon2001}, 
\citet{Jonsson2006}, and \citet{Bianchi2008}.
The system volume is segmented into cells with the number of ``photons'' 
injected
per cell determined according to the ratio of the cell stellar luminosity
to the system stellar luminosity for a given number of total injected 
particles $N_{total}$. 
The parallelisation is done using OpenMP\footnote[18]{http://openmp.org/wp/} 
with one thread per cell, injecting 
all particles for the cell.
Each photon emitted within a cell is released with an isotropic
angular distribution uniformly over the cell 
with frequency sampled from the stellar luminosity spectrum at that location.
Following emission, the first interaction is forced to increase sampling 
efficiency \citep{Gordon2001}. 
The interaction length is sampled according to the dust optical depth
in the emitted direction, and the photon is propagated that distance.
The interaction is either a scattering or absorption (or the photon is
lost from the system if the interaction length is outside the boundary).
The probability for a scattering or absorption depends on the frequency
dependence of the scattering and absorption cross section for the assumed
grain mixture.
If the photon is scattered, the new direction of the photon is calculated 
according to the dust grain type (determined by the relative
sizes of the scattering cross sections for the assumed grain mixture)
using the Henyey-Greenstein angular distribution function \citep{Henyey1941}.
If the photon is absorbed, the grain type that absorbed the photon is 
determined from the relative sizes of the absorption cross sections 
of the grain types in the assumed mixture.
If the absorbing grain type is ``large'', that is, it reemits in thermal
equilibrium, we use the ``temperature correction'' 
method of \citet{BjorkmanWood2001} where the absorbing dust at the 
new location is heated by the photon, raising its temperature. 
To enforce radiative equilibrium the dust immediately reemits a photon, 
where the reemitted frequency is chosen from a distribution that corrects
the temperature of the dust prior to its new temperature following the 
absorption of the photon.
If the absorbing grain type undergoes stochastic heating (e.g., the nanograin
components of the dust mixture) we cannot treat it 
using this method, and so the amount of luminosity absorbed in the cell 
is recorded and the photon is removed from the system (to be 
dealt with in a subsequent step, as described below).
The scattered or absorbed/reemitted photons propagate
in the system until they either escape, or are fully absorbed.
This process is then repeated for each injected stellar photon.

To treat the photons absorbed on the grains undergoing stochastic heating, 
the frequency-dependent 
absorbed luminosity by these grains is used to compute the stochastic 
heating emissivity for each cell using the ``thermal continuous'' method 
described by \cite{Draine2001}.
The procedure used above for injection of stellar photons is then followed, 
except the stochastic heating emissivity distribution 
is used in place of the stellar luminosity distribution.
With this method, we achieve particle conservation better than 
$\sim 10^{-4}$ after a single pass.

We record the intensity distribution of the ISRF at selected points in 
the Galactic volume used for the Monte Carlo calculation.
The intensities as a function of frequency at each location 
are recorded as HEALPix images \citep{Gorski:2005}.
The low-energy photon 
number density at each location is the integrated intensity over 
4$\pi$ sr.
The ISRF intensity and/or number density 
at any location in the Galaxy is obtained by linearly interpolating amongst the
sampling positions used in the Monte Carlo simulation.

The full data set for the ISRF used in the current paper is available from the 
\GP{} website (see http://galprop.stanford.edu/code.php for access 
instructions).

\section{CR propagation results}
\label{appendix:CR}
\begin{figure*}
\plottwo{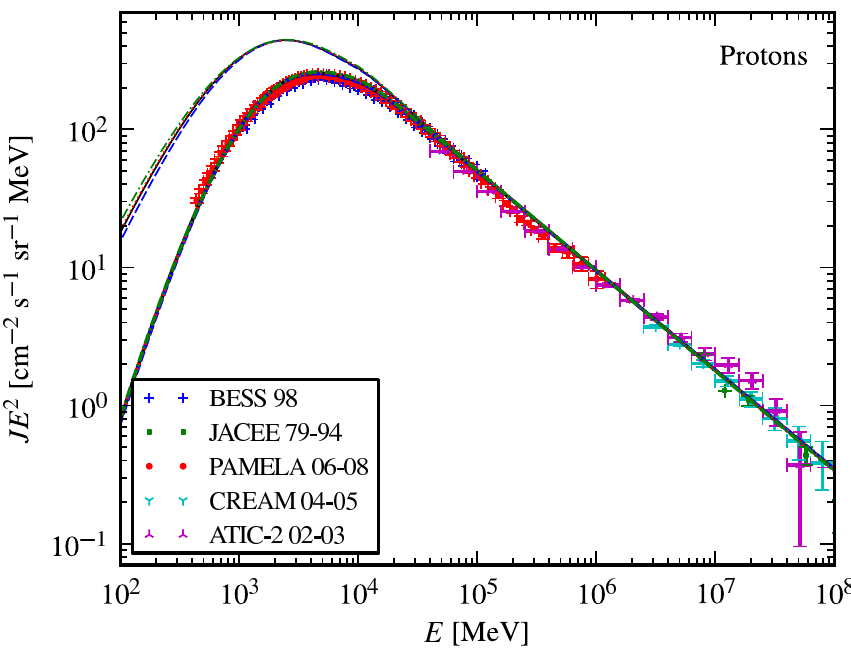}{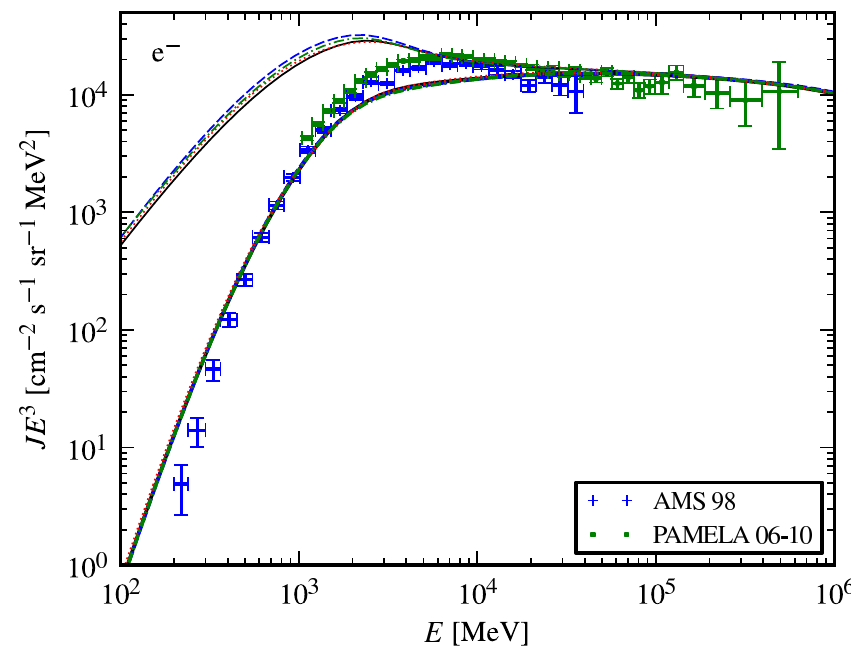}\\
\plottwo{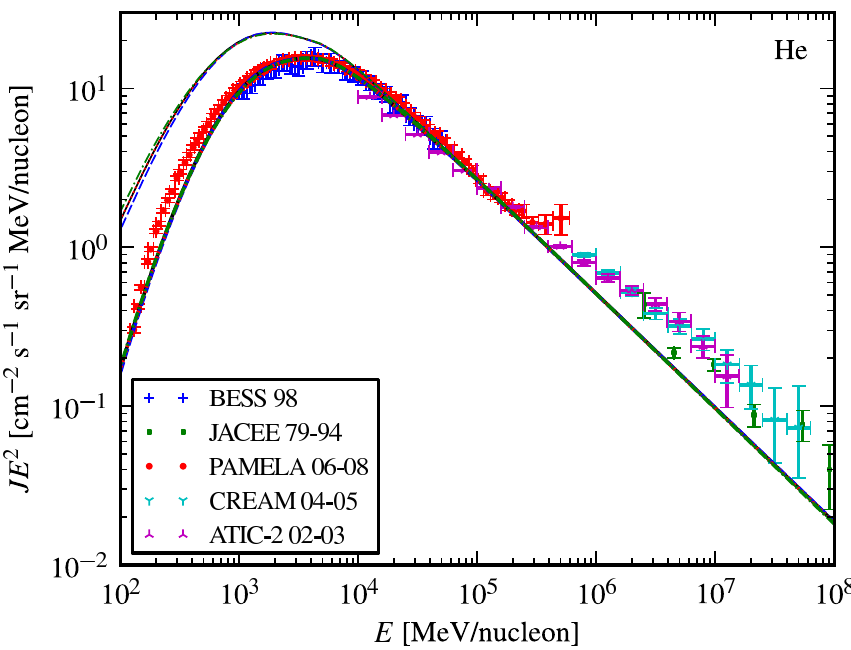}{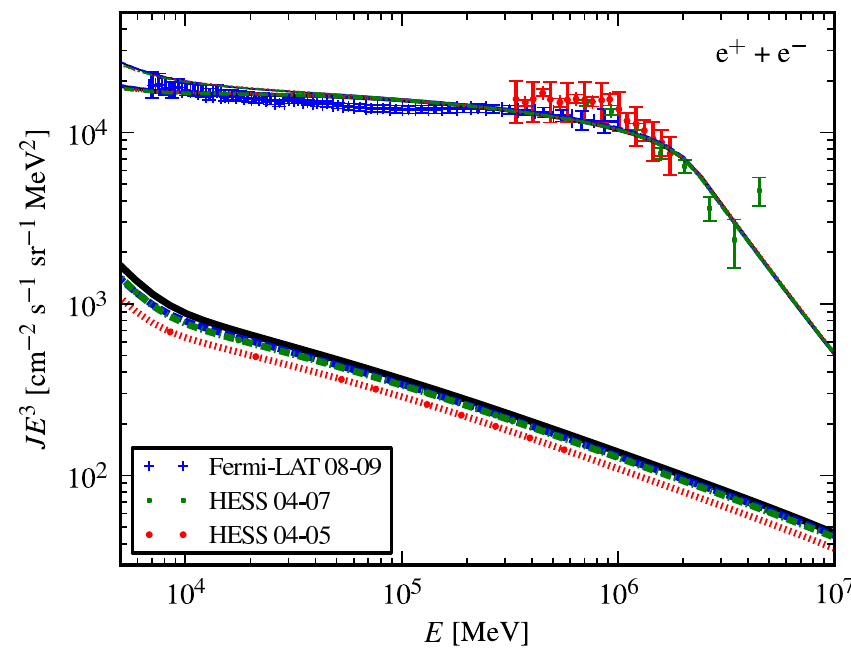}
\caption{Comparison of model \model{S}{4}{20}{150}{5} (black solid
curve), \model{L}{6}{20}{\infty}{5} (blue dashed
curve), \model{Y}{10}{30}{150}{2} (red dotted curve), and \model{O}{8}{30}{\infty}{2} (green dash-dotted curve) to CR observations.
Shown are protons (top left), helium (top right), electrons (bottom left), and
electrons and positrons (bottom right).
In addition to the data we used for the CR fit (see Section~\ref{sec:CRs}) we also show data from PAMELA \citep{2011PhRvL.106t1101A,Pamela:protons}, ATIC-2 \citep{2008ICRC....2...31W}, and CREAM \citep{2011ApJ...728..122Y}
Error bars for the x-axis indicate bin width and error bars for the y-axis
include systematic error.
Models are corrected for solar modulation with the appropriate modulation potential found
either from the CR fits and shown in Figure~\ref{fig:CRmodulation} or with the
fixed values given in Section~\ref{sec:CRs}.
Because \fermilat{} and \hess{} cannot discriminate between positrons and 
electrons the model total electron + positron spectrum is compared to the data.
The positron component of the model is shown as the thick line, which is 
corrected with a modulation potential of 300 MV.
The interstellar spectra are shown as the thin curves.
\label{fig:CRspectrumPrimary}}
\end{figure*}

\begin{figure}
\epsscale{0.53}
\plotone{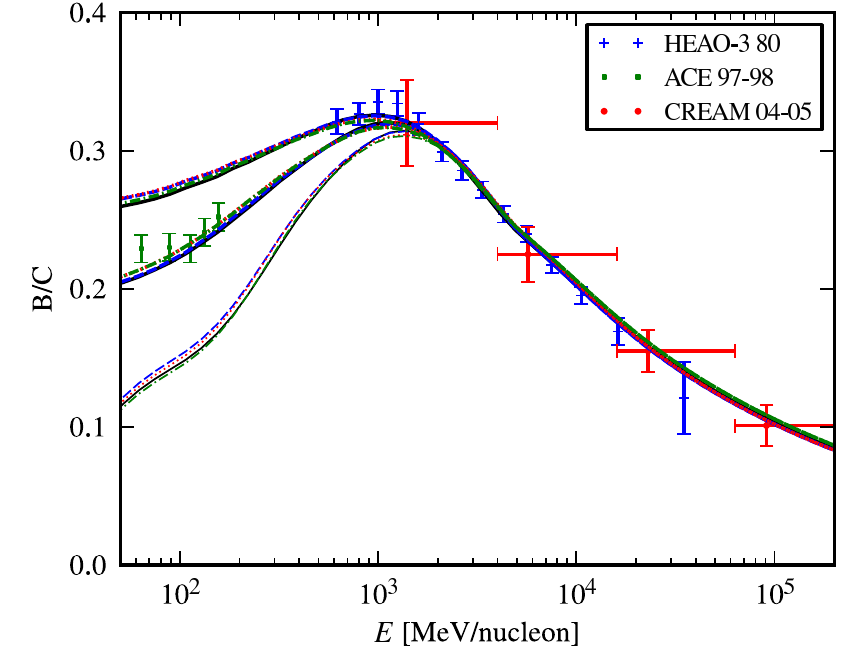}\\
\plotone{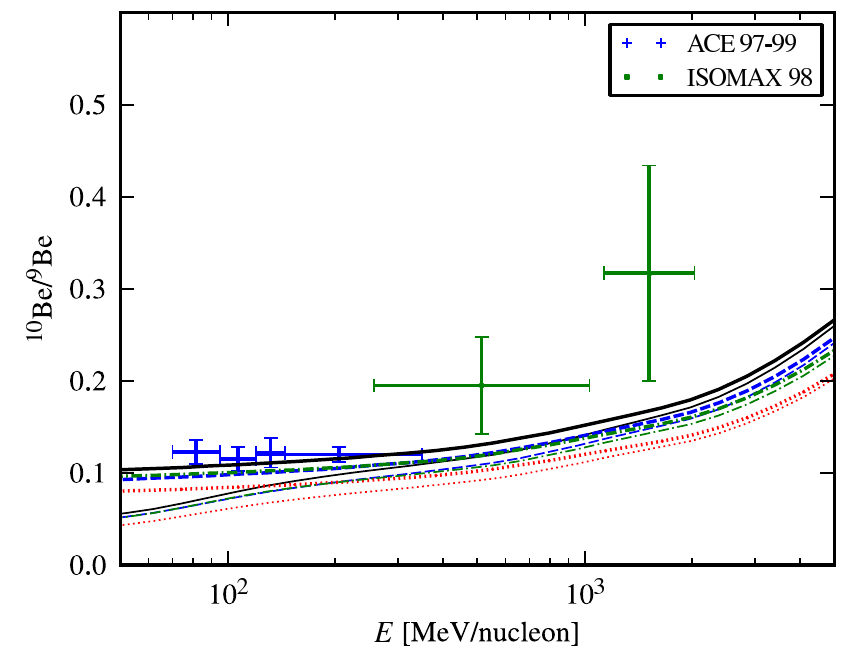}
\caption{Comparison of model \model{S}{4}{20}{150}{5} (black solid
curve), \model{L}{6}{20}{\infty}{5} (blue dashed
curve), \model{Y}{10}{30}{150}{2} (red dotted curve), and \model{O}{8}{30}{\infty}{2} (green dash-dotted curve) to
CR observations of B/C ratio (top), and $^{10}$Be/$^9$Be ratio (bottom).
In addition to the data we used for the CR fit (see Section~\ref{sec:CRs}) we also show data from CREAM \citep{2010ApJ...715.1400A}, ACE \citep{2001ApJ...563..768Y}, and ISOMAX \citep{2004ApJ...611..892H}.
Error bars for the x-axis indicate bin width and error bars for the y-axis
include systematic error.
Models are corrected for solar modulation with the modulation potential shown
in Figure~\ref{fig:CRmodulation}.
$^{10}$Be/$^9$Be ratio uses modulation for ACE.
For a direct comparison, we also show the interstellar spectrum of the 
components as thin curves.
\label{fig:CRspectrumSecondary}}
\end{figure}

\begin{figure}
\plottwo{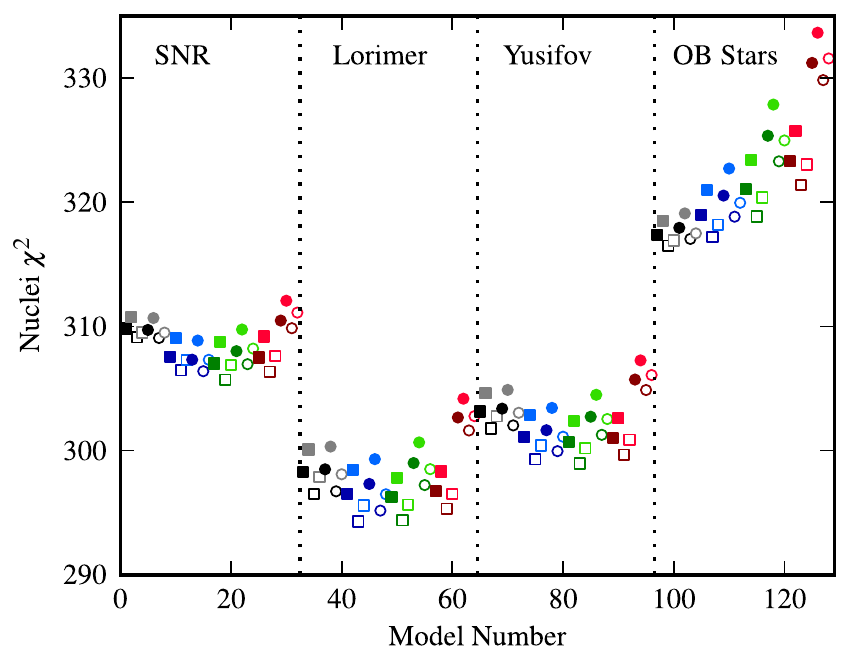}{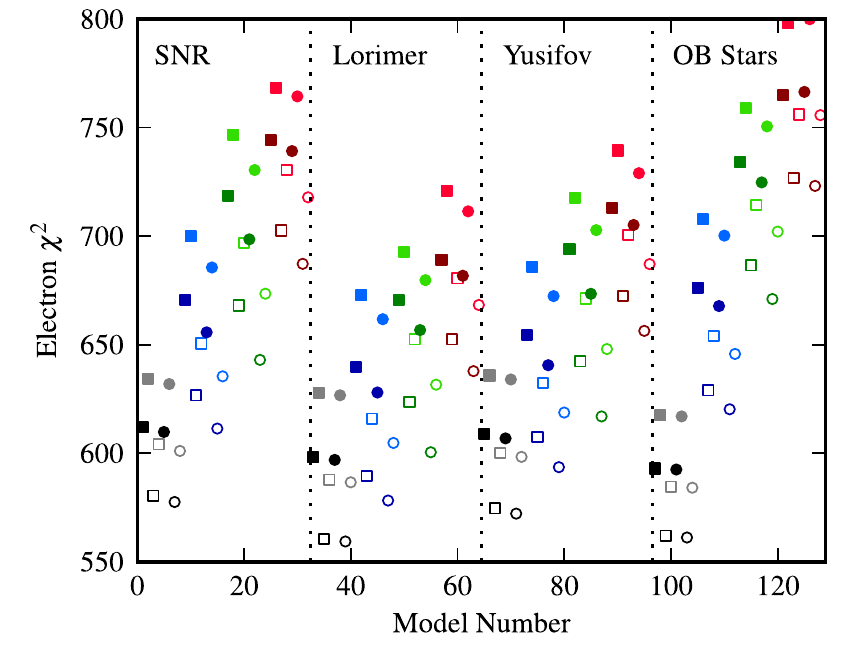}
\caption{Resulting $\chi^2$ values from our fit to the nuclei data (left) and electron data (right).
The numbers of degrees of freedom is 131 and 95 for the nuclei and electron fit respectively.
\label{fig:CRchisq}}
\end{figure}

The $\chi^2$ values resulting from the CR fit are shown 
in Figure~\ref{fig:CRchisq}. 
The nuclei part of the fit results in $\chi^2/{\rm d.o.f}\approx 300/131$ for 
both pulsar source distributions, increasing 
to $\chi^2/{\rm d.o.f}\approx 340/131$ for the OB star distribution. 
The largest contribution to the $\chi^2$ value comes from low-energy 
protons and high-energy nuclei.
Note that the $\chi^2$ value is not strongly dependent on the halo 
size and gas model, and all models provide a good representation of the 
nuclei data that were fitted.
This can be seen in Figure~\ref{fig:CRspectrumPrimary} and 
Figure~\ref{fig:CRspectrumSecondary} that compare CR observations to a 
few selected models.
Figures for all models are in the online supplementary material.

\begin{figure}
\epsscale{0.33}
\plotone{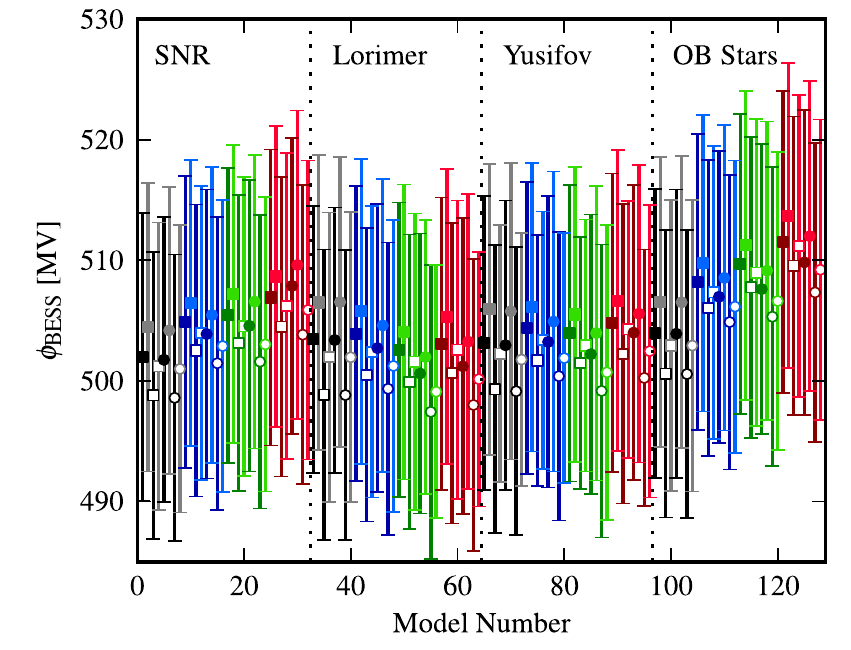}\\
\plotone{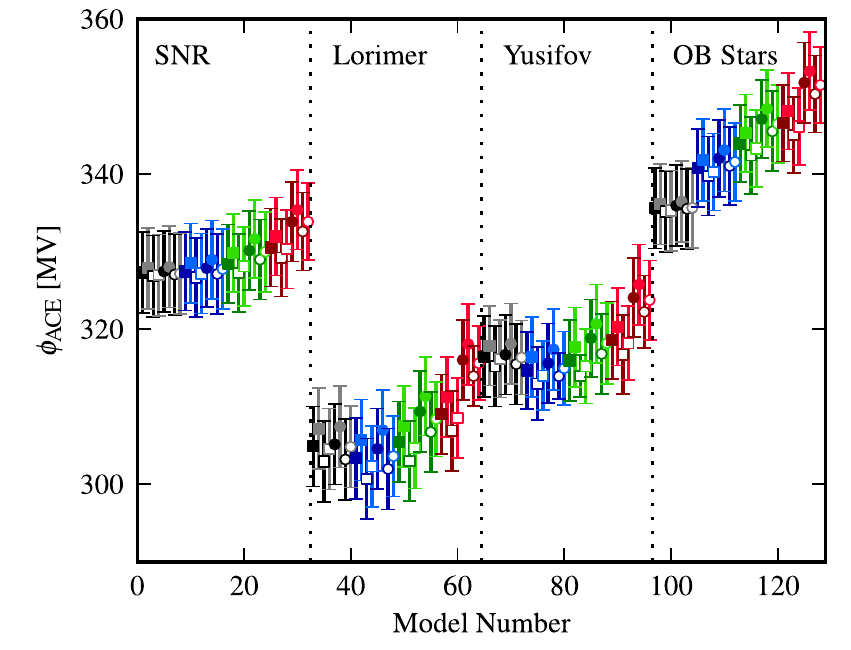}\\
\plotone{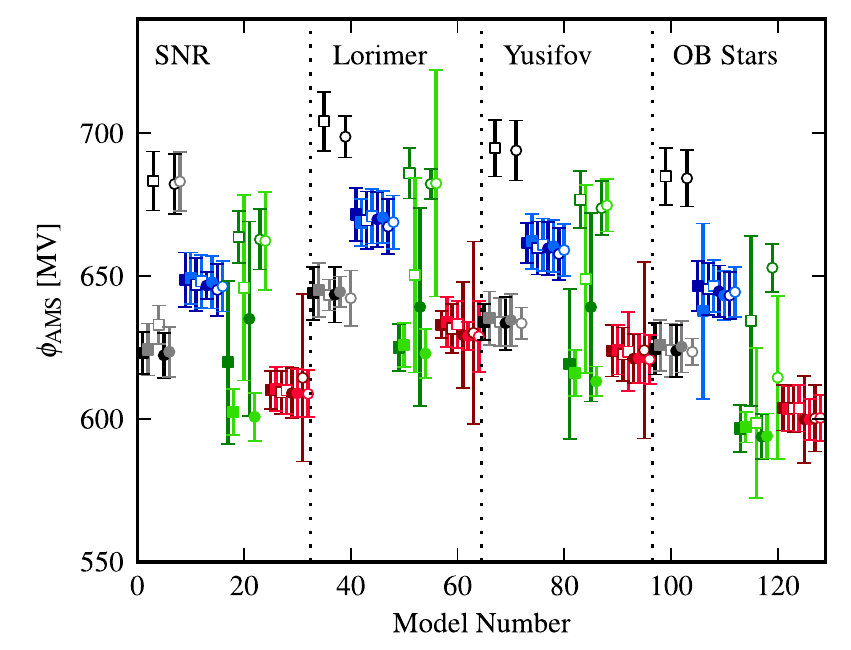}
\caption{The modulation parameter found from the best-fit to CR data.
Top: BESS, middle: ACE, bottom: AMS.
See Figure~\ref{fig:gammaLikelihood} for legend.
\label{fig:CRmodulation}}
\end{figure}

The models incorporated the effect of solar modulation using the force-field
approximation (see Section~\ref{sec:CRs}), with 
the value of the modulation potential determined in the fit 
for each of the instrument given in
Figure~\ref{fig:CRmodulation}.  Our results are compatible with other estimates of the modulation potential 
for the observational epochs considered \citep{CRISModulation}.

\begin{figure}
\plotone{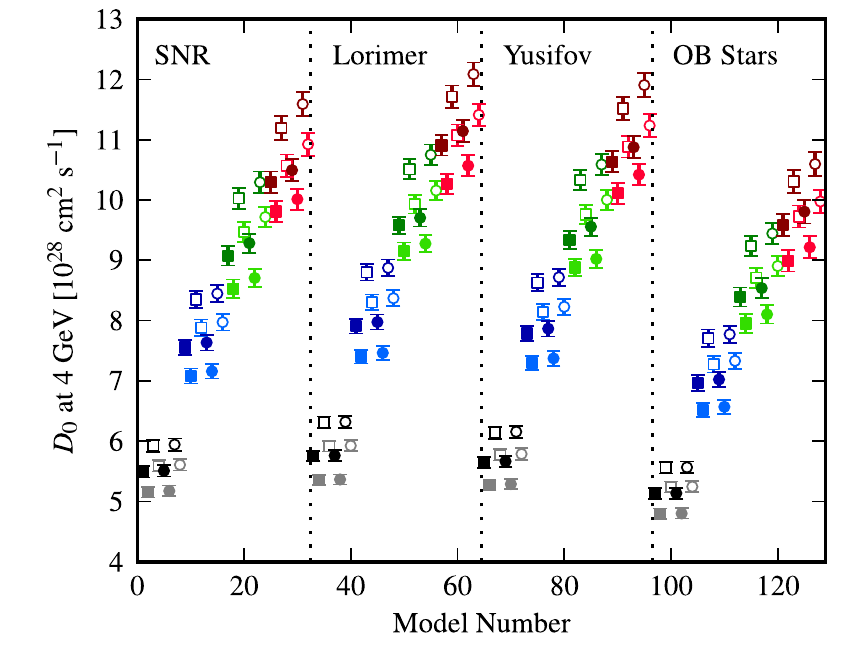}\\
\plotone{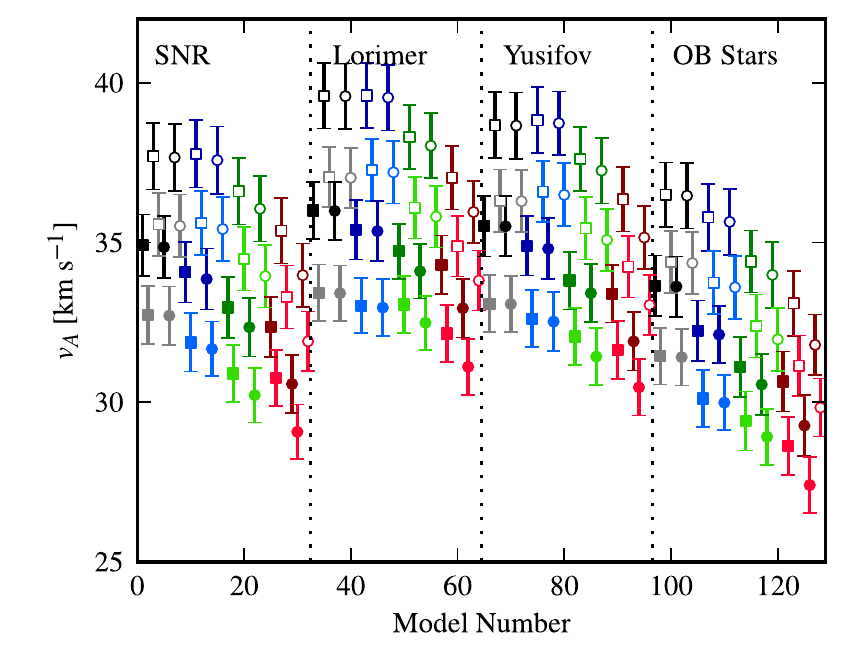}
\caption{The resulting propagation parameters from the fit to CR-nuclei data.
Top shows $D_0$ and bottom shows $\valf$.
\label{fig:CRparametersNuc}}
\end{figure}

The difference in $\chi^2$ for the different CR source distributions is 
largely due to differences at low energies, below $\sim 10$~GeV.
Note that there is a correlation between the modulation potential of the
ACE experiment given in Figure~\ref{fig:CRmodulation} and the 
nuclei $\chi^2$ value, and also between the nuclei $\chi^2$ 
and $\gamma_{p,1}$ shown in Figure~\ref{fig:CRparametersNuc}.
While the difference is statistically significant, the use of force-field 
approximation for the modulation and the constraints of the propagation model 
do not warrant model selection based on the CR fit.
The accuracy of the numerical solution of the propagation equation can
also make significant changes to the $\chi^2$ value.  For numerical
fitting of the CR spectrum we had to compromise between accuracy and
speed.  Note that the changes in the model prediction giving rise to the
differences in $\chi^2$ are very small, as demonstrated in
Figure~\ref{fig:CRspectrumPrimary}.

The electron fit is considerably worse, 
giving $\chi^2/{\rm d.o.f}\approx 650/95$ for $z_h=4$~kpc up to $\chi^2/{\rm d.o.f}\approx 850/95$ for $z_h=10$~kpc.
There is a strong dependence on both halo size and \Xco{} 
value\footnote[19]{The \ebv{} magnitude cut and $T_S$ 
value affect the propagation only through the \Xco{} value.}, 
which is expected because 
secondary electrons and positrons comprise a significant 
fraction ($\approx 50$\%) of the total at low energies.
A larger halo also increases the IC cooling of the 
electrons \citep{LuminosityPaper}.
The high $\chi^2$ values are mostly due to the large fraction of 
secondaries at low energies, making the AMS and \fermilat{} spectra 
incompatible, but also due to the convex shape of the observed \fermilat{}
electron spectrum.
This is expected because \citet{Ackermann:electrons} found that the
form of injection spectrum used in the current analysis does not reproduce all 
the features of the \fermilat{} total CR electron data.
The low-energy part could be improved by increasing the modulation 
potential for the \fermilat{} observations, but this is not well motivated 
given the low level of solar activity during the LAT sky survey observations analysed here.
Because the IC and bremsstrahlung components are subdominant 
in the energy range where the statistics for studying the DGE are the greatest, 
this does not affect our results significantly.

\begin{figure*}
\epsscale{0.93}
\plottwo{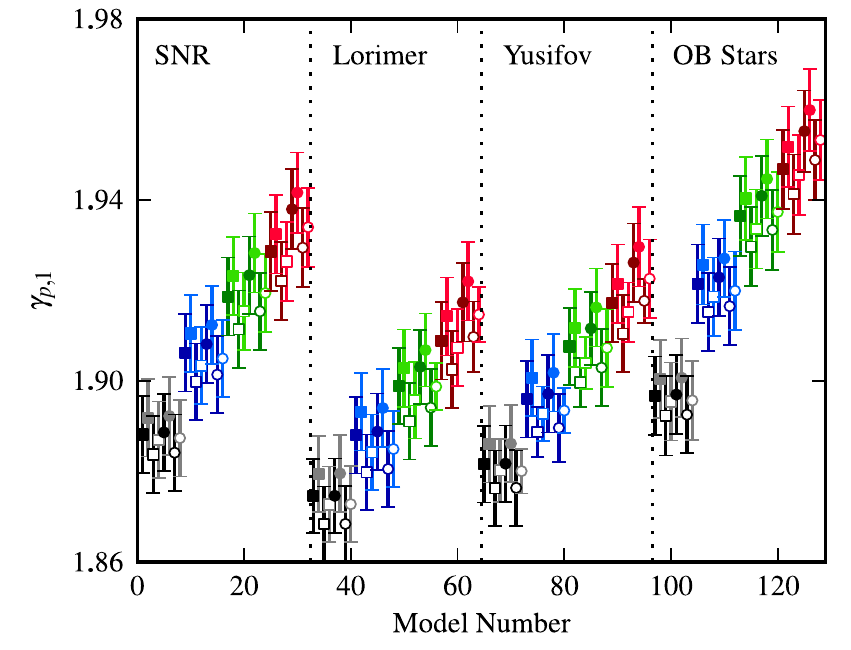}{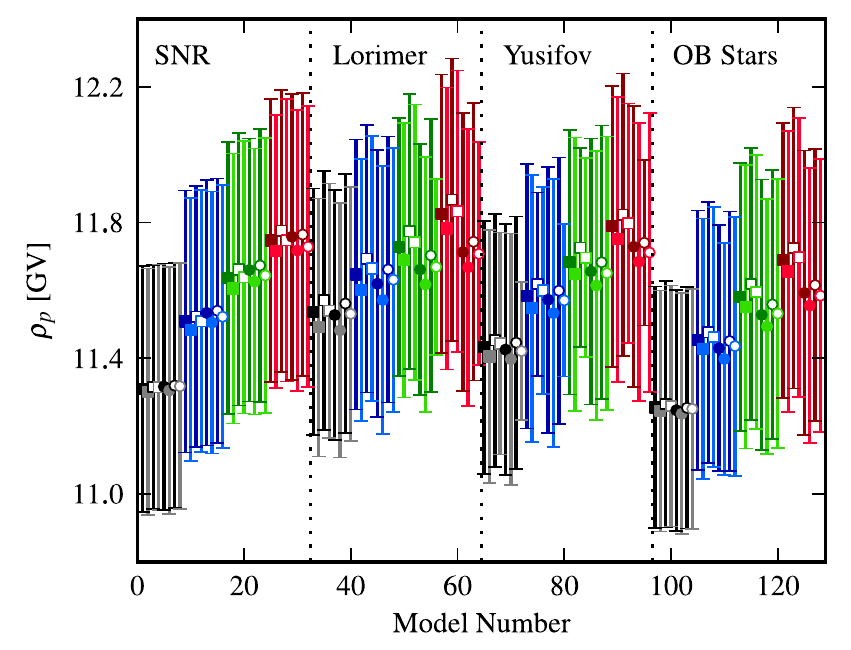}\\
\plottwo{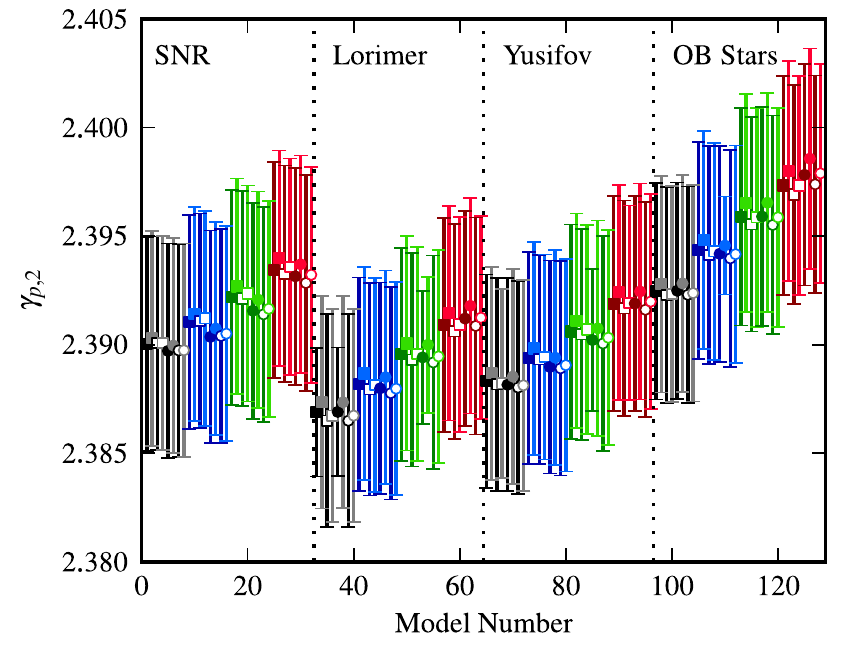}{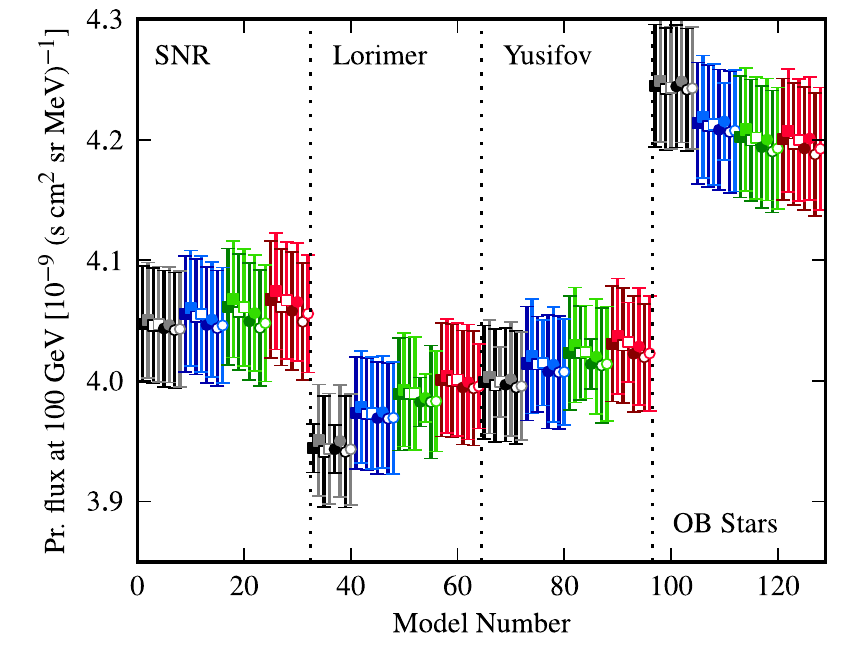}
\caption{The resulting propagation parameters from the nuclei fit.
Shown are low energy nuclei index (top left), high energy nuclei 
index (top right), nuclei
break rigidity (bottom left), and proton normalisation (bottom right).
See Figure~\ref{fig:gammaLikelihood} for legend.
\label{fig:CRparametersNuc2}}
\end{figure*}

\begin{figure}
\epsscale{0.93}
\plottwo{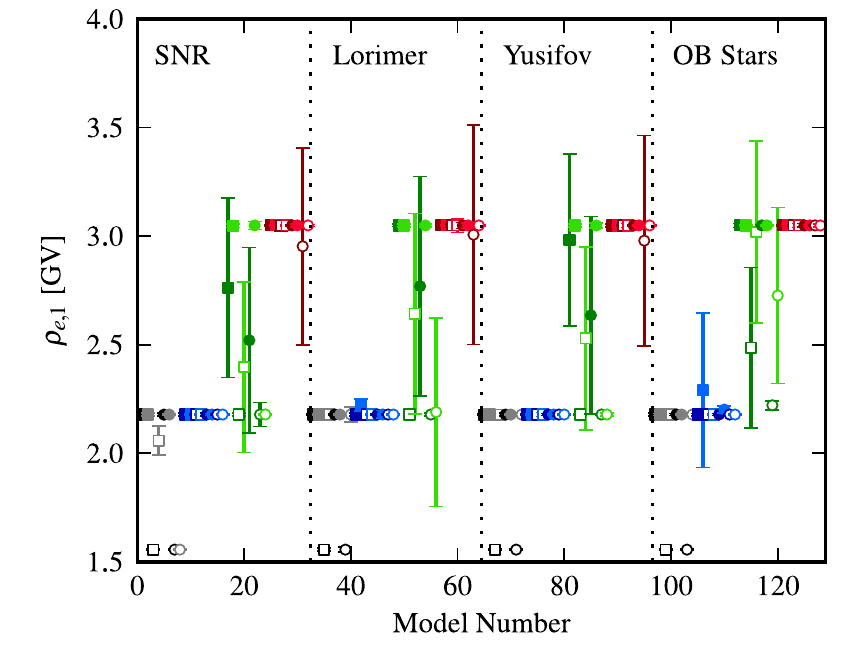}{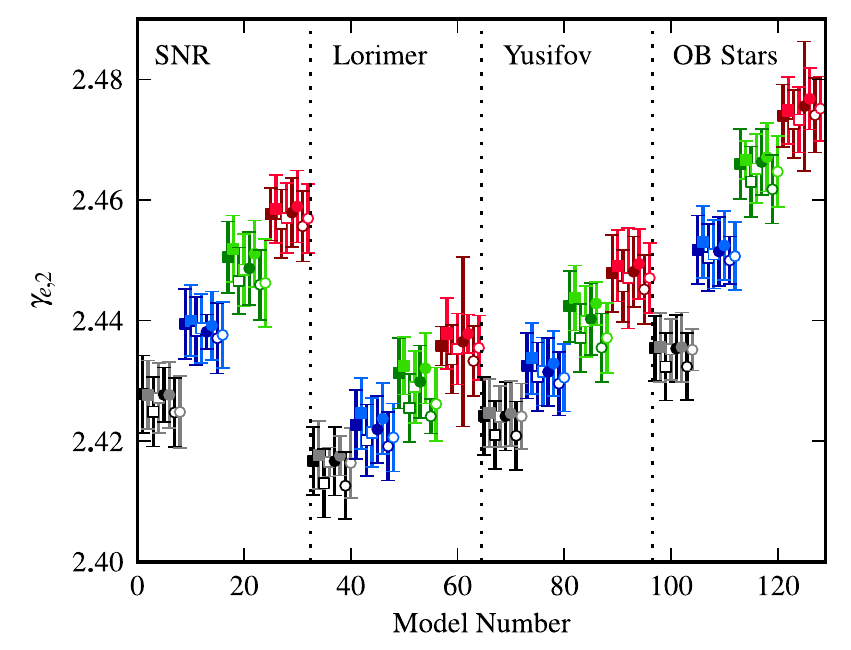}\\
\plottwo{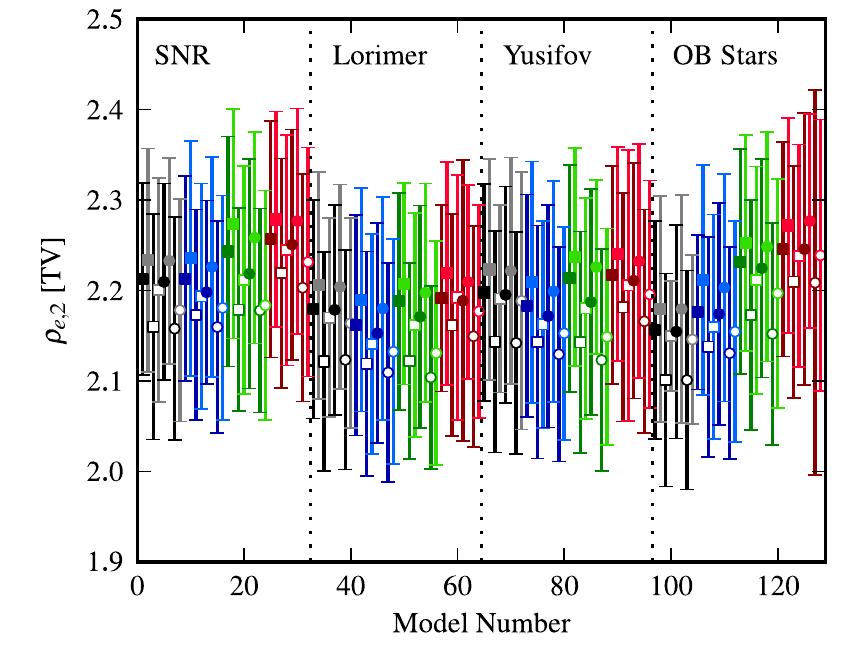}{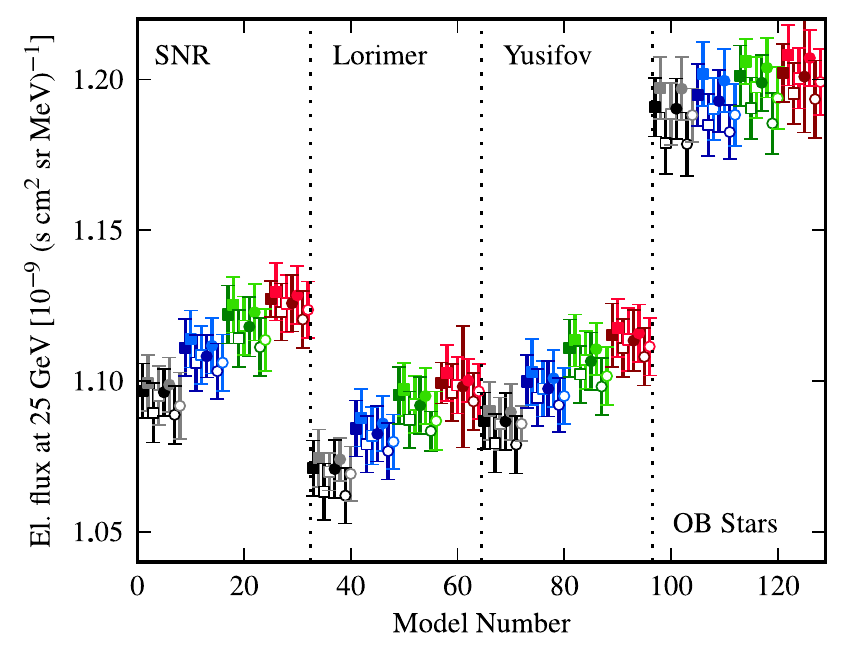}
\caption{The resulting propagation parameters from the electron fit.
Shown are low-energy break (top left), high-energy break (top right),
power-law index (bottom left), and electron normalisation (bottom right).
Note that the error bars on many of the low-energy break points are
underestimated, the minimiser seems to find a very narrow dip in the $\chi^2$
function, possibly associated with the discontinuity in the AMS data at $\sim
2$~GeV (see Figure~\ref{fig:CRspectrumPrimary}).
See Figure~\ref{fig:gammaLikelihood} for legend.
\label{fig:CRparametersEl}}
\end{figure}

Figure~\ref{fig:CRparametersNuc}, Figure~\ref{fig:CRparametersNuc2}, and 
Figure~\ref{fig:CRparametersEl} show the
resulting parameters from the fit to CR data.
Most of these are in agreement with parameters found in earlier studies 
of CR propagation \citep{2004ApJ...613..962S, Trotta2010}.
The main difference is the electron injection spectrum, which has now been 
measured accurately up to 1~TeV by the \fermilat{} \citep{2009PhRvL.102r1101A}.
The spectrum found in our analysis is now more akin to the one used in
the optimized model by \citet{2004ApJ...613..962S} although the break energy
at $\sim 3$~GV found in this analysis is much lower than the 20~GV break
assumed in the optimized model.  The proton spectrum is also
slightly different, with a break rigidity at $\sim 11.5$~GV instead of 9~GV in
\citet{2004ApJ...613..962S} and spectral indices of $\sim1.9$ and $\sim 2.40$
below and above the break, respectively, instead of 1.98 and 2.42 used in the
conventional model in 
\citet{2004ApJ...613..962S}.  Our proton spectrum is much closer to the one
determined in the more recent analysis by \citet{Trotta2010}.
Most of these differences can be attributed to a different CR source 
distribution and the inclusion of \Xco$(R)$ in the propagation calculation.
The values of the propagation parameters $D_0$ and $\valf$ agree reasonably
well with the older analysis if one chooses similar models for the comparison,
since the value of $D_0$ and $\valf$ are correlated with $z_h$ and the 
gas distribution.
Note that only the \Xco{} values were changed for the propagation; 
the \hi{} distribution is constant.
This causes an overestimate of the gas densities in the optically thin 
assumption, because the analytical model is not corrected for $T_S$ or 
dust variations, and \Xco{} compensates to a certain degree for the 
change in \hi{} column density in the \gray{} fit.
This overestimate enhances the variations in $D_0$ and $\valf$ but 
does not significantly affect the \gray{} analysis.
Note also that the injection spectrum softens with increasing $z_h$, both 
for nuclei and electrons, although the variations are barely larger 
than the error bars.
The error bars on the data are statistical only and no attempt has been 
made to assess the systematic error.
The largest systematic uncertainty is likely associated with the 
absolute energy scale of the data \citep{Ackermann:electrons}.  
This can potentially change the location of the break energies along with 
the intensity of the modelled spectrum.
A 10\% change in absolute energy would shift the normalisation 
by $\sim$30\%, which directly translates to a change in the intensity 
of the DGE model.

\clearpage

\clearpage

\begin{deluxetable}{rcccc}
\tablecaption{The gas-to-dust ratio determined from a linear fit to the \hi{} 
and CO component.
The \Xco{} is determined from the gas-to-dust ratios under the assumption 
that the dust-to-proton ratio is the same for both \hi{} and H$_2$.
See text for details.
\label{tab:gas-to-dustRatio}}
\tablehead{
\colhead{$T_S$\tablenotemark{a}}&
\colhead{\ebv{} cut\tablenotemark{b}} &
\colhead{\hi{} ratio\tablenotemark{c}} &
\colhead{CO ratio\tablenotemark{d}} &
\colhead{\Xco\tablenotemark{e}}
}
\startdata
150     & 2 & 74.37 & 19.52 & 1.91\\
150     & 5 & 73.00 & 21.87 & 1.67\\
100,000 & 2 & 61.39 & 21.13 & 1.45\\
100,000 & 5 & 59.99 & 23.78 & 1.26\\
\enddata
\tablenotetext{a}{In units of K.  $T_S = 10^5$~K is equivalent to the optically thin 
assumption.} 
\tablenotetext{b}{In units of mag.}
\tablenotetext{c}{In units of $10^{20}$ cm$^{-2}$ mag$^{-1}$.}
\tablenotetext{d}{In units of K (km s$^{-1}$) mag$^{-1}$.}
\tablenotetext{w}{In units of $10^{20}$ cm$^{-2}$ (K (km s$^{-1}$) )$^{-1}$.}
\end{deluxetable}

\begin{deluxetable}{lcc}
	\tablecaption{Boundaries of Galactocentric annuli used in gas maps.
\label{tab:AnnuliBoundaries}}
\tablehead{
\colhead{Annulus} & 
\colhead{$R_{\rm min}$} & 
\colhead{$R_{\rm max}$}\\
\colhead{\#} & 
\colhead{[kpc]} & 
\colhead{[kpc]}
}
\startdata
1   & 0   &  1.5  \\
2   & 1.5 &  2.0  \\
3   & 2.0 &  2.5  \\
4   & 2.5 &  3.0  \\
5   & 3.0 &  3.5  \\
6   & 3.5 &  4.0  \\
7   & 4.0 &  4.5  \\
8   & 4.5 &  5.0  \\
9   & 5.0 &  5.5  \\
10  & 5.5 &  6.5  \\
11  & 6.5 &  7.0  \\
12  & 7.0 &  8.0  \\
13  & 8.0 & 10.0  \\
14  &10.0 & 11.5  \\
15  &11.5 & 16.5  \\
16  &16.5 & 19.0  \\
17  &19.0 & 50.0  \\
\enddata
\end{deluxetable}

\begin{deluxetable}{rccccc}
\tablecaption{The mapping between model numbers (SSZZRTC+1) and model input
parameters.  SS stands for CR source distribution, ZZ for $z_h$, R for $R_h$,
T for $T_S$, and C for the \ebv{} magnitude cut.
\label{tab:numberMapping}}
\tablehead{
\colhead{Value}&
\colhead{SS} &
\colhead{ZZ} &
\colhead{R} &
\colhead{T} &
\colhead{C} 
}
\startdata
00 & SNR      & 4~kpc  & 20~kpc & 150~K          & 2 mag \\
01 & Lorimer  & 6~kpc  & 30~kpc & Optically Thin & 5 mag \\
10 & Yusifov  & 8~kpc  &        &                &       \\
11 & OB stars & 10~kpc &        &                &       \\
\enddata
\end{deluxetable}

\end{document}